# Public Health Insurance of Children and Maternal Labor Market Outcomes

Konstantin Kunze*

April 15, 2025

**Abstract**

This paper exploits variation resulting from a series of federal and state Medicaid expansions between 1977 and 2017 to estimate the effects of children's increased access to public health insurance on the labor market outcomes of their mothers. The results imply that the extended Medicaid eligibility of children leads to positive labor supply responses of single mothers and to negative labor supply responses of married mothers. The analysis of mechanisms suggests that extended children's Medicaid eligibility positively affects take-up of Medicaid and health of children.

---

*Postdoctoral Scholar, Department of Economics, University of Rochester; kkunze@ur.rochester.edu.
I am deeply indebted to Marianne Bitler and Marianne Page for their invaluable support and guidance. This paper has benefited from the comments and feedback of Gonzalo E. Basante-Pereira, Matthew Bombyk, Colin Cameron, Scott Carrell, Grace Cooper, Sarah Flood, Christian vom Lehn, Santiago Pérez, Monica Singhal, Geoffrey Schnorr, Ivan Strahof, Jenna Stearns, Kari Williams, and seminar participants at the University of California Davis. I thank Sarah Cohodes, Ammar Farooq, Linda Giannarelli, Tal Gross, Xavier Jaravel, Adriana Kugler, Lara Shore-Sheppard, Kendall Swenson, Likun Tian, and Laura Wherry for sharing data used in some analyses of this project or insights about institutional knowledge.

# 1 Introduction

Medicaid – one of the largest government programs in the United States – provided public health insurance to 37.5% of all children ages 0 to 18 in 2019 (CPS, 2022). The literature demonstrates that access to public health insurance during childhood leads to positive short- and long-term effects on their outcomes (Buchmueller et al., 2016). Existing work also documents possible family spillovers of child's access to health insurance on parental outcomes (Aouad, 2021). Hence, it is natural to ask whether and to what extent children's access to Medicaid can affect parental outcomes.

In this paper, I answer this question and study the effects of extended child Medicaid eligibility on labor market outcomes of mothers between 1977 and 2017. During the analysis period, Medicaid expansions increased access to health insurance service for many low-income children. However, there was substantial variation in Medicaid eligibility by state, year, and age of children. I exploit this variation using the simulated eligibility strategy first developed by Currie and Gruber (1996a,b). This approach only uses variation in the generosity of eligibility for public health insurance laws to determine eligibility for a common national sample of children. Thus it abstracts away from characteristics of the children or family that may be correlated with both Medicaid eligibility and the outcome of interest. To account for the eligibility which can vary by age for each child within each family, I use a family-level and marital-status-specific simulated eligibility measure.

The main analyses are based on the Annual Social and Economic Supplement to the Current Population Survey (CPS) from 1977 to 2017. Using these data, I estimate the effect of children's access to Medicaid on the contemporaneous labor market outcomes of mothers (hours worked per week, weeks worked per year, labor force participation, earned income, occupational choice). To understand the mechanisms of estimated effects on parental labor market outcomes, I examine Medicaid take-up and health of all children in the family as well as marital outcomes of mothers. I also provide a back of the envelope cost-benefit calculation for the Medicaid expansions which encompasses costs associated with Medicaid expansions, the net tax liabilities, and the cost of altered social safety net programs. This adds the effects generated by the spillovers onto mother's labor supply and marriage and implications for previous cost-benefit analyses focused on children alone.

The relationship between children's access to Medicaid and parental labor market outcomes is important from the perspective of the well-being of the children and from the perspective of policy makers weighting costs and benefits of program changes. Children's well-being crucially depends on material resources and time investments from their parents. Since parents invest financial means and their time into raising their children, parental labor supply is an important factor in the cognitive and physical development of children (Heckman and Mosso, 2014). From the perspective of the policy maker, on the one hand, increased labor supply can recoup some of the costs associated with implementation of Medicaid through increased tax revenue and on the other hand, negative labor supply responses may have implications for program design.

Empirical analysis is necessary since there are many potential mechanisms through which children's Medicaid eligibility can affect the labor market outcomes of their parents in different and unclear ways. On the one hand, access to Medicaid can improve the health of children and hence lead to increased parental labor supply (Eriksen et al., 2021). On the other hand, extended Medicaid eligibility, which effectively translates into an increase in income, can result in reduced parental labor supply because parents face lower financial burden associated with uninsured children or out-of-pocket expenses for private insurance (Gross and Notowidigdo, 2011). Given that Medicaid is a means-tested program, parents might also solely adjust the intensive margin to qualify for Medicaid coverage of their children.[1] The incentives for married and single mothers are vastly dif-ferent because married mothers have access to partner's income and single mothers rely on their income. Therefore the same changes in children's Medicaid coverage may affect labor supply decisions of married and single mothers differently.

[1] Pei (2017), however, does not find evidence for strategic labor supply adjustments of parents in order to gain Medicaid eligibility for their children.

I first show that expanding Medicaid eligibility to children affects contemporaneous maternal labor market outcomes differently for single and married mothers. Standard labor supply measures of single mothers increase contemporaneously by 3% to 6% relative to baseline average labor supply as a result of making one additional child per family eligible for Medicaid. The elasticities of labor supply responses range between 0.03 and 0.06. In contrast, married mothers decrease labor force participation by 2% relative to the baseline mean with elasticity of point estimate around 0.01. The earnings effects for single and married mothers are concentrated in the lower end of the earnings distributions, consistent with Medicaid being targeted at low-income families. If I assume that the only channel through which the expansions affect earnings are through the increases in maternal labor supply generated by the expansions, I can use the estimated changes in take-up by children to generate a treatment-on-treated (TOT) estimate of the change in earned income. This involves dividing my intent-to-teat (ITT) labor supply results by take-up estimates and taking into account changes in simulated eligibility. The TOT estimates are 1.5 times bigger than the ITT estimates and represent at least half of the food spending in low income families shown in the literature Page and Kantor (2025).

While increased labor supply of single mothers is driven by increased work in managerial, professional, technical, sales, and administrative occupations; exit from the labor force of married mothers is driven by a reduced employment in managerial, professional, service, and manufacturing occupations. Using different measures of occupational mobility, I also show that there is no clear relationship between maternal job-to-job transition and extended Medicaid eligibility of children.

The back of the envelope calculation of Medicaid's return on investment suggests that the labor supply responses alone cannot offset the costs of expanding Medicaid since the federal and state benefit payments exceed the federal and state tax revenues. The reduction in governmental payments for public assistance, education assistance, school lunch, and energy subsidy is able to compensate for roughly 33% of costs associated with extending Medicaid to one additional child.

I then provide empirical evidence on effects of children's Medicaid coverage and health. These estimates are important in understanding the mechanism for changes in parental outcomes, because expanded eligibility translates into improved health of children or reduced financial distress through a corresponding increase in program take-up. The results suggest that extended child Medicaid eligibility leads to a family-level marginal take-up rate of 32% for single and 41% for married mothers. I also find that more children in the family report excellent health and fewer children report good, fair, and poor health as a results of extending Medicaid eligibility. The magnitude of the increase in number of children reporting excellent health is 15% (5%) relative to the baseline mean in one-parent (two-parent) families.

Using a sample of one year panels from linking individuals across interviews 1-4 and 5-8 of the CPS with baseline marital status and estimates from existing literature on effects of marital outcomes on labor supply, I show that changes in marriage are not likely to bias estimates on maternal labor market outcomes. This finding alleviates the endogeneity concern of using maternal marital status in the construction of simulated eligibility and analyzing single and married mothers.

This study contributes to three strands of literature. First, it connects to the literature analyzing the effect of child's access to Medicaid on labor supply responses of parents. Earlier work by Ham and Shore-Sheppard (2005a) replicated the work of Yelowitz (1995) by incorporating important institutional features and estimated imprecise effects on parental labor supply using the eligibility of youngest child in the family. In contemporaneous work focusing on maternal well-being and using the simulated eligibility approach, Grossman et al. (2025) accounts for the Medicaid eligibility of all children in the family and document positive effects on marital status and negative effects on labor force participation.[2] I also advance the literature in several ways. Using a larger data set and more cohorts, I focus on the eligibility of all children in the family and analyze all possible labor market outcomes such as extensive and intensive margin responses as well as occupational choice. Finally, I also examine program's return on investment by incorporating maternal tax payments and participation in social programs which has been omitted from cost-benefit analysis in studies analyzing children's Medicaid expansions.

Second, this study complements the literature on child Medicaid eligibility and parental occupational mobility. To date few studies have examined the relationship between public health insurance of children and occupational mobility of their parents (Bansak and Raphael 2008; Farooq and Kugler 2020). These studies find that expanded access to Medicaid for children is associated with higher separation rates and occupational mobility towards better paid jobs. Incorporating the recent advancements in the literature on occupational mobility (vom Lehn et al., 2022), I show that there is no systematic relationship between increased Medicaid generosity for children and maternal occupational mobility.

Third, this study adds to literature that employs the simulated eligibility approach.[3] There is a substantial difference across the studies in the method used to construct simulated eligibility, which can be summarized based on two main criteria: type (e.g., fixed or annual simulated eligibility) and structure (e.g., state, year, and age or state, year, age, and marital status). I show that the results are not sensitive to the choice of the simulated eligibility type, but may be sensitive to the structure of the eligibility measure. Moreover, the analysis demonstrates the importance of the correct simulated eligibility structure when the goal is to capture group-specific effects and eligibility is likely to be different

[2] In comparison to my paper, they focus on different cohorts (parents born between 1957 and 1964) because they rely in part on data from the National Longitudinal Survey of Youth – 1979 Cohort.

[3] Among others, Currie and Gruber (1996a,b); Cutler and Gruber (1996); Zavodny and Bitler (2010); Gross and Notowidigdo (2011); Cohodes et al. (2016); East et al. (2023); Brown et al. (2019); Miller and Wherry (2019).

across these groups (e.g., marital status).

The rest of the paper is structured as follows. Section 2 provides the history and evolution of the Medicaid program. I introduce the simulated eligibility measures and the empirical approach in Section 3. Data sources and sample selection are described in Section 4. Section 5 discusses the results, and Section 6 provides the conclusions.

## 2 History of the Medicaid Program

Medicaid is a joint state and federal program that was signed into law in 1965 as Title XIX of the Social Security Amendments.[4] In the beginning of the analysis period Medicaid eligibility for non-disabled children was originally restricted to single-parent families receiving cash welfare payments under the Aid to Families with Dependent Children (AFDC) program or eligibility under three additional state optional programs.[5] The income eligibility thresholds varied by state and family size, most of which were below the federal poverty line (FPL). These stringent eligibility requirements meant that only a few children of working mothers were eligible for Medicaid and if a women was to leave welfare her child would not be covered by Medicaid. Hence, in order for children to remain eligible for Medicaid, mothers were given the incentive not to participate in the labor force and cut their working hours.

In the mid-1980s legislation started to gradually separate Medicaid and AFDC by expanding eligibility to children not qualifying for AFDC. Initially Medicaid eligibility was extended for children under five years of age who were born after September 30, 1983 and who were living in families that met the financial, but not the family structure require-ments for AFDC through the 1984 Deficit Reduction Act. Omnibus Reconciliation Act (OBRA) 1986 and 1987 as well as Medicare Catastrophic Coverage Act and Family Support Act further weakened the link between Medicaid and AFDC by allowing and requiring states to increases the income limits for Medicaid eligibility for children belonging to certain age and birth cohorts.

[4]The focus of this review is legislation targeted at the eligibility of children between 1977 and 2017. Table B.1 provides an overview of federal and state legislation for this period. Sources for this section include Gruber (2003) and Buchmueller et al. (2016). See appendix section B for a detailed explanation of legislative changes during the analysis period.

[5]AFDC-Unemployed Parent (AFDC-UP) program extended Medicaid eligibility to families with an unemployed primary earner, the Ribicoff Children program provided Medicaid eligibility to children who did not meet the family structure requirements but met the income and resource requirements for AFDC, and the Medicaid's Medically Needy program extended Medicaid eligibility to those with high medical expenses whose income exceeded the maximum threshold but family structure satisfied the AFDC requirements.

Additional federal expansions were embedded in OBRA 1989 and 1990 - until then the largest expansions in US history. OBRA 1989 expanded Medicaid eligibility to pregnant women and children up to age six with family incomes below 133% of FPL and OBRA 1990 required states to cover children born after September 30, 1983 with family incomes below the FPL. These children remained eligible until the age of 18. By introducing Temporary Assistance for Needy Families (TANF) program, Personal Responsibility and Work Opportunity Reconciliation Act (PRWORA) of 1996 removed the link between AFDC and Medicaid completely since TANF did not provide Medicaid eligibility automatically. However "Section 1931 eligibility" required states to cover families that would have been eligible under AFDC before the welfare reform. The next milestone in the evolvement of the Medicaid program was the Balanced Budget Act (BBA) in 1997. BBA created the State Children's Health Insurance Program (SCHIP), allowing states to cover uninsured children in families ineligible for Medicaid and providing continuous coverage for up to twelve months regardless of increases of child's family income.

The policy changes between 1977 and 2017 had a large effect on Medicaid eligibility of children. Figure 1 documents that the fraction of eligible children increased substantially from 0.12 in 1979 to 0.46 in 2017 with the biggest increase in eligibility around the rollout of OBRA 1989 and 1990 as well as the introduction of SCHIP in 1997. The Medicaid expansions affected marital-status groups differently. Throughout the analysis period, the fraction of children with single mothers eligible for Medicaid is higher than the fraction of children with two parents. Moreover, the increase in eligibility of children with single mothers was stronger after legislations targeted at lower-income families (e.g., OBRA 1990). The difference in eligibility between marital-status groups underlines the importance to use a marital-status-specific eligibility measure in the heterogeneous analysis.

# 3 Methodology

## 3.1 Simulated Eligibility

Sociodemographic characteristics can affect the number of children who are eligible for Medicaid independent of legislative changes as well as outcomes of children and their parents resulting in an endogenous measure of children's actual Medicaid eligibility. For instance, improved economic conditions may increase average income for certain groups of the population and hence reduce the number of children who are income eligible for public health insurance. At the same time, changes in economic environment may also affect outcomes of parents and their children. To address this type of potential endogeneity, I follow the simulated eligibility approach first developed by Currie and Gruber (1996a,b) and Cutler and Gruber (1996). The goal of the simulated eligibility strategy is to create a measure of eligibility abstracting from omitted variables that may be correlated with both children's actual eligibility and parental or child outcomes, so that identification is based only on legislative variation.

Since the primary focus is on parental outcomes, one has to account for the eligibility of each child in the family. Using eligibility of only one child in a family might underestimate the effects of extended Medicaid eligibility because having multiple children in the family eligible for Medicaid might affect parental outcomes in a way that is not fully captured by the eligibility of a single child. To construct family's total simulated eligibility, I first construct child-level eligibility measure.[6] To construct the child's own simulated eligibility measure, I use all children of age 0-18 in each year of the analysis period. Using this national data set, I calculate the child-level simulated eligibility as the fraction of eligible children in each state, year, age, and marital status group by leaving out children from the state for which the eligibility is being imputed.[7] To obtain the family-level eligiblity measure, I sum the simulated eligibility fractions across all children in a family which mimics the number of eligible children in a family. The total simulated eligibility measure therefore ranges from 0 to the maximum number of children in a family and is on average 0.65 eligible children per family.

[6] Appendix B describes in detail the construction of simulated eligibility measures, how eligibility is imputed, and the legislative rules used to impute eligibility. Medicaid eligibility is determined using calculators from East et al. (2023).

Marital status plays an important role in the analysis of Medicaid expansions. On the one hand, children with single and married mothers might respond differently to the same level of simulated eligibility, which can be captured by analyzing heterogeneous responses across marital status groups. On the other hand, one- and two-parent families are distinguished by different levels of simulated eligibility because of systematic differences in characteristics relevant for eligibility determination resulting in a measurement error in simulated eligibility if not accounted for. Figure 4 explains the importance of a marital-status-specific simulated eligibility measure for subgroup analysis. Subfigure (a) shows that the marital-status-specific simulated eligibility traces out the actual eligibility very well for both groups. On the other hand, the non-marital-status-specific simulated eligibility is not close to the actual eligibility of any group shown in subfigure (b). The group-specific simulated eligibility measure addresses this concern by allowing the Medicaid eligibility to vary by marital status.[8] A potential critique of using marital status to construct the simulated eligibility is that marital status is not a fixed characteristic like race and ethnicity but can be affected by expanded eligibility and is also a determinant of Medicaid eligibility.[9] In section 5.4.1, however, I show that marriage responses to expanded Medicaid eligibility is unlikely to bias the estimates.

[7]The child-specific simulated eligibility measure is in spirit of earlier literature that examines the effects of child Medicaid eligibility on child's health insurance coverage and health outcomes (e.g., Cutler and Gruber 1996; Currie and Gruber 1996a,b). The main difference to the eligibility measure used in these studies is the marital status component.

[8]This argument applies to analysis of other subgroups as well. For instance, if the focus of the analysis is to estimate differential effects by race and ethnicity, the the simulated eligibility measure should be constructed by using race and ethnicity.

[9]Marital status was an important eligibility criteria for AFDC primarily in the beginning of the analysis period before PRWORA (see section B.1).

## 3.2 Empirical Approach

I estimate the effects of increased Medicaid eligibility on outcomes of children (insurance coverage and health status) and labor market outcomes of their mothers by running a mother-level regression and regressing the outcome of interest on simulated eligibility measure as well as a set of controls.[10] Specifically, I begin by estimating the model of the following functional form:

$$y_{jstm} = \beta_0 + \beta_1 SIMT_{jstm} + X'_{jstm}\beta_2 + Z'_{st}\beta_4 + \delta_s + \gamma_t + \tau_y + \tau_o + \tau_d + \varepsilon_{jstm} \quad (1)$$

where the dependent variable is outcome of parent *j*, in state *s*, calendar year *t*, and marital status group *m*. $SIMT_{jstm}$ is the total simulated Medicaid eligibility - the treatment variable of interest. From the policy maker's perspective, the coefficient $\beta_1$ on total simulated eligibility captures the effect of an additional child in the family becoming eligible for Medicaid. Equation 1 includes state of residence fixed effects ($\delta_s$) which capture fixed differences in outcomes of parents and their children across states and calendar year fixed effects ($\gamma_t$) to account for potential national changes over time. I also include indicators for age of the youngest child ($\tau_y$), age of the oldest child ($\tau_o$), and difference between youngest and oldest child in the family ($\tau_d$) to account for fixed differences in outcomes of parents and their children across children's ages. Alternative specification add interactions between state, year, and children's age fixed effects. The vector of parent-level control variables, $X_{jstm}$, contains indicators for marital status, race and ethnicity, parental age, and number of children in the family.[11] In the baseline model, I follow the literature and include a vector of annual state-specific economic and policy character-istics, $Z_{st}$, which incorporates the unemployment rate, the minimum wage, inflation-adjusted maximum welfare benefits for a family of 4, state-level EITC amounts measured as a percentage of the federal EITC, implementation of six types of welfare waivers, and implementation of any waiver or TANF. In specifications for the full sample, all control variables and fixed effects are interacted with a marital status indicator. The regression is weighted by maternal weights and standard errors are clustered at the state level.

[10]This approach may be problematic if children's Medicaid eligibility affects parental fertility. Existing evidence, however, suggests that access to Medicaid does not affect fertility (e.g., Zavodny and Bitler 2010; DeLeire et al. 2011; East et al. 2017).

[11]As explained in section 3.1 marital status is not an endogenous control in this context. Additionally, the estimated effects from models including and excluding marital status as a covariate are very similar.

The specification shown in equation 1 exploits plausibly exogenous variation that results from the state and federal Medicaid expansions during the analysis period. There are three main sources of variation - across states because of state differences in AFDC eligibility limits prior to the expansions and difference in state's implementation of optional expansions, over time as the expansions were implemented with different pieces of legislation, and across children's age since younger children are more likely to be eligible.[12] Figures 2 and 3 summarize the underlying variation. Figure 2 shows the difference in total simulated eligibility between 1977 and 2017 for each state for single-child and multiple-child families. While Medicaid eligibility increased over time, there is substantial heterogeneity across states. In some states Medicaid expansions resulted in an average increase in simulated eligibility on the order of two simulated eligible children per family whereas in other states on the order of only 0.2 simulated eligible children per family. This pattern is quite similar across single-child and multiple-child families implying that differences in the number of children per family is not driving the difference in simulated eligibility. Figure 3 shows the national trend in simulated eligibility between 1977 and 2017 by child's age for single-child and multiple-child families. All age groups saw a substantial increase in simulated Medicaid eligibility between 1977 and 2017. While families with younger children were mostly affected during the first half of the analysis period, families with older children were affected during the second half of the analysis period suggesting that the Medicaid program became more generous for older children over time.

[12] Changes in children's characteristics are also contributing to the identifying variation since the simulated eligibility measures are constructed using data from the year for which the eligibility is imputed. Results are, however, robust to alternative eligibility measures that abstract from this possibly non-exogenous part of the identifying variation (see section 5.4.5).

Two general difference-in-difference-in-difference identifying assumptions in equation 1 are invoked for the validity of the empirical approach. The first identifying assumption is that no shock differentially affects Medicaid generosity and outcomes of children and their parents in the same state, during the same year, and with the same number of children of the same age. Hence omitted variables specific to parents with the same number of children of the same age and state of residence that change over time and are correlated with both Medicaid legislation and outcomes of children or their parents would invalidate this empirical strategy. The second identifying assumption requires that public health insurance eligibility rules are not set based on outcomes of parents and their children. The simulated eligibility approach will therefore fail if states phase in Medicaid expansions because of changing trends in parental or child outcomes. I discuss both identifying assumptions in section 5.4.2.

# 4 Data

## 4.1 Current Population Survey

To analyze contemporaneous parental labor market responses, educational attainment, and marital outcomes, child's insurance coverage and health status, as well as to predict Medicaid eligibility I use data from survey years 1977 to 2018 of the Annual Social and Economic (ASEC) supplement to the Current Population Survey (CPS) obtained from the integrated public use microdata series (Flood et al. 2020). The CPS is a nationally representative survey interviewing approximately 60,000 households per month. Each household is interviewed for four months, left out for eight months, and then interviewed for four months before the household exits the survey. This rotation pattern allows to link households and create a short panel. The main analysis is based on the repeated cross section and uses only the second observation. In some supplemental analyses, I make use of the short panel and analyze the linked sample.

The ASEC supplement provides a comprehensive body of data containing information on individuals' demographic characteristics, employment, health insurance coverage, and income. CPS captures information on the family composition, educational attainment and demographic characteristics at the interview date, health insurance coverage at any time during the previous calendar year, income during the previous calendar year, and labor supply measures either with reference to last week or previous calendar year.[13] Hence, simulated eligibility is measured contemporaneously to the reference period of the outcome variable of interest. With respect to labor market responses, I analyze usual hours worked per week, labor force participation last week, weeks worked last year, annual earnings last year, and occupational choice last year.[14] Medicaid coverage and health of children is captured by number of covered children and number of children with different levels of health (excellent, very good, good, fair, poor) per family. In supplementary analysis, I examine parental educational attainment (no high school, high school, some college, college or more) and marital outcomes (married, never married, ever married, divorced).[15] To capture working age mothers with children residing in the household, the sample is restricted to children age 0-18 with parents age 20-64.

Table 1 shows summary statistics of the 785,390 women with children included in the main analysis sample and 232,775 included in the linked sample. In the cross-sectional sample, roughly 24% are single, 64% are white non-Hispanic, 13% are black non-Hispanic, 4% are Asian non-Hispanic, 16% are Hispanic, 2% are other race, and 49% completed high school or less. With 1.9 children per family, the fraction of mothers with children of different ages is very similar - roughly one third of the sample have children under age 6, age 6-11, or age 12-18. Around 17% are blow FPL, 39% are between 100% and 300% of FPL, and 44% of families are above 300% of FPL. With exception of number of eligible children, racial origin, and poverty status, single and married women with children are quite similar - single mothers are more likely to have more eligible children, be black and have income below FPL. The difference in number of eligible children is quite substantial. While single mothers have 1.21 eligible children per family, married mothers have 0.51 eligible children per family. With few exceptions, the cross-sectional and linked samples are quite similar. The linked sample consists of fewer Hispanic mothers, and families with few eligible children mainly due to higher number of older children and slightly higher income.

[13]When answering questions about health insurance coverage, some respondents might ignore the reference period and instead answer based on their status at the time of the interview (Klerman et al. 2009, Ziegenfuss and Davern 2011). The results are, however, very similar when the simulated eligibility is assigned as of last year or as of interview date.

[14]Hours worked per week and weeks worked per year include zeros. I mainly focus on labor supply as of previous year. However, labor force participation is only measured as of last week.

[15]CPS redefined the education variable from years of education to degree receipt in 1992. To attain comparable educational categories across the whole analysis period, I use the method proposed by Jaeger (1997).

## 4.2 Supplemental Data

I use supplemental data from various sources. State-level minimum wage, welfare benefits, and Earned Income Tax Credit (EITC) amounts are obtained from U.S. Department of Labor, Urban Institute, and Tax Policy Center respectively. State-level unemployment rate, Consumer Price Index, compensation of employees, and number of total nonfarm employees come from the U.S. Bureau of Labor Statistics. I obtain federal poverty guidelines and information about implementation of state welfare waivers from the Office of the Assistant Secretary for Planning and Evaluation. The source for state-level income per capita is the U.S. Bureau of Economic Analysis. Data on Medicaid spending comes from the Medicaid Statistical Information Statistics (MSIS) maintained by Centers for Medicare & Medicaid Services.[16]

[16]Since children's Medicaid cost is at the state and year level for all children under 21, I disaggregate the cost to obtain a parent-level measure. First, I allocate the aggregated cost to each age based on the fraction of eligible children of a given age and divide by the total population in a given state, year and age. The child-level per capita cost of each child in a family is then summed up to obtain total Medicaid cost per family. MSIS data is only available for the 1980-2012 period. Hence the sample that uses MSIS data is restricted to 1981-2013.

# 5 Results

## 5.1 Labor Market Outcomes

I begin by estimating the effect of extended eligibility of children on maternal usual hours worked per week, weeks worked per year, and labor force participation. Table 2 presents the estimated treatment effects of marital-status-specific total simulated eligibility on maternal labor market outcomes. For the full sample, the estimated effects on usual hours worked per week, labor force participation, and weeks worked per year are close to zero and not statistically significant at conventional levels. The sub-sample analysis, however, shows that the estimated effects for single and married mothers are statistically significant at conventional levels and reveal substantial differences for these groups. While single mothers increase labor supply by 3-6% (elasticity: 0.032-0.057) relative to the baseline mean as a results of increasing the number of simulated eligible children in the family by one child, married mothers decrease labor supply by 2-3% (elasticity: 0.007-0.011) relative to the baseline mean.[17] In addition, the estimates on usual hours worked per week, weeks worked per year, and labor force participation are statistically different between single and married mothers at conventional levels.

The effects on labor supply of mothers translate into changes in annual earnings. The top panel of table 4 shows that the elasticity of estimated effects on total earnings of single and married mothers is 0.7 and 0.03, respectively. The effects on total earned income in one- and two-parent families are driven by wage earnings and not by self-employment earnings (see middle and bottom panel of table 4). Figure 5 shows that the effects of extended Medicaid eligibility on earnings are concentrated in the lower part of the earnings distribution. I follow Kuka and Shenhav (2024) and estimate a series of regressions

[17] As shown in table 3, for single mothers the effects on hours worked are driven by working any hours and transitioning from part- to full-time employment ($\geq$ 35hours) and for married mothers the effects are driven by reducing any hours worked and full-time employment. Weeks worked per year are affected across the whole distribution in response to extended Medicaid eligibility (see figure A.1).

where each dependent variable is an indicator equals to one if earnings are greater than X, where X is (0,5000,...,150000). For reference average minimum, median, and maximum eligibility limits during the analysis period are labeled. I find a significant increase in density up to the median eligibility cutoff for single mothers where between 5% and 55% of children in the CPS ASEC sample are eligible for Medicaid. For married mothers the earnings effects are concentrated between the median and maximum eligibility cutoff where between 2% and 7% of children are eligible for Medicaid. The effects above the median (maximum) eligibility limit for single (married) mothers are small and statistically insignificant. Importantly, as shown in figure A.2, this is not because there are no mothers with earnings in these regions.

The differences in labor supply responses of single and married mothers reflect different incentives in one- and two-parent families. Single mothers are the only breadwinner in the family and need to exploit any opportunity to generate earned income. Hence once children's health improves as a result of access to Medicaid, mothers would supply more labor. In contrast married mothers are not constrained due to partner's income and could reduce earned income once the cost for children's health insurance disappear. Therefore the insurance value of Medicaid could play a bigger role for married mothers and the health of children seems to drive the effects for single mothers.

Extended Medicaid eligibility of children also affects the type of occupations women are choosing. In table A.1, I show that single mothers are more likely to work in managerial, professional, technical, sales, and administrative occupations, but less likely to work in farming, forestry, fishing, and service occupations as a response to expanding Medicaid eligibility.[18] For two-parent families, additional eligible children in the family, result in lower probability of mothers working in all occupations except in technical, sales, and administrative occupations. The change in occupational compositions of women is driven by change in occupations across the whole wage distribution for single mothers and by occupations above the 50th percentile for married mothers (see table A.3).

[18]To estimate the effect of simulated eligibility on parental occupational choice, individuals not in the labor force are assigned a zero for the occupation indicators. Hence the estimates should be viewed as a combination of occupational choice through job entry as well as through occupational mobility. Appendix C provides a detailed description of occupational classification.

I also analyze whether extended Medicaid eligibility of children affects the occupational mobility of their mothers. Theoretically the insurance value of children's Medicaid coverage could remove the job lock and allow parents to switch occupations since the potential negative shock associated with the risky occupational switch is ameliorated due to children's coverage. Analyzing the period between 1996 and 2012, existing literature has shown empirically that parents are more likely to switch occupations and move to better paid jobs thereby ruling out compensating differentials (Farooq and Kugler, 2020). This work, however, uses longitudinal occupational mobility which is characterized by a high degree of measurement error and significant correlation with observable characteristics in contrast to retrospective mobility that is merely downward biased (vom Lehn et al., 2022).[19] Table A.4 shows the effects on job-to-job transitions using retrospective and longitudinal measure of one-, two-, and three-digit occupations and table A.5 shows additional occupational mobility measures such as transition to three-digit occupations with higher wages, variance in wages, separation rates, and educational requirements. Only longitudinal mobility for single mothers is increasing by children's Medicaid eligibility. An additional eligible child in the family increases maternal occupational mobility by 10%-22%. Most of the estimates for the retrospective mobility are negative and only estimates of the one- and two-digit occupational mobility for the full sample are statistically significant at conventional levels. Most of the estimated effects for different types of occupational mobility shown in table A.5 have also opposite signs for longitudinal and retrospective occupational mobility. These results suggest that the choice of the occupational mobility measure is crucial and that the relationship between children's Medi-caid eligibility and parental occupational mobility is not clearly pointing in one direction. Moreover, the different results for the longitudinal occupational mobility in comparison to Farooq and Kugler (2020) imply that the analysis period and empirical approach also drive the effects.

[19]Retrospective occupational mobility is obtained by comparing last year's with last week's occupations and longitudinal occupational mobility is obtained by comparing last week's occupation in first and second spell of the survey.

To understand the magnitude of the effects on labor supply, I take into account that the implied effect is one additional eligible child per family and that the average increase in total simulated eligibility from the beginning to the end of the analysis period represents only 0.4 (0.6) eligible children per family for single (married) mothers. To account for the difference of family-level eligibility between 1977 and 2017, I therefore scale the estimated effects on labor supply by the average increase in total simulated eligibility. Moreover, since not every child enrolls in Medicaid, the results so far should be interpreted as intent-to-treat estimates where treatment is defined as program participation. To convert intent-to-treat (ITT) to treatment-on-treated (TOT) estimates, I divide the scaled estimated effect by the corresponding family-level Medicaid take-up rates (see section 5.2.1). In general, the TOT scale factor for labor supply measures is quite similar for single (1.6) and for married (1.5) mothers. For reference, the scaled estimated effects represent 17-38% and 6-13% of the difference in labor supply measures between the beginning and end of the analysis period for one- and two-parent families, respectively. As another point of reference, I compare the changes in earned income to the the food expenditure during the analysis period. Page and Kantor (2025) show that low-income families spend 30% on food and I find that single (married) mothers increase (decrease) earned income by 12% (18%) relative to the baseline mean.

## 5.2 Mechanisms

### 5.2.1 Children's Medicaid Coverage

I first analyze if extended Medicaid eligibility translates into increased Medicaid coverage children since program take-up is the primary channel in understanding the re-lationship between child Medicaid eligibility and parental labor market decisions. Table 5 presents results for the estimated effects of marital-status-specific simulated eligibility on public health insurance coverage of children. The top panel shows the results using the traditional measure of Medicaid coverage in CPS. The point estimate for the full sample implies that the number of covered children per family increases by roughly one-third of a child as a result of one more child per family becoming eligible which is equivalent to a take-up rate of 37% among newly eligible children or an elasticity of 0.5 $(0.37 * 0.68 \div 0.48)$. Column two and three show the effects on Medicaid coverage by maternal marital status. The take-up of children with a single mother (elasticity of 0.3) is smaller and statistically different ($p<0.05$) than the take-up of children with a married mother (elasticity of 0.6).[20] One potential explanation are the greater barriers to Medicaid enrollment (e.g., insufficient knowledge about the programs, confusion about the eligibility, difficulties with the application) for low-income families (Stuber and Bradley 2005). To check the robustness to imputation and methodological changes in CPS, I estimate the effect of family-level Medicaid eligibility on Medicaid coverage using a harmonized measure of Medicaid coverage constructed by State Health Access Data Assistance Center (SHADAC).[21] Shown in bottom panel of table 5, the estimates are quantitatively and qualitatively very similar to the model that uses the traditional CPS measure albeit slightly smaller since Medicaid coverage enhanced by SHADAC is only available for 1987-2014.

### 5.2.2 Children's Health Status

Next, I analyze if extended Medicaid eligibility affects health of children as another part of the mechanism for maternal labor supply responses. Beginning in 1996, CPS asks the respondents to rate their current health on a five-point scale (excellent, very good, good, fair, or poor health). To capture the family-level health of children I estimate the effect of extended Medicaid eligibility on number of children with excellent, very good, good, fair, or poor health (see table 6). As a result of expanding Medicaid eligibility more children per family report to be in excellent health and fewer children report to be in poor, fair, or good health. The improvement in excellent health is stronger for children in one-parent families due to lower baseline health and bigger point estimate. The positive and statistically significant effects on children's health status at the higher end of the health distribution suggest that increased take-up translates into better health of children.

[20] The average number of children per family as well as the distribution of number of children per family is similar in one- and two-parent families. On average single (married) mothers have 1.8 (1.9) children. Similarly, 50% and 62% of one- and two-parent families have more than one child.

[21] For 1979-1986, CPS imputed health insurance for children age 0-14. In addition, during the analysis period the collection of health insurance coverage in CPS underwent multiple methodological changes (SHADAC 2009).

#### 5.2.3 Maternal Educational Attainment

Since children's access to Medicaid affects educational outcomes of mothers, I check whether extended Medicaid eligibility may affect labor supply of mothers through changes in maternal educational attainment. As shown in table A.6, single mothers are less likely to drop out of high school (13%) and to graduate from college (16%) as well as more likely to graduate from high school (11%) and attend some college (6%). The effects on some college attendance are not considered for this calculation since the estimates are not robust to inclusion of additional controls as shown in figure A.6. Back of the envelope calculation suggests that children's Medicaid is less likely to affect maternal labor market outcomes through educational attainment since Medicaid expansions result in larger effects on college non-completion than on high school graduation and earnings premium is higher for college than for high school (Goldin and Katz, 2007). For married non of the estimates are robust to alternative specifications (see figure A.6).

### 5.3 Return on Investment

To examine program's return on investment, I compare the estimated effect of extended Medicaid eligibility on maternal tax payments with the estimated effect on chil-dren's Medicaid cost.[22] The results of this analysis provided in table A.7 show that the total cost for Medicaid increases by $1,058 ($p < 0.01$) for single mothers and by $971 ($p < 0.01$) for married mothers as a result of making one additional child per family eligible for Medicaid. At the same time, extended Medicaid eligibility leads to a statically significant decrease of federal and state tax liabilities for single mothers and an imprecisely estimated increase for married mothers. When I consider 19% of Federal Insurance Contributions Act (FICA) tax as social benefit, the estimates of net tax liabilities are very similar.[23] The imprecise effects for married mothers on net tax liabilities are partially driven by increased tax benefits as a result of labor supply responses (e.g., benefits from EITC). I also analyze if children's access to Medicaid affects family-level government transfers which is another source of public expenditures (see table A.8). While benefits from welfare, educational assistance, school lunch and energy subsidy decrease, supplemental security income increase. Program receipt is however often mismeasured in surveys due to nonresponse, imputation, and measurement error potentially resulting in biased estimates (Meyer et al., 2015). Therefore the estimates on program participation should be take with caution. The results of this analysis suggest that government is not able to recover cost of expenditures on the Medicaid program just through additional taxes collected. Hendren and Sprung-Keyser (2020), however, calculate the marginal value of public funds associated with Medicaid expansion to pregnant women and infants between 1979 and 1992 and find that Medicaid has paid for itself when all benefits (e.g., improved health of children) are accounted for.

[22] I calculate net tax liabilities under US federal and state income tax laws using a simulation program (TAXSIM 32 available at http://www.nber.org/taxsim) provided by the National Bureau of Economic Research (Feenberg and Coutts, 1993). In two-parent families, I assume that the father is the primary earner.

[23] Since individuals can recover some of the FICA contributions in form of Medicare and Social Security benefits, Heller and Mumma (2023) suggest treating 19% of FICA as social benefit.

## 5.4 Robustness

### 5.4.1 Marital Status

Given that maternal marital status could respond to extended Medicaid eligibility of children, I test whether composition bias is driving the estimated effects. First, the effects of children's Medicaid eligibility on marital outcomes of mothers show no clear pattern in any direction. While the results show that mothers are more likely to be married (probability of being never married decreases, the probability of being married and ever married increase), the likelihood of being divorced also increases as a results having an eligible child (see figure A.3).

Second, I use the linked sample and construct the simulated eligibility measure based on last year's marital status that is not affected by current year's simulated eligibility. The results of this analysis for main labor supply measures are shown in table A.11. The table also shows estimates for the linked data set using the current year's marital status to confirm that mothers in the linked and cross-sectional sample respond similarly to Medicaid legislations. The estimated effects of child's Medicaid eligibility on maternal labor market outcomes using the linked sample are generally in line with labor supply responses found in the cross-sectional sample.

Third, I use estimates of earnings gaps due to marriage from existing literature in a back of the envelop calculation to derive bounds for endogenous sample selection. Most of the literature documents negative marriage earnings gaps for women (Waldfogel 1997, 1998).[24] Given that mothers are 1.0% (0.9%) more likely to get married (divorced) as a result of extended Medicaid eligibility and the lowest effect size for maternal labor supply is 3.2% (2.0%) for single (married) mothers, marital status response could result in endogenous sample selection if marriage (divorce) changes labor supply by at least 320%

[24] An exception is the recent work by Juhn and McCue (2016) that does not finds lower earnings among married women from 1966-1975 birth cohort. However, even in the 1966-1975 birth cohort married women with children that are most relevant for my analysis have 35 percent lower earnings than their single counterparts.

(222%) which by far exceeds the estimated effects in the literature.

### 5.4.2 Identifying Assumption

Two general difference-in-difference identifying assumptions in equation 1 are invoked for the validity of the empirical approach. The first identifying assumption is that no shock differentially affects Medicaid generosity and outcomes of children and their parents in the same state, during the same year, and with the same number of children of the same age. Hence omitted variables specific to parents with the same number of children of the same age and state of residence that change over time and are correlated with both Medicaid legislation and outcomes of children or their parents would invalidate this empirical strategy. To address this potential confounder, I estimate a version of equation 1 with pairwise interactions between state, year, and children's age fixed effects. Figures A.4 and A.5 show five models that test the first identifying assumption. Model one is the baseline model shown in equation 1. Model two adds state-by-youngest child's age, state-by-oldest child's age, and state-by-difference in age between oldest and youngest child fixed effects to the baseline model. Model three adds state-by-year fixed effects to the baseline model. Model four adds year-by-youngest child's age, year-by-oldest child's age, and year-by-difference in age between oldest and youngest child fixed effects to the baseline model. Model five adds state-by-oldest child's age, state-by-youngest child's age, state-by-difference in age between oldest and youngest child, state-by-year, year-by-youngest child's age, year-by-oldest child's age, year-by-difference in age between oldest and youngest child fixed effects to the baseline model. With exception of models four and five for some labor supply measures, most of the estimates are very similar across the different models. Models four and five, however, absorb almost all of the identifying variation in the total simulated eligibility (see table A.9)

The second identifying assumption requires that public health insurance eligibility rules are not set based on outcomes of parents and their children. The simulated eligibility approach will therefore fail if states phase in Medicaid expansions because of changing trends in parental or child outcomes. To test the validity of this identifying assumption, I regress the maximum Medicaid eligibility limits for children age 0-18 in a given state and year on contemporaneous and lagged (first and second order) state-level characteristics.[25] I use Medicaid eligibility limits as opposed to state-level annual simulated eligibility since state governments set eligibility levels and do not control simulated eligibility as a policy parameter. The results shown in table A.10, suggest that generosity of Medicaid is not affected by outcomes of parents and their children or other state-level policy determinants. Baughman and Milyo (2009) also show that state Medicaid expansions are not driven by percentage of uninsured children in the state and Farooq and Kugler (2020) find no evidence that state demographic and economic characteristics are affecting Medicaid generosity. In comparison to Baughman and Milyo (2009) and Farooq and Kugler (2020), I show that state-level policies discussed in the literature (e.g., Miller and Wherry 2019) have no effect on Medicaid expansions. In addition, the second identifying assumption has been invoked repeatedly in the simulated eligibility literature.[26]

### 5.4.3 Measurement

Existing literature shows that labor supply measures in the CPS are prone to errors due to imputation.[27] Shown in table A.13 the average earnings estimates are not driven by nonresonse biases or disclosure avoidance methods. To check for nonresonse biases, I follow Hirsch and Schumacher (2004) and Bollinger and Hirsch (2006) and drop imputed nonrespondents and reweigh the sample with inverse-probability weights to restore population representatives. I apply cell mean replacement topcodes introduced in

[25] For the years prior to state expansions (1979-1987), I use the maximum eligibility threshold for Medically Needy Program or AFDC. As an approximation for AFDC eligibility, I use the average ratio of the needs standard to the corresponding poverty guideline across all family sizes. For the years 1988-2014, I use the maximum state-level Medicaid eligibility levels across all ages 0-18. Since states expanded eligibility for different age groups, it is more consistent to use eligibility limits across a broad age group and not focus on narrow defined age groups (e.g., children age 0-5).

[26] See for instance Currie and Gruber (1996a,b); Cutler and Gruber (1996); Gross and Notowidigdo (2011); Cohodes et al. (2016); East et al. (2023); Brown et al. (2019); Miller and Wherry (2019)

[27] Borjas and Hamermesh (2024) discuss the quality of hours and weeks worked and Bollinger et al. (2019) analyze the accuracy of earnings.

1996 from Larrimore et al. (2008) and rank proximity swap topcodes used starting in 2011 from Census Bureau to the earlier period to test if top coding methods are driving the results. I also show that usual hours worked per week, weeks worked per year, and labor force participation are robust to excluding imputed observations and that using hours worked last week gives qualitatively and quantitatively similar results for single mothers (see table A.12). In contrast, the estimated effects on usual hours worked per week and weeks worked per year for married mothers are statistically insignificant when the imputed observations are excluded.

#### 5.4.4 Maternal Eligibility

A small proportion of women were directly affected by extended Medicaid eligibility during the analysis period because some legislations expanded Medicaid to pregnant women and children at the same time. Since Medicaid eligibility for pregnant women was also applicable for their newborns until the first birthday, parents with children age zero might change labor supply as a result of direct effect of extended Medicaid eligibility and not as a result of spillovers from their children. To understand if parental labor market outcomes are driven by direct effects of Medicaid, I test if the estimated effects are sensitive to using maternal eligibility for zero-year old children by using two common measures in the Medicaid literature of maternal eligibility and dropping children of age zero.[28] The results of this analysis are shown in tables A.14. All estimates across different specifications are very similar in terms of magnitude and significance, suggesting that direct effects of expansions are not driving labor supply responses.

[28] Following Currie and Gruber (1996a,b), the first measure is constructed by using all women of reproducible age (15-44) in each calendar year across the full sample period. Using this national data set, I calculate the fraction of eligible women in each state, year, and race-ethnicity group. The second state-year-race-ethnicity maternal eligibility measure is obtained by using mothers with children of age zero. Similarly to children's simulated eligibility, I leave out women from the state for which the simulated eligibility is being imputed. The maternal eligibility measures are then assigned to zero-year old children based on the state, year, and maternal marital status.

#### 5.4.5 Simulated Eligibility Type

Since the simulated eligibility measure used through out the analysis is constructed using all children from the year for which the simulated eligibility is estimated, one might be concerned that characteristics used to determine eligibility (e.g. family structure or family income) may respond to Medicaid expansions.[29] To account for this potential endogeneity, I construct alternative simulated eligibility measures that use children from period before the analysis starts. To obtain the simulated fixed eligibility measures, I use all children from 1977 CPS ASEC and inflate the income to the year for which the eligibility is imputed. Changes in the national and regional Consumer Price Index for All Urban Consumers (CPI-U) as well as average wages are used to adjust the income.[30] The demographic characteristics of children in 1977 might however not reflect demographic characteristics of children observed in later years of the analysis period and inflation or wage growth might not fully capture changes in income over time. Using a fixed national data set from pre-analysis period to create simulated eligibility might therefore result in a mismeasured simulated eligibility for later years of the analysis period. Since changes in socio-demographic characteristics are also correlated with changes in the structure of the labor market and hence parental labor market outcomes, using a fixed eligibility measure might as well result in biased estimates.[31] Hence, the annual and fixed eligibility measures have advantages and disadvantages.

Figure 6 documents the differences between the simulated eligibility measures. The changes across the simulated annual eligibility - the measure used throughout the analysis - and actual eligibility track quite well. The trends in simulated fixed eligibility mea-sures, however, deviate from trends in actual eligibility, especially towards the end of the analysis period. The eligibility measures constructed by using CPI-U perform worse than the measure constructed by using average wages. The results are, however, very similar across specifications using the different types of the simulated eligibility measure for single mothers.[32] Similarly to to the analysis of non-imputed observations, hours worked per week and weeks worked per year of married mother are not precisely estimated suggesting that the effects on labor supply are driven by the extensive margin.

[29]Policy endogeneity can arise because of a response to federal or state-level expansions. For instance, Deficit Reduction Act of 1984 could affect labor market outcomes of parents and hence the family income. Using family income of children observed in years after 1984 to determine eligibility in the same year would result in a biased estimate of the simulated eligibility measure.

[30]Average wages are calculated as the ratio of compensation of employees to total non-farm employees.

[31]Since immigration has a major influence on the size and demographic structure of the US population, immigration can be one factor leading to a change of socio-demographic characteristics of individuals observed in later years of the analysis period.

# 6 Conclusion

United States has witnessed a substantial increase in public health insurance coverage of children between 1977 and 2017. Despite the extensive literature studying the consequences of expanded Medicaid coverage, spillover effects on other family members have been under-studied. This paper presents new evidence on the effects of children's Medicaid eligibility on parental labor market outcomes. To identify the effects of Medicaid eligibility the empirical strategy exploits legislative variation at the state, year, and age of the child level which resulted from Medicaid expansions between 1977 and 2017. To address endogeneity of actual eligibility I use the simulated eligibility strategy by estimating reduced form impacts of simulated Medicaid eligibility of children on labor market outcomes of their parents.

I demonstrate that extended Medicaid eligibility of children leads to increased maternal labor supply at the extensive and intensive margin for single mothers and a decrease at the extensive margin for married mothers. Analyzing the earnings across the distribution, I show that the changes in labor supply decisions are driven by low-income women consistent with Medicaid program targeting low-income families. I also show that important first-order mechanism measures such as Medicaid take-up and health of children increases as a response to extending Medicaid eligibility. Given the different labor supply responses of single and married mothers to extended Medicaid eligibility of their children insurance value seems more important for two-parent families and the ability to invest more time into the labor market time for single-parent families.

[32] Estimated effects from models using simulated annual, simulate fixed (CPI), simulated fixed (RCPI), and simulated fixed (WAGE) eligibility are shown in tables A.15.

This work may emphasize the following policy implications. First, the findings of this study may have implications for the overall generosity of Medicaid eligibility since the general equilibrium effects may exceed the direct benefits of the public health insurance coverage. Second, focusing on disadvantaged population may provide guidance about targeting Medicaid to certain groups, for example about making Medicaid more generous for low-income families since the effects are concentrated in the low end of the earning distribution.

# Figures and Tables

Figure 1:
National Variation in Medicaid Eligibility by Child's Race & Ethnicity

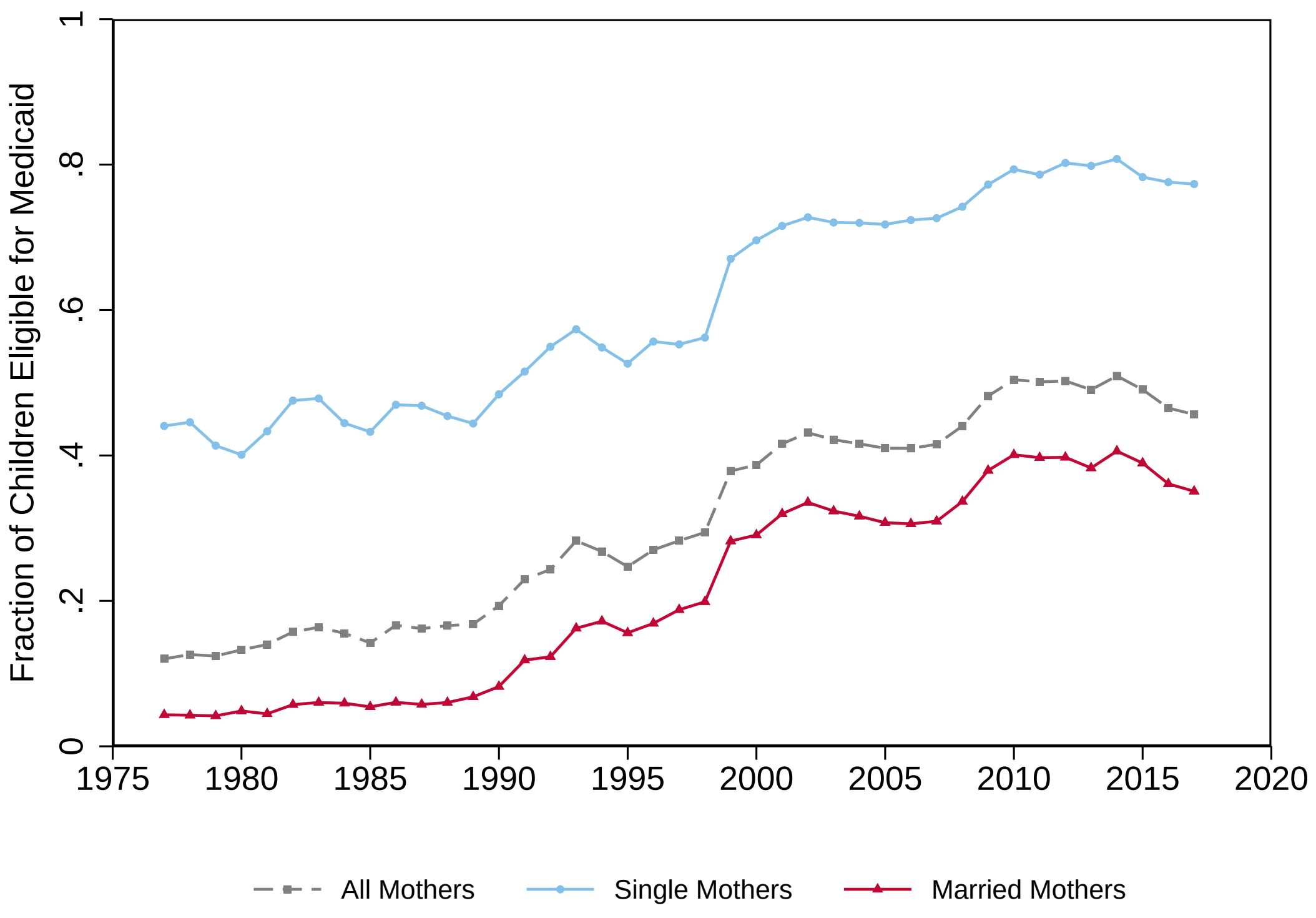

*Notes:* This figure shows the fraction of eligible children between 1977 and 2017. The data is from CPS ASEC 1978-2018. The sample is restricted to mothers age 20-64 with children age 0-18. Arizona is not included because the state did not adopt a Medicaid program until 1982.

Figure 2:
State Variation in Total Simulated Eligibility

(a) Single-Child Families

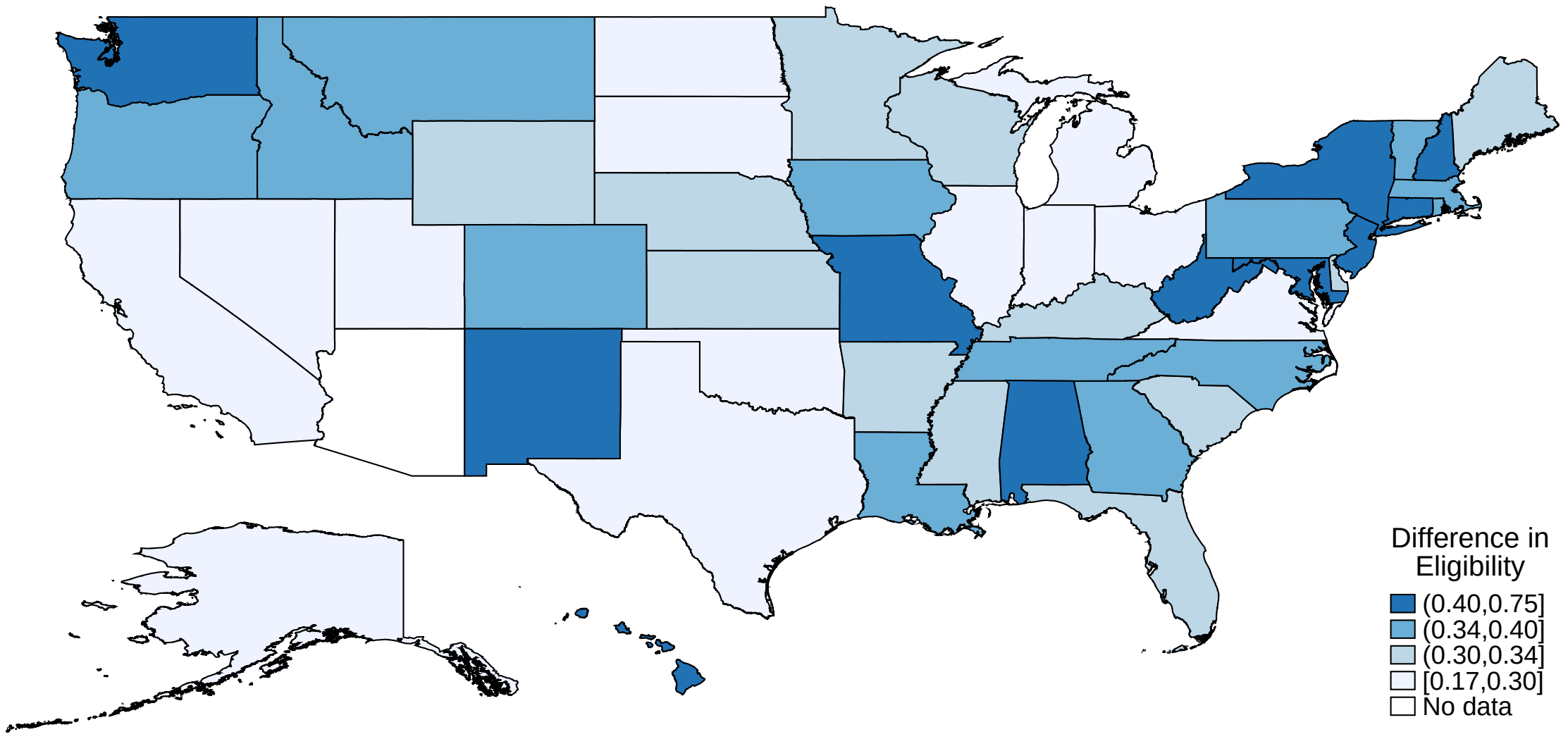

(b) Multiple-Child Families

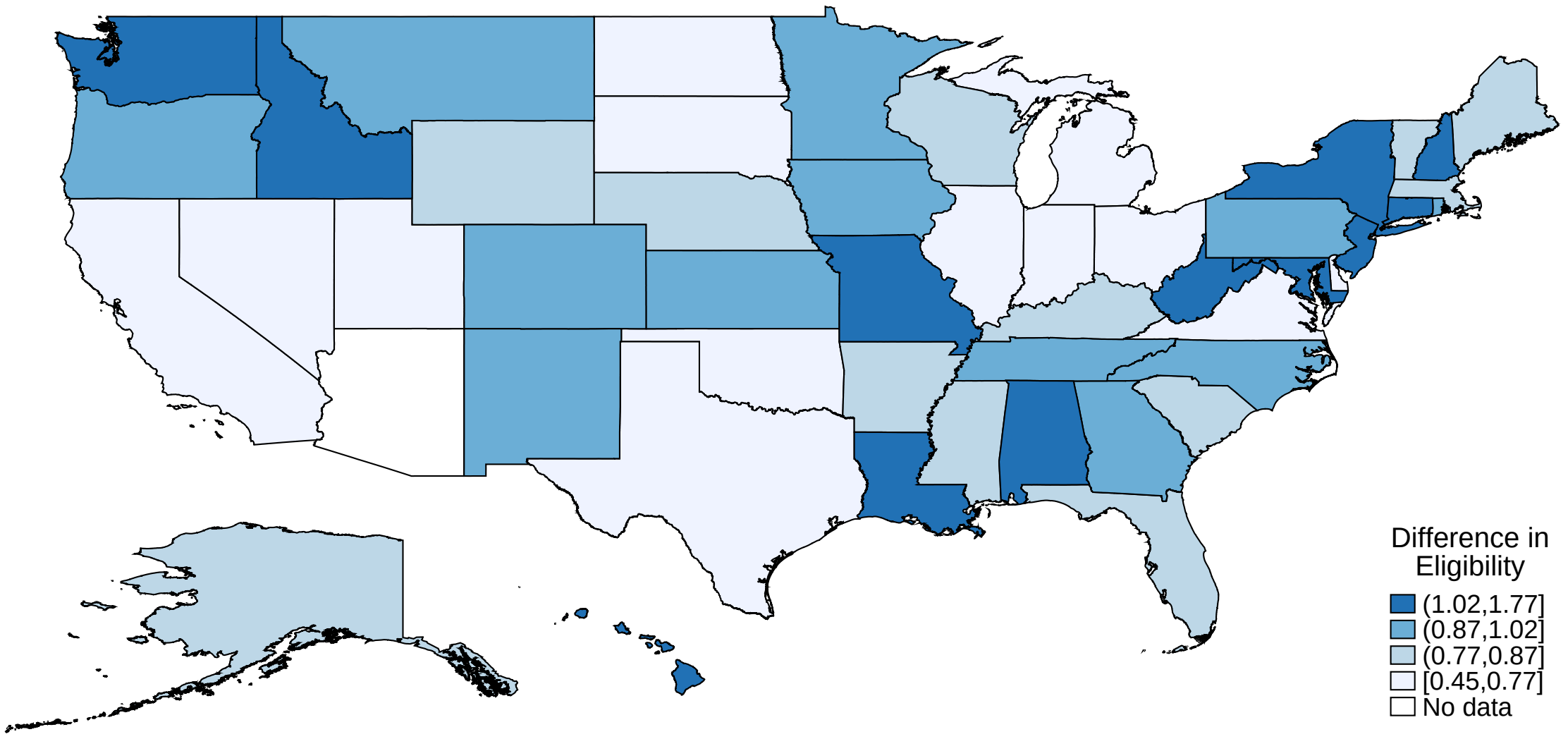

*Notes:* These figures show the difference in marital-status-specific total simulated eligibility between 1977 and 2017 for (a) single- and (b) multiple-child families in each state. The quartiles represent the difference in total simulated eligibility between 1977 and 2017. These years are the start and end of the analysis period. The data is from CPS ASEC 1978-2018. The sample is restricted to mothers age 20-64 with children age 0-18. Arizona is not included because the state did not adopt a Medicaid program until 1982.

Figure 3:
National Variation in Total Simulated Eligibility

(a) Single-Child Families

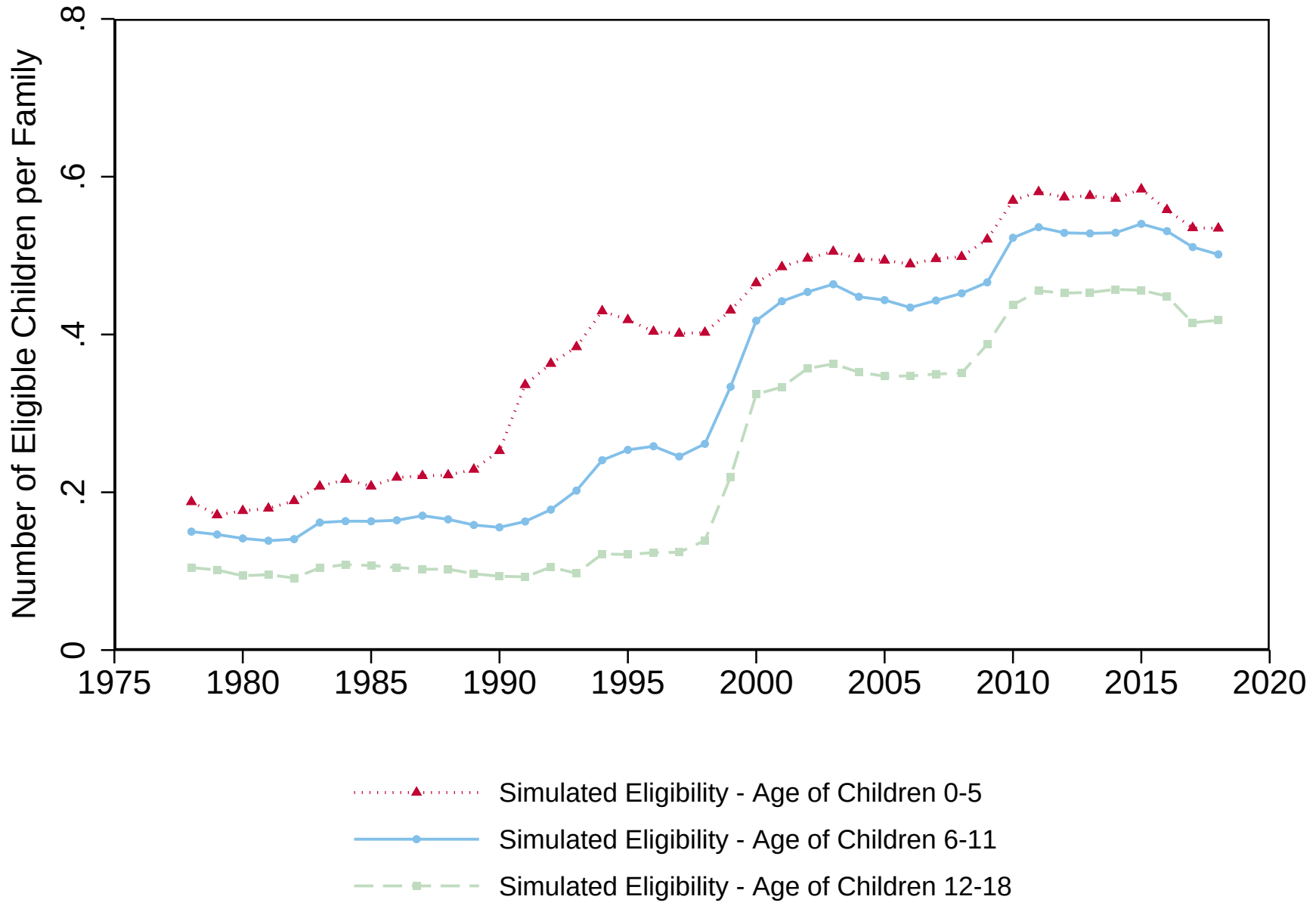

(b) Multiple-Child Families

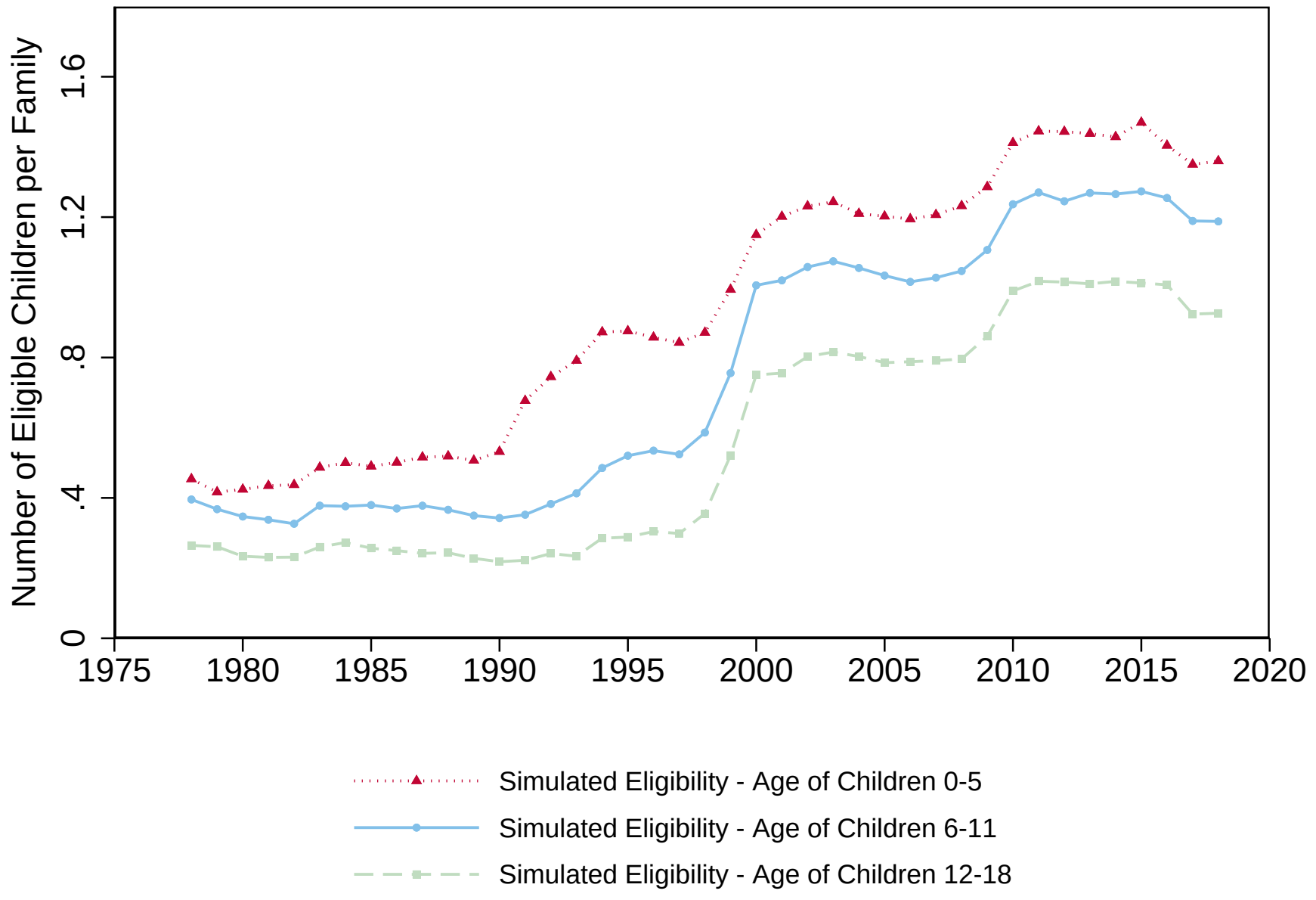

*Notes:* These figures show the marital-status-specific total simulated eligibility between 1977 and 2017 for (a) single- and (b) multiple-child families by child's age. The data is from CPS ASEC 1978-2018. The sample is restricted to mothers age 20-64 with children age 0-18. Arizona is only included after 1982 because the state did not adopt a Medicaid program until 1982.

Figure 4:
Group-Specific vs. Non-Group-Specific Simulated Eligibility Measure

(a) Marital-Status-Specific Simulated Eligibility

(b) Non-Marital-Status-Specific Simulated Eligibility

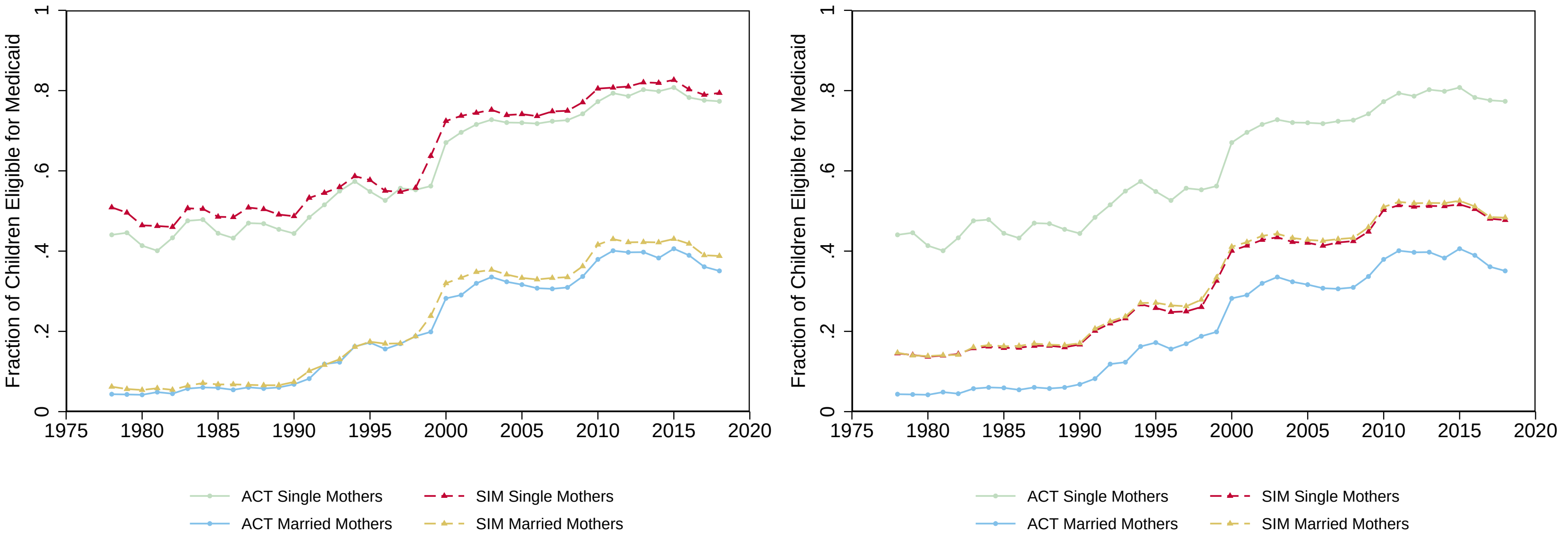

*Notes:* This figure shows marital-status-specific simulated eligibility measure and non-marital-status-specific simulated eligibility measure for children with single and married mothers between 1977 and 2017. ACT and SIM refers to actual and simulated eligibility, respectively. The data is from CPS ASEC 1978-2018. The sample is restricted to mothers age 20-64 with children age 0-18. Arizona is not included because the state did not adopt a Medicaid program until 1982.

Figure 5:
Effect of Total Simulated Eligibility on Maternal Annual Earnings ($2020)

(a) Sincle Mothers (b) Married Mothers

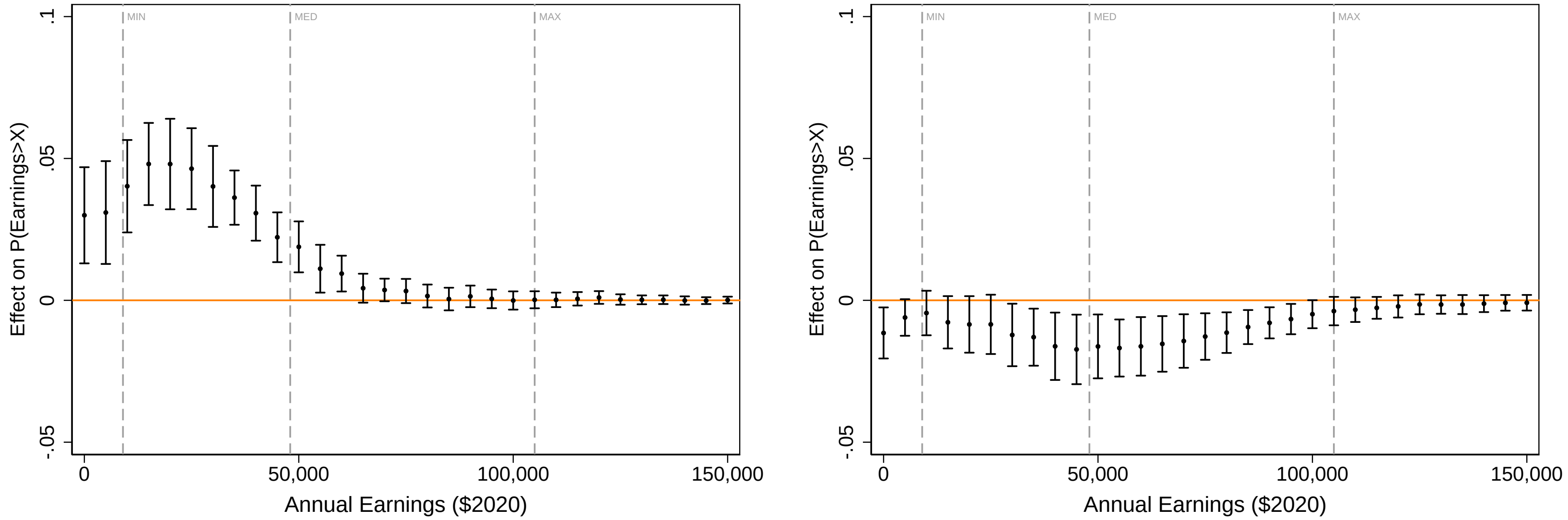

*Notes:* These figures show the coefficients and 95% confidence intervals from regressions estimating the effect of marital-status-specific total simulated eligibility on annual earnings ($2020) last year of single and married mothers. Each point estimate and confidence interval is obtained from a different regression where the dependent variable is an indicator equals to one if maternal annual earnings ($2020) were at least as great as X (0,5000,...,150000) last year. All models include maternal-level controls (indicators for maternal age, state of residence, calendar year, age of the youngest child, age of the oldest child, difference in age between oldest and youngest child, and number of children) and state-level controls (unemployment rate, minimum wage, inflation-adjusted maximum welfare benefit for a family of 4, state-level EITC, implementation of six types of welfare waivers, implementation of any waiver or TANF). Regressions are weighted with maternal survey weights. The data is from CPS ASEC 1978-2018. The sample is restricted to mothers age 20-64 with children age 0-18.

Figure 6:
National Variation in Alternative Simulated Eligibility Measures

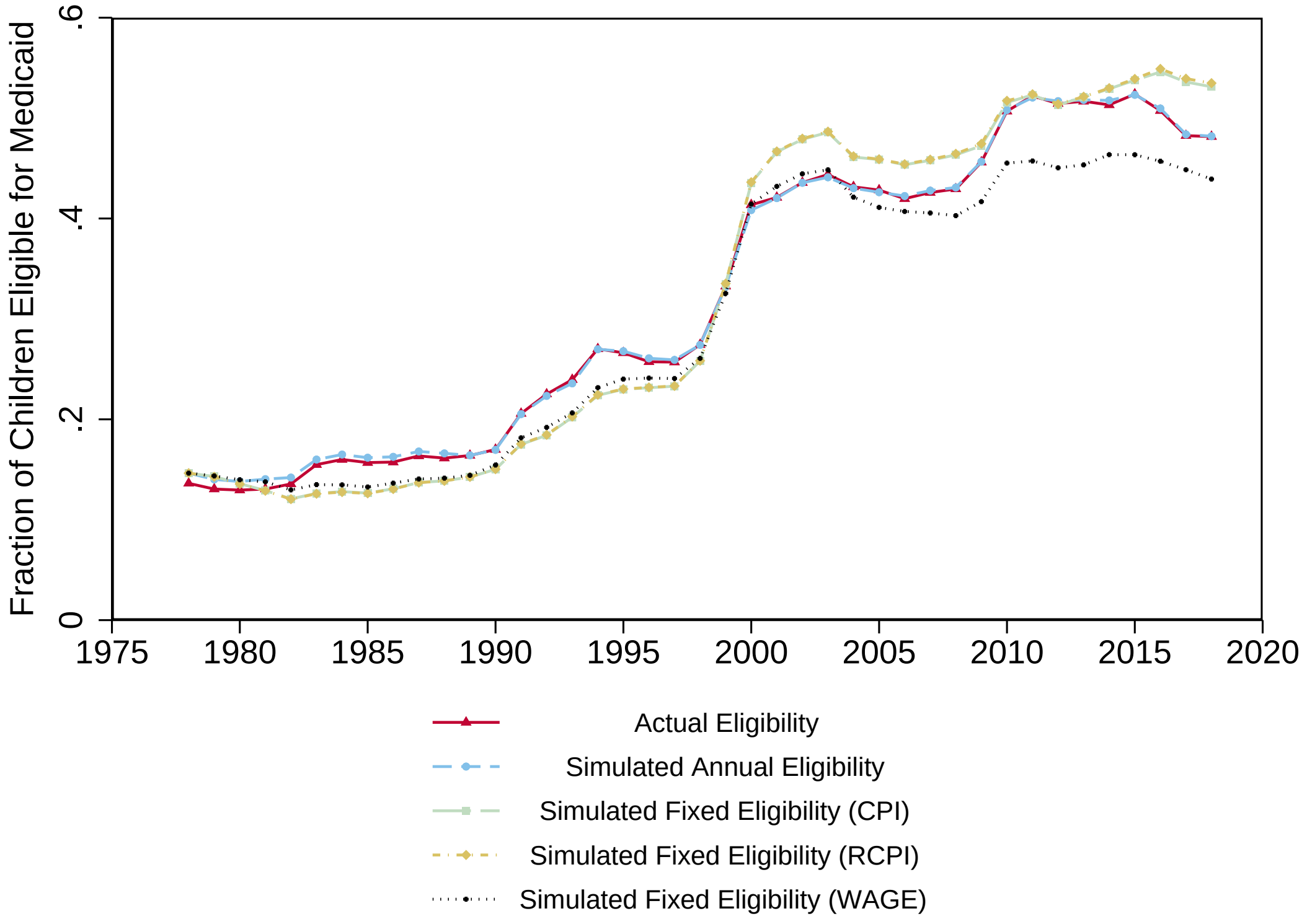

*Notes:* This figure shows different marital-status-specific total simulated eligibility measures between 1977 and 2017. CPI, RCPI, and WAGE refers to consumer price index, regional consumer price index, and wage per capita, respectively. The data is from CPS ASEC 1978-2018. The sample is restricted to mothers age 20-64 with children age 0-18. Arizona is not included because the state did not adopt a Medicaid program until 1982.

Table 1:
Summary Statistics of Maternal Characteristics

| | Cross-Sectional Dataset | | | Linked Dataset | | |
|---|---|---|---|---|---|---|
| | All | Single | Married | All | Single | Married |
| Maternal Age | 36.32 | 35.12 | 36.73 | 37.17 | 35.49 | 37.70 |
| | (8.59) | (9.10) | (8.37) | (8.19) | (8.87) | (7.89) |
| Fraction with High School or Less | 0.49 | 0.57 | 0.46 | 0.47 | 0.55 | 0.45 |
| | (0.50) | (0.49) | (0.50) | (0.50) | (0.50) | (0.50) |
| Fraction White | 0.64 | 0.47 | 0.70 | 0.69 | 0.50 | 0.75 |
| | (0.48) | (0.50) | (0.46) | (0.46) | (0.50) | (0.43) |
| Fraction Black | 0.13 | 0.30 | 0.08 | 0.14 | 0.33 | 0.08 |
| | (0.34) | (0.46) | (0.27) | (0.35) | (0.47) | (0.27) |
| Fraction Asian | 0.04 | 0.02 | 0.05 | 0.04 | 0.02 | 0.05 |
| | (0.20) | (0.14) | (0.21) | (0.20) | (0.14) | (0.21) |
| Fraction Hispanic | 0.16 | 0.18 | 0.16 | 0.11 | 0.13 | 0.10 |
| | (0.37) | (0.39) | (0.36) | (0.31) | (0.33) | (0.30) |
| Fraction Other Race | 0.02 | 0.03 | 0.02 | 0.02 | 0.02 | 0.02 |
| | (0.16) | (0.16) | (0.15) | (0.15) | (0.16) | (0.15) |
| Number of Children per Family | 1.90 | 1.77 | 1.94 | 1.82 | 1.69 | 1.86 |
| | (0.99) | (1.00) | (0.99) | (0.94) | (0.93) | (0.94) |
| Number of Eligible Children per Family | 0.69 | 1.21 | 0.51 | 0.55 | 1.09 | 0.38 |
| | (1.16) | (1.25) | (1.07) | (1.05) | (1.19) | (0.94) |
| Fraction with Children under Age 6 | 0.34 | 0.32 | 0.35 | 0.29 | 0.28 | 0.29 |
| | (0.42) | (0.42) | (0.42) | (0.40) | (0.41) | (0.40) |
| Fraction with Children Age 6-11 | 0.29 | 0.30 | 0.28 | 0.29 | 0.31 | 0.29 |
| | (0.36) | (0.39) | (0.36) | (0.37) | (0.39) | (0.36) |
| Fraction with Children Age 12-18 | 0.37 | 0.38 | 0.36 | 0.41 | 0.41 | 0.42 |
| | (0.43) | (0.44) | (0.43) | (0.45) | (0.45) | (0.44) |
| Fraction under 100% of FPL | 0.17 | 0.38 | 0.10 | 0.14 | 0.36 | 0.07 |
| | (0.37) | (0.48) | (0.30) | (0.35) | (0.48) | (0.25) |
| Fraction 100%-133% of FPL | 0.07 | 0.11 | 0.05 | 0.06 | 0.10 | 0.04 |
| | (0.25) | (0.31) | (0.23) | (0.23) | (0.30) | (0.21) |
| Fraction 133%-185% of FPL | 0.11 | 0.14 | 0.10 | 0.10 | 0.14 | 0.09 |
| | (0.31) | (0.35) | (0.29) | (0.30) | (0.35) | (0.28) |
| Fraction 185%-300% of FPL | 0.22 | 0.20 | 0.23 | 0.22 | 0.21 | 0.23 |
| | (0.41) | (0.40) | (0.42) | (0.42) | (0.41) | (0.42) |
| Fraction above 300% of FPL | 0.44 | 0.18 | 0.53 | 0.48 | 0.19 | 0.57 |
| | (0.50) | (0.38) | (0.50) | (0.50) | (0.39) | (0.49) |
| Number of Observations | 785,390 | 188,439 | 596,951 | 232,775 | 51,220 | 181,555 |

*Notes:* The table shows demographics characteristics of mothers observed in CPS ASEC 1977-2018 using a repeated crossection and CPS ASEC 1979-2018 using linked observations. For the cross-sectional cohort, the sample is restricted to second observation. Number of eligibile children per family refers to number of actual eligibile children. The sample is restricted to mothers age 20-64 with children age 0-18. Means are weighted with maternal cross-sectional and linking-adjusted survey weights. Standard deviations are shown in parenthesis.

Table 2:
Effect of Total Simulated Eligibility on Maternal Labor Supply

| | All | Single | Married |
|---|---|---|---|
| | Usual Hours Worked per Week | | |
| SIMT | -0.09 | 1.39*** | -0.56** |
| | ( 0.18) | ( 0.32) | ( 0.24) |
| Observations | 785,386 | 188,438 | 596,948 |
| Adjusted $R^2$ | 0.08 | 0.09 | 0.07 |
| Mean Y - Baseline | 20.21 | 24.74 | 19.11 |
| Mean Y - Overall | 25.58 | 28.05 | 24.74 |
| | Weeks Worked per Year | | |
| SIMT | -0.02 | 1.20*** | -0.42** |
| | ( 0.17) | ( 0.45) | ( 0.19) |
| Observations | 785,386 | 188,438 | 596,948 |
| Adjusted $R^2$ | 0.09 | 0.12 | 0.09 |
| Mean Y - Baseline | 23.33 | 27.20 | 22.39 |
| Mean Y - Overall | 31.51 | 33.03 | 30.99 |
| | Labor Force Participation | | |
| SIMT | -0.00 | 0.02*** | -0.01*** |
| | ( 0.00) | ( 0.01) | ( 0.00) |
| Observations | 788,824 | 187,830 | 600,994 |
| Adjusted $R^2$ | 0.08 | 0.08 | 0.07 |
| Mean Y - Baseline | 0.51 | 0.63 | 0.49 |
| Mean Y - Overall | 0.68 | 0.73 | 0.66 |

*Notes:* This table shows results from regressions estimating the effect of marital-status-specific total simulated eligibility on maternal usual hours worked per week last year, weeks worked last year, and labor force participation last week. Usual hours worked per week and weeks worked last year include zeros. All models include maternal-level controls (indicators for maternal age, state of residence, calendar year, age of the youngest child, age of the oldest child, difference in age between oldest and youngest child, and number of children) and state-level controls (unemployment rate, minimum wage, inflation-adjusted maximum welfare benefit for a family of 4, state-level EITC, implementation of six types of welfare waivers, implementation of any waiver or TANF). In models using the full sample a marital status indicator is included and all controls are interacted with a marital status indicator. Regressions are weighted with maternal survey weights. Standard errors in parentheses are clustered at the state level. The data is from CPS ASEC 1977-2018. The sample is restricted to mothers age 20-64 with children age 0-18. *** $p < 0.01$, ** $p < 0.05$, * $p < 0.10$.

Table 3:
Effect of Total Simulated Eligibility on Maternal Usual Hours Worked per Week

| | All | Single | Married |
|---|---|---|---|
| | | Positive Hours | |
| SIMT | -0.0027 | 0.0246*** | -0.0115** |
| | (0.0041) | (0.0087) | (0.0045) |
| Observations | 785,386 | 188,438 | 596,948 |
| Adjusted $R^2$ | 0.0641 | 0.0779 | 0.0578 |
| Mean Y - Baseline | 0.6004 | 0.6589 | 0.5863 |
| Mean Y - Overall | 0.7145 | 0.7470 | 0.7035 |
| | | Part-Time Employment | |
| SIMT | -0.0036 | -0.0251*** | 0.0033 |
| | (0.0037) | (0.0046) | (0.0047) |
| Observations | 785,386 | 188,438 | 596,948 |
| Adjusted $R^2$ | 0.0277 | 0.0156 | 0.0263 |
| Mean Y - Baseline | 0.2022 | 0.1207 | 0.2219 |
| Mean Y - Overall | 0.2007 | 0.1585 | 0.2151 |
| | | Full-Time Employment | |
| SIMT | 0.0009 | 0.0497*** | -0.0148** |
| | (0.0050) | (0.0076) | (0.0071) |
| Observations | 785,386 | 188,438 | 596,948 |
| Adjusted $R^2$ | 0.0666 | 0.0729 | 0.0552 |
| Mean Y - Baseline | 0.3982 | 0.5382 | 0.3643 |
| Mean Y - Overall | 0.5138 | 0.5885 | 0.4884 |

*Notes:* This table shows results from regressions estimating the effect of marital-status-specific total simulated eligibility on likelihood of mothers working any hours last year, working part time last year (>0 and <35 hours per week), and working full time last year ($\geq$ 35 hours per week). All models include maternal-level controls (indicators for maternal age, state of residence, calendar year, age of the youngest child, age of the oldest child, difference in age between oldest and youngest child, and number of children) and state-level controls (unemployment rate, minimum wage, inflation-adjusted maximum welfare benefit for a family of 4, state-level EITC, implementation of six types of welfare waivers, implementation of any waiver or TANF). In models using the full sample a marital status indicator is included and all controls are interacted with a marital status indicator. Regressions are weighted with maternal survey weights. Standard errors in parentheses are clustered at the state level. The data is from CPS ASEC 1977-2018. The sample is restricted to mothers age 20-64 with children age 0-18. *** $p < 0.01$, ** $p < 0.05$, * $p < 0.10$.

## Table 4:
## Effect of Total Simulated Eligibility on Maternal Annual Earnings

| | All | Single | Married |
|---|---|---|---|
| | | Annual Total Earnings | |
| SIMT | -684* | 1,445*** | -1,369** |
| | (399) | (377) | (572) |
| Observations | 785,386 | 188,438 | 596,948 |
| Adjusted $R^2$ | 0.0754 | 0.0878 | 0.0719 |
| Mean Y - Baseline | 13,502 | 17,985 | 12,417 |
| Mean Y - Overall | 24,877 | 23,658 | 25,291 |
| | | Annual Wage Earnings | |
| SIMT | -780** | 1,334*** | -1,462*** |
| | (364) | (380) | (522) |
| Observations | 785,386 | 188,438 | 596,948 |
| Adjusted $R^2$ | 0.0723 | 0.0832 | 0.0693 |
| Mean Y - Baseline | 13,069 | 17,513 | 11,994 |
| Mean Y - Overall | 23,910 | 22,911 | 24,250 |
| | | Annual Self-Employment Earnings | |
| SIMT | 97 | 110 | 92 |
| | (63) | (76) | (85) |
| Observations | 785,386 | 188,438 | 596,948 |
| Adjusted $R^2$ | 0.0043 | 0.0056 | 0.0039 |
| Mean Y - Baseline | 433 | 472 | 423 |
| Mean Y - Overall | 966 | 748 | 1,041 |

*Notes:* This table shows results from regressions estimating the effect of marital-status-specific total simulated eligibility on maternal annual total earnings ($2020) last year, wage earnings ($2020) last year, and self-employment earnings ($2020) last year. All models include maternal-level controls (indicators for maternal age, state of residence, calendar year, age of the youngest child, age of the oldest child, difference in age between oldest and youngest child, and number of children) and state-level controls (unemployment rate, minimum wage, inflation-adjusted maximum welfare benefit for a family of 4, state-level EITC, implementation of six types of welfare waivers, implementation of any waiver or TANF). In models using the full sample a marital status indicator is included and all controls are interacted with a marital status indicator. Regressions are weighted with maternal survey weights. Standard errors in parentheses are clustered at the state level. The data is from CPS ASEC 1977-2018. The sample is restricted to mothers age 20-64 with children age 0-18. *** $p < 0.01$, ** $p < 0.05$, * $p < 0.10$.

Table 5:
Effect of Total Simulated Eligibility on Total Medicaid Coverage of Children

| | All | Single | Married |
|---|---|---|---|
| | Total Medicaid Coverage (Census) | | |
| SIMT | 0.37*** | 0.25*** | 0.41*** |
| | (0.06) | (0.05) | (0.07) |
| Observations | 747,470 | 181,122 | 566,348 |
| Adjusted $R^2$ | 0.35 | 0.45 | 0.21 |
| Mean Y - Baseline | 0.25 | 0.77 | 0.10 |
| Mean Y - Overall | 0.48 | 0.89 | 0.34 |
| Mean SIMT - Overall | 0.68 | 1.18 | 0.50 |
| | Total Medicaid Coverage (SHADAC) | | |
| SIMT | 0.25*** | 0.20*** | 0.27*** |
| | (0.05) | (0.05) | (0.06) |
| Observations | 463,565 | 112,520 | 351,045 |
| Adjusted $R^2$ | 0.36 | 0.48 | 0.21 |
| Mean Y - Baseline | 0.31 | 0.84 | 0.13 |
| Mean Y - Overall | 0.47 | 0.90 | 0.32 |

*Notes:* This table shows results from regressions estimating the effect of marital-status-specific total simulated eligibility on total actual eligibility and total Medicaid coverage. All models include maternal-level controls (indicators for maternal age, state of residence, calendar year, age of the youngest child, age of the oldest child, difference in age between oldest and youngest child, and number of children) and state-level controls (unemployment rate, minimum wage, inflation-adjusted maximum welfare benefit for a family of 4, state-level EITC, implementation of six types of welfare waivers, implementation of any waiver or TANF). In models using the full sample a marital status indicator is included and all controls are interacted with a marital status indicator. Regressions are weighted with maternal survey weights. Standard errors in parentheses are clustered at the state level. The data is from CPS ASEC 1977-2018. The sample is restricted to mothers age 20-64 with children age 0-18. *** $p < 0.01$, ** $p < 0.05$, * $p < 0.10$.

Table 6:
Effect of Total Simulated Eligibility on Child's Health

| | All | Single | Married |
|---|---|---|---|
| | | Excellent Health | |
| SIMT | 0.077*** | 0.107*** | 0.069** |
| | (0.027) | (0.038) | (0.030) |
| Observations | 481,046 | 115,332 | 365,714 |
| Adjusted $R^2$ | 0.206 | 0.146 | 0.208 |
| Mean Y - Baseline | 0.969 | 0.729 | 1.048 |
| Mean Y - Overall | 0.980 | 0.762 | 1.055 |
| | | Very Good Health | |
| SIMT | -0.012 | 0.007 | -0.017 |
| | (0.022) | (0.041) | (0.023) |
| Observations | 481,046 | 115,332 | 365,714 |
| Adjusted $R^2$ | 0.096 | 0.104 | 0.093 |
| Mean Y - Baseline | 0.564 | 0.586 | 0.557 |
| Mean Y - Overall | 0.558 | 0.538 | 0.564 |
| | | Good Health | |
| SIMT | -0.051** | -0.092** | -0.040** |
| | (0.020) | (0.035) | (0.019) |
| Observations | 481,046 | 115,332 | 365,714 |
| Adjusted $R^2$ | 0.085 | 0.106 | 0.071 |
| Mean Y - Baseline | 0.314 | 0.394 | 0.288 |
| Mean Y - Overall | 0.302 | 0.380 | 0.275 |
| | | Fair Health | |
| SIMT | -0.011** | -0.020 | -0.008** |
| | (0.004) | (0.012) | (0.003) |
| Observations | 481,046 | 115,332 | 365,714 |
| Adjusted $R^2$ | 0.016 | 0.018 | 0.010 |
| Mean Y - Baseline | 0.044 | 0.077 | 0.033 |
| Mean Y - Overall | 0.035 | 0.056 | 0.028 |
| | | Poor Health | |
| SIMT | -0.004** | -0.002 | -0.004* |
| | (0.002) | (0.004) | (0.002) |
| Observations | 481,046 | 115,332 | 365,714 |
| Adjusted $R^2$ | 0.004 | 0.005 | 0.002 |
| Mean Y - Baseline | 0.007 | 0.012 | 0.006 |
| Mean Y - Overall | 0.006 | 0.010 | 0.005 |

*Notes:* This table shows results from regressions estimating the effect of marital-status-specific total simulated eligibility on number of children in the family with excellent, very good, good, fair, or poor health. All models include maternal-level controls (indicators for maternal age, state of residence, calendar year, age of the youngest child, age of the oldest child, difference in age between oldest and youngest child, and number of children) and state-level controls (unemployment rate, minimum wage, inflation-adjusted maximum welfare benefit for a family of 4, state-level EITC, implementation of six types of welfare waivers, implementation of any waiver or TANF). In models using the full sample a marital status indicator is included and all controls are interacted with a marital status indicator. Regressions are weighted with maternal survey weights. Standard errors in parentheses are clustered at the state level. The data is from CPS ASEC 1996-2018. The sample is restricted to mothers age 20-64 with children age 0-18. *** $p < 0.01$, ** $p < 0.05$, * $p < 0.10$.

# Appendix

## A Supplemental Figures and Tables

Figure A.1:
Effect of Total Simulated Eligibility on Maternal Labor Supply

(a) Weeks Worked per Year

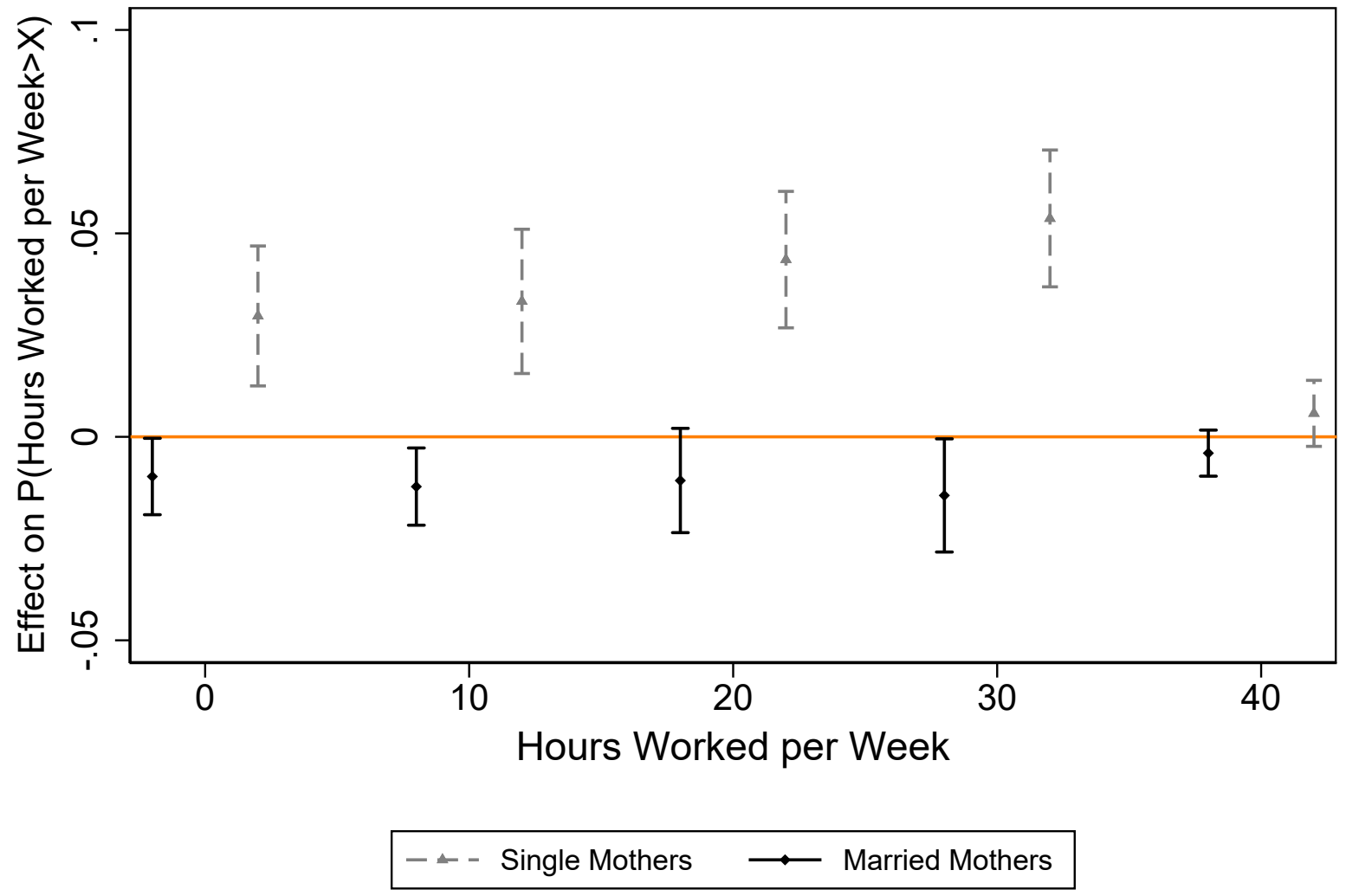

(b) Hours Worked per Week

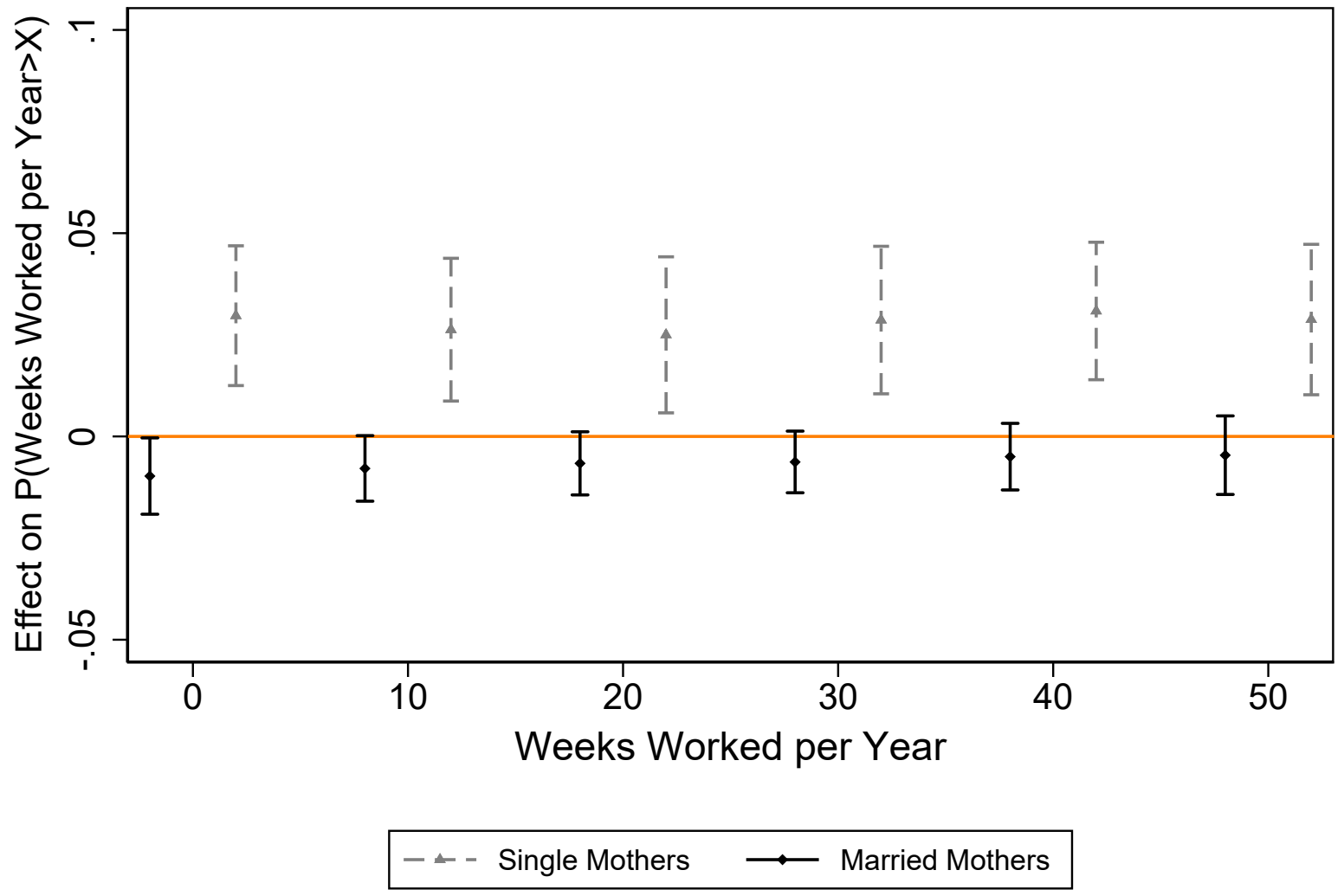

*Notes:* These figures show the coefficients and 95% confidence intervals from regressions estimating the effect of marital-status-specific total simulated eligibility on maternal weeks worked last year and usual hours worked per week last year. Each point estimate and confidence interval is obtained from a different regression where the dependent variable is an indicator equals to one if a mother worked more than X (0,10,...,50) weeks last year or X (0,10,...,40) usual hours worked per week last year. All models include maternal-level controls (indicators for maternal age, state of residence, calendar year, age of the youngest child, age of the oldest child, difference in age between oldest and youngest child, and number of children) and state-level controls (unemployment rate, minimum wage, inflation-adjusted maximum welfare benefit for a family of 4, state-level EITC, implementation of six types of welfare waivers, implementation of any waiver or TANF). Regressions are weighted with maternal survey weights. The data is from CPS ASEC 1978-2018. The sample is restricted to mothers age 20-64 with children age 0-18.

Figure A.2:
Distribution of Maternal Earnings

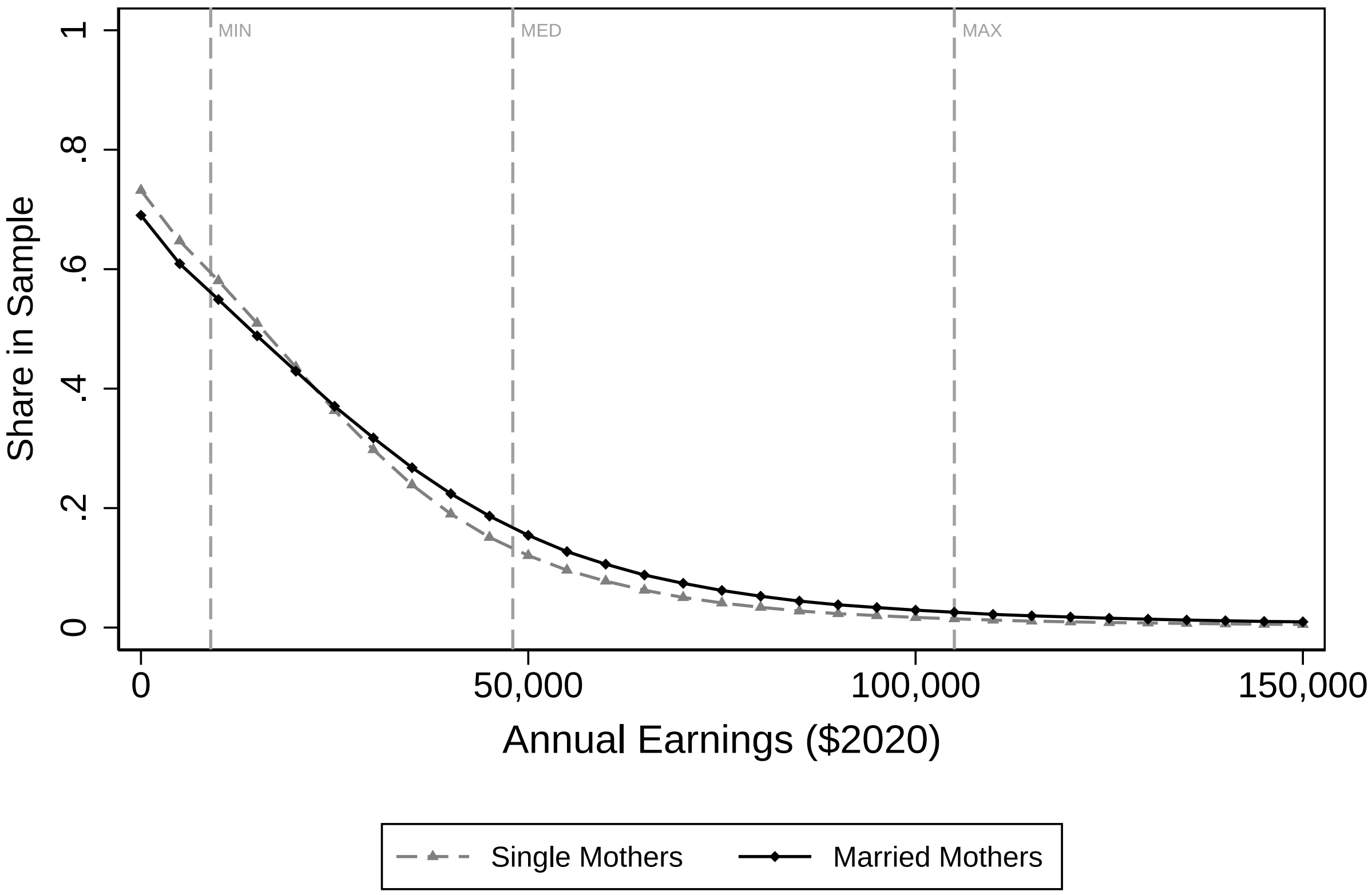

*Notes:* This figure shows the distribution of maternal earnings. The data is from CPS ASEC 1978-2018. The sample is restricted to mothers age 20-64 with children age 0-18.

Figure A.3:
Effect of Total Simulated Eligibility on Maternal Marital Outcomes
Robustness to Identifying Assumption

(a) Married

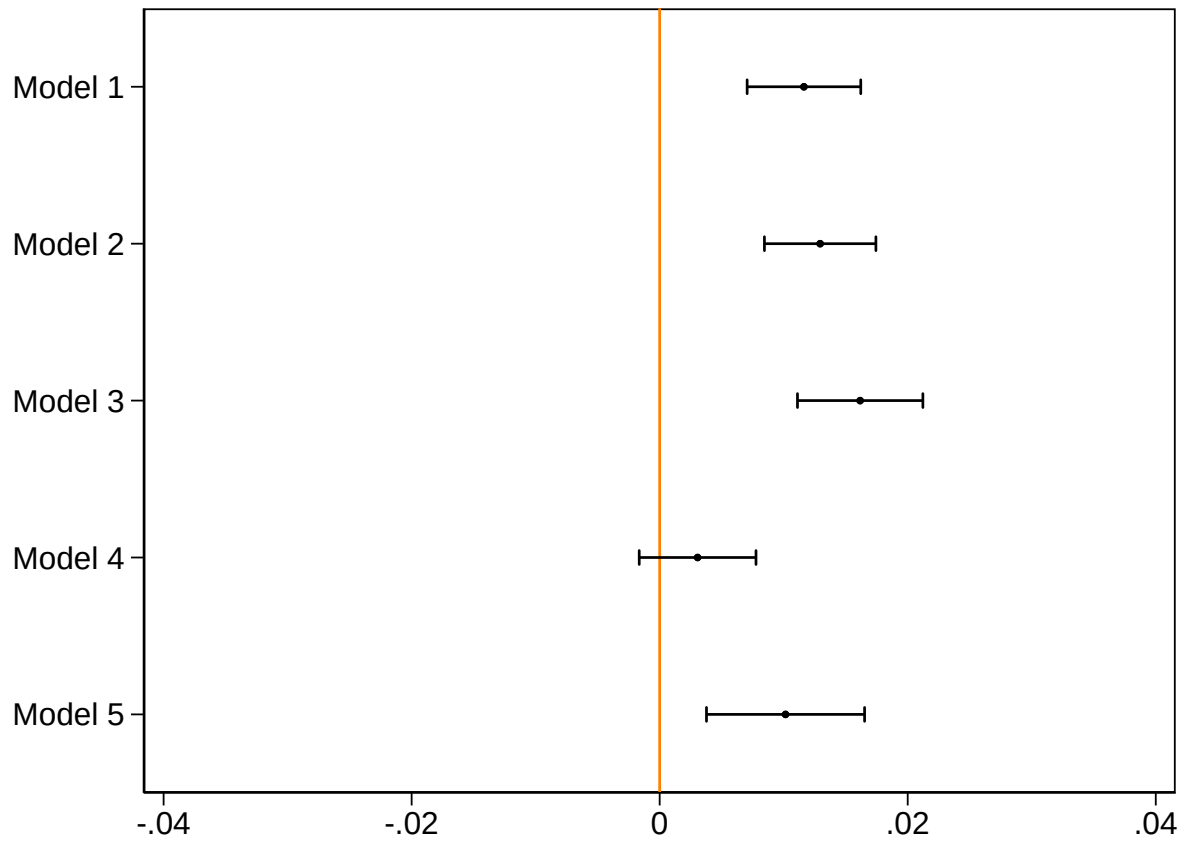

(b) Divorced

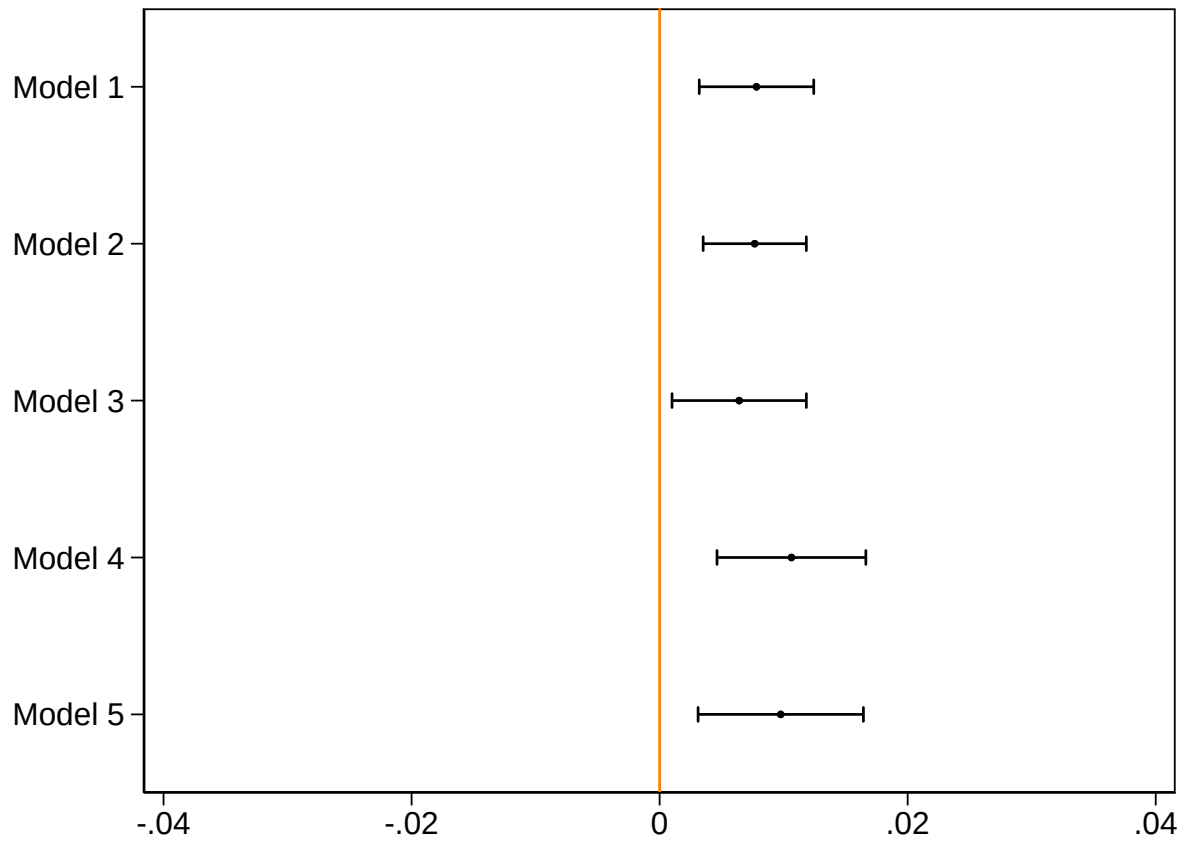

(c) Ever Married

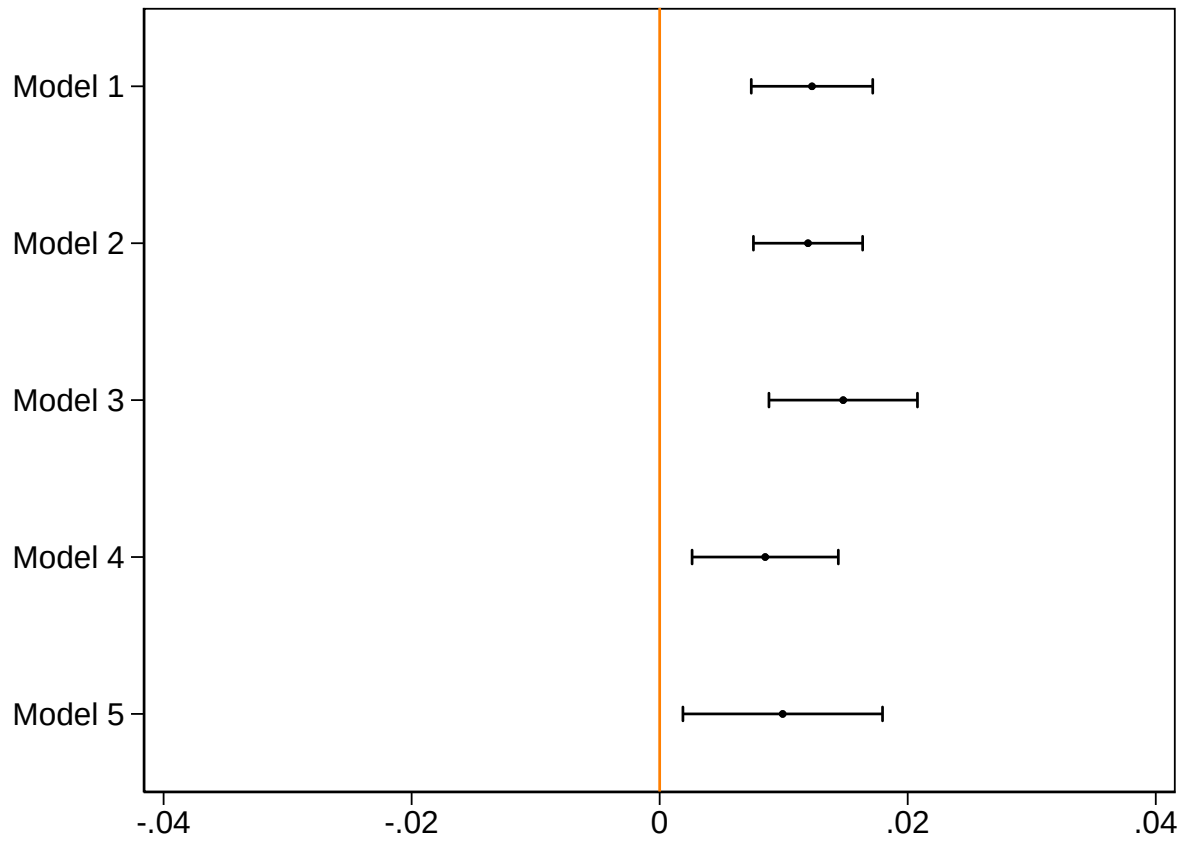

(d) Never Married

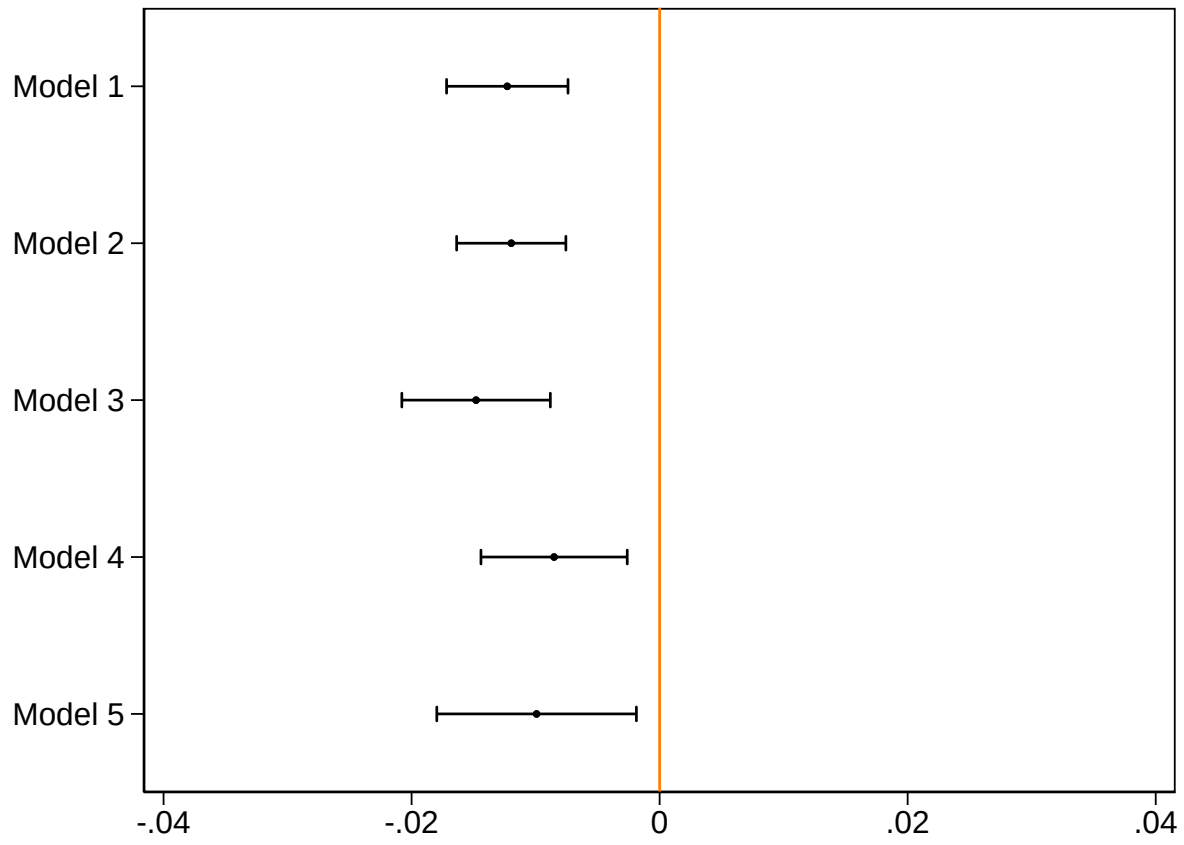

*Notes:* These figures show the coefficients and 95% confidence intervals from regressions estimating the effect of marital-status-specific total simulated eligibility on maternal marital outcomes. All models include maternal-level controls (indicators for maternal age, state of residence, calendar year, age of the youngest child, age of the oldest child, difference in age between oldest and youngest child, and number of children). All models except model 3 and 5 contain state-level controls (unemployment rate, minimum wage, inflation-adjusted maximum welfare benefit for a family of 4, state-level EITC, implementation of six types of welfare waivers, implementation of any waiver or TANF). Model 2 adds state-by-youngest child's age, state-by-oldest child's age, and state-by-difference in age between oldest and youngest child fixed effects to the baseline model. Model 3 adds state-by-year fixed effects to the baseline model. Model 4 adds year-by-youngest child's age, year-by-oldest child's age, and year-by-difference in age between oldest and youngest child fixed effects to the baseline model. Model 5 adds state-by-oldest child's age, state-by-youngest child's age, state-by-difference in age between oldest and youngest child, state-by-year, year-by-youngest child's age, year-by-oldest child's age, year-by-difference in age between oldest and youngest child fixed effects to the baseline model. The data is from CPS ASEC 1977-2018. The sample is restricted to mothers age 20-64 with children age 0-18.

Figure A.4:
Effect of Total Simulated Eligibility on Maternal Labor Supply
Robustness to Identifying Assumption

(a) Labor Force Participation

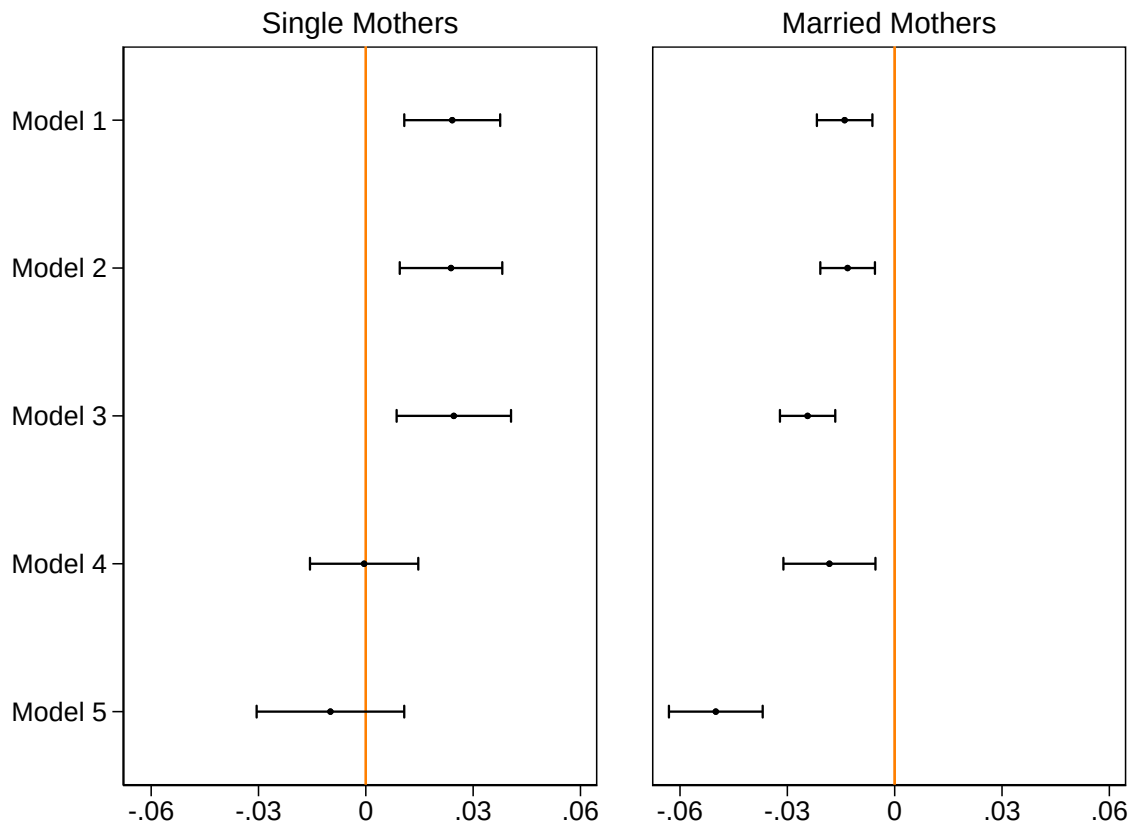

(b) Hours Worked per Week

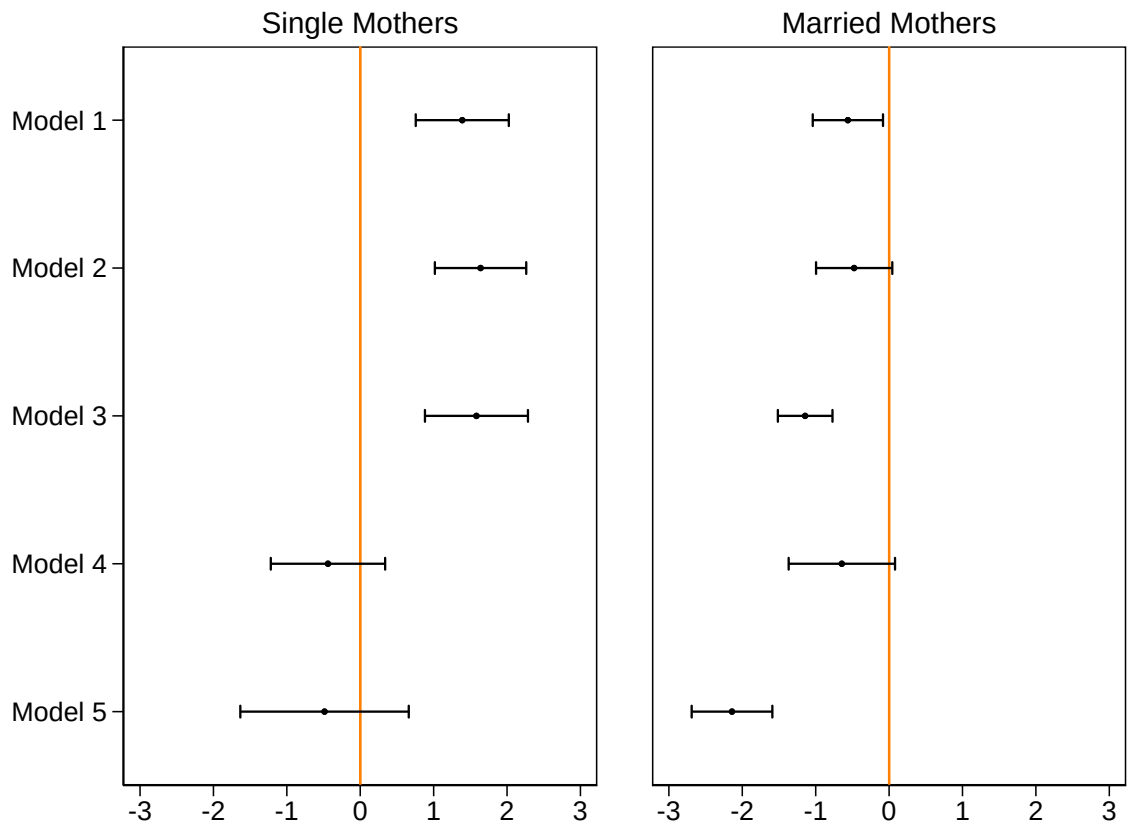

(c) Weeks Worked per Year

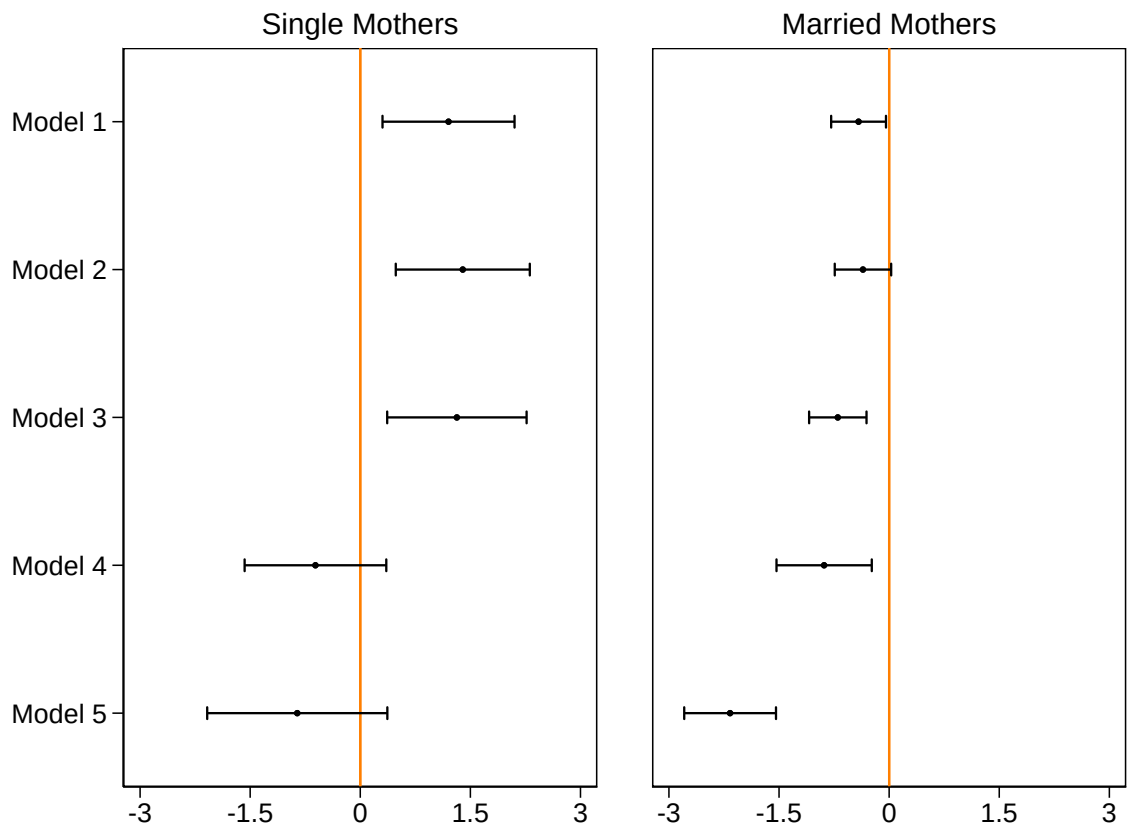

(d) Annual Earnings ($2020)

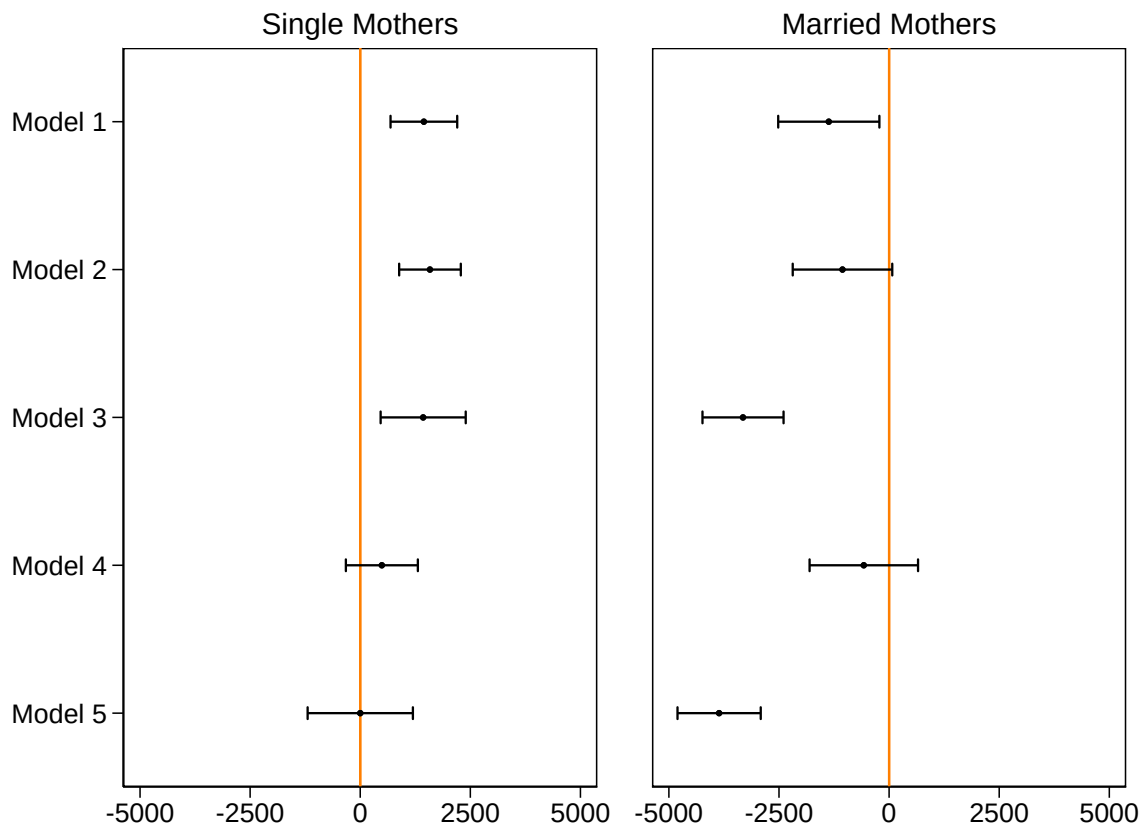

*Notes:* These figures show the coefficients and 95% confidence intervals from regressions estimating the effect of marital-status-specific total simulated eligibility on maternal labor force participation last week, usual hours worked per week last year weeks worked last year, and annual earnings ($2020) last year. All models include maternal-level controls (indicators for maternal age, state of residence, calendar year, age of the youngest child, age of the oldest child, difference in age between oldest and youngest child, and number of children). All models except model 3 and 5 contain state-level controls (unemployment rate, minimum wage, inflation-adjusted maximum welfare benefit for a family of 4, state-level EITC, implementation of six types of welfare waivers, implementation of any waiver or TANF). Model 2 adds state-by-youngest child's age, state-by-oldest child's age, and state-by-difference in age between oldest and youngest child fixed effects to the baseline model. Model 3 adds state-by-year fixed effects to the baseline model. Model 4 adds year-by-youngest child's age, year-by-oldest child's age, and year-by-difference in age between oldest and youngest child fixed effects to the baseline model. Model 5 adds state-by-oldest child's age, state-by-youngest child's age, state-by-difference in age between oldest and youngest child, state-by-year, year-by-youngest child's age, year-by-oldest child's age, year-by-difference in age between oldest and youngest child fixed effects to the baseline model. The data is from CPS ASEC 1977-2018. The sample is restricted to mothers age 20-64 with children age 0-18.

Figure A.5:
Effect of Total Simulated Eligibility on Take-Up & Health of Children
Robustness to Identifying Assumption

(a) Medicaid Coverage

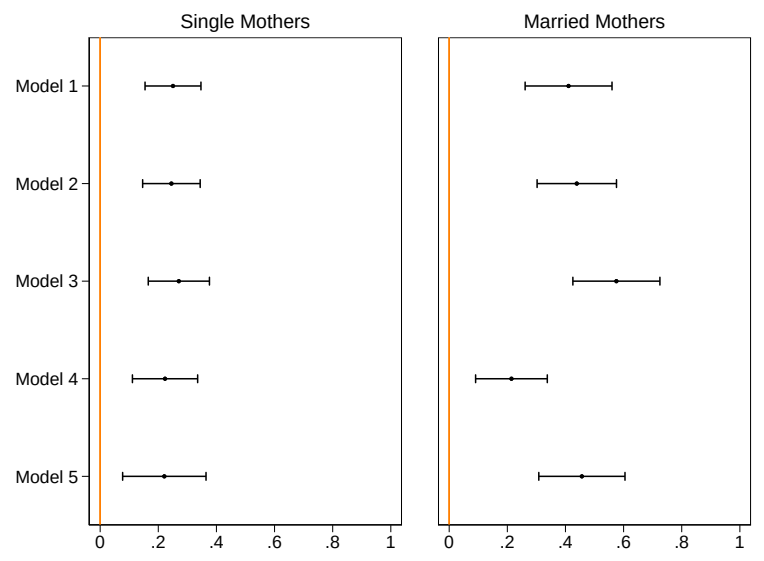

(b) Excellent Health

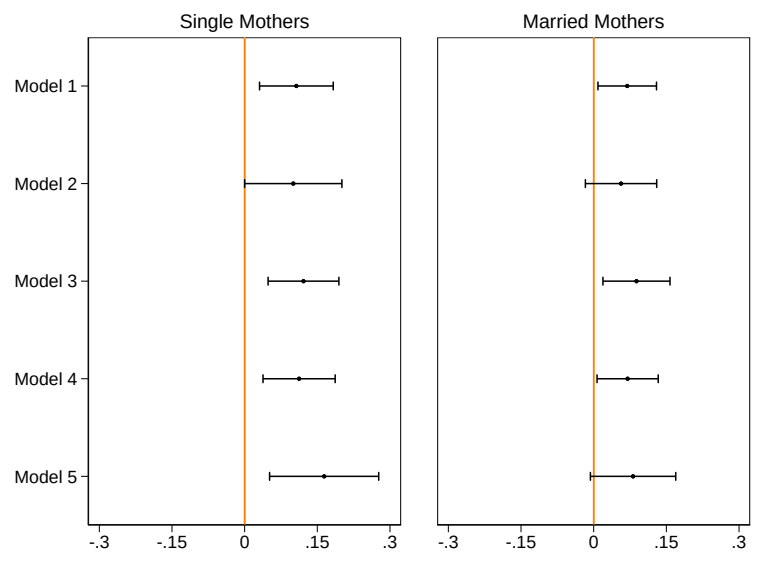

(c) Very Good Health

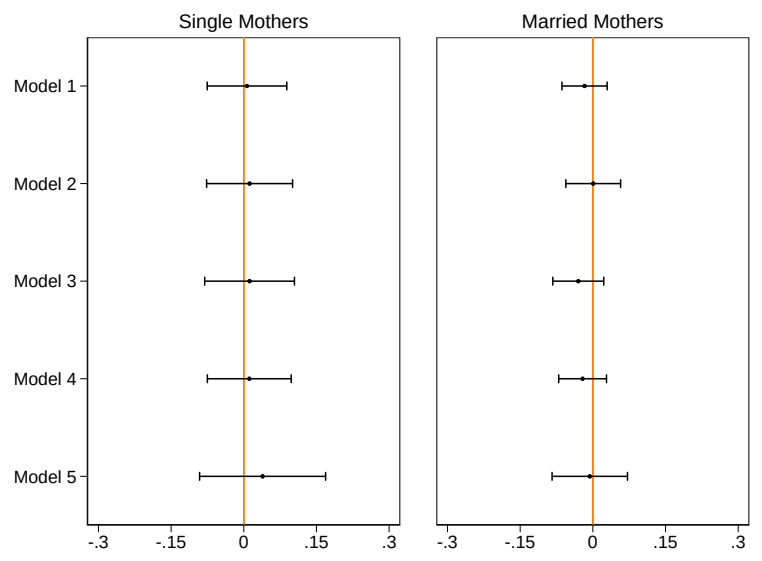

(d) Good Health

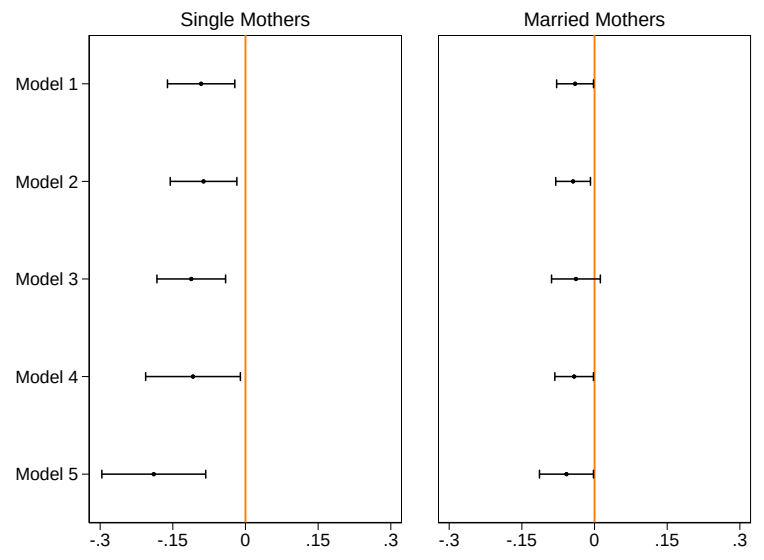

(e) Fair Health

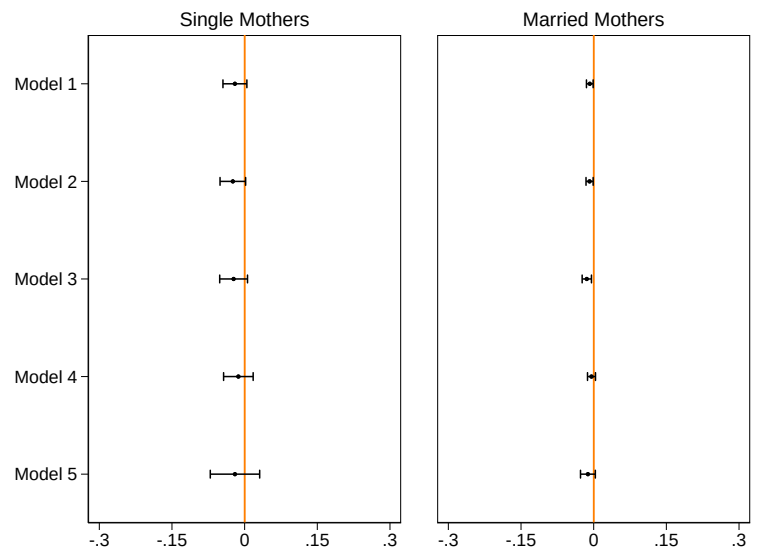

(f) Poor Health

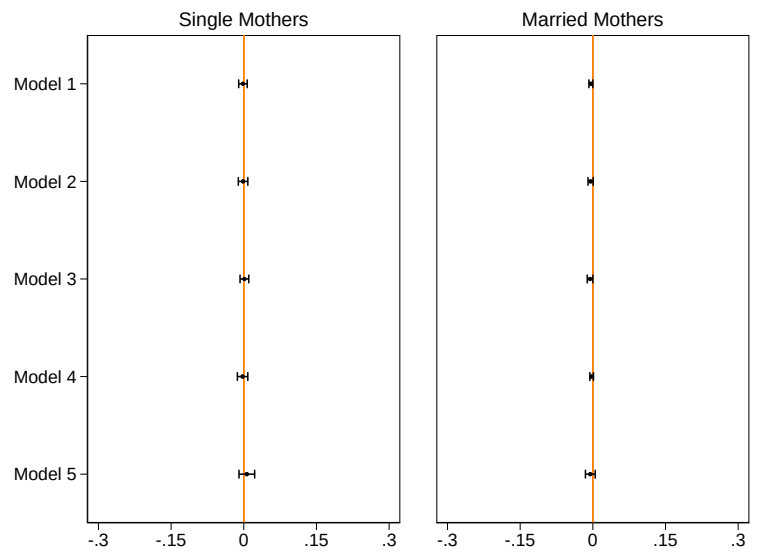

*Notes:* These figures show the coefficients and 95% confidence intervals from regressions estimating the effect of marital-status-specific total simulated eligibility on Medicaid coverage and health of children (indicator for excellent, very good, good, fair, and poor health). All models include maternal-level controls (indicators for maternal age, state of residence, calendar year, age of the youngest child, age of the oldest child, difference in age between oldest and youngest child, and number of children). All models include maternal-level controls (indicators for maternal age, state of residence, calendar year, age of the youngest child, age of the oldest child, difference in age between oldest and youngest child, and number of children). All models except model 3 and 5 contain state-level controls (unemployment rate, minimum wage, inflation-adjusted maximum welfare benefit for a family of 4, state-level EITC, implementation of six types of welfare waivers, implementation of any waiver or TANF). Model 2 adds state-by-youngest child's age, state-by-oldest child's age, and state-by-difference in age between oldest and youngest child fixed effects to the baseline model. Model 3 adds state-by-year fixed effects to the baseline model. Model 4 adds year-by-youngest child's age, year-by-oldest child's age, and year-by-difference in age between oldest and youngest child fixed effects to the baseline model. Model 5 adds state-by-oldest child's age, state-by-youngest child's age, state-by-difference in age between oldest and youngest child, state-by-year, year-by-youngest child's age, year-by-oldest child's age, year-by-difference in age between oldest and youngest child fixed effects to the baseline model. The data is from CPS ASEC 1996-2018. The sample is restricted to mothers age 20-64 with children age 0-18.

## Figure A.6:
## Effect of Total Simulated Eligibility on Maternal Educational Attainment Robustness to Identifying Assumption

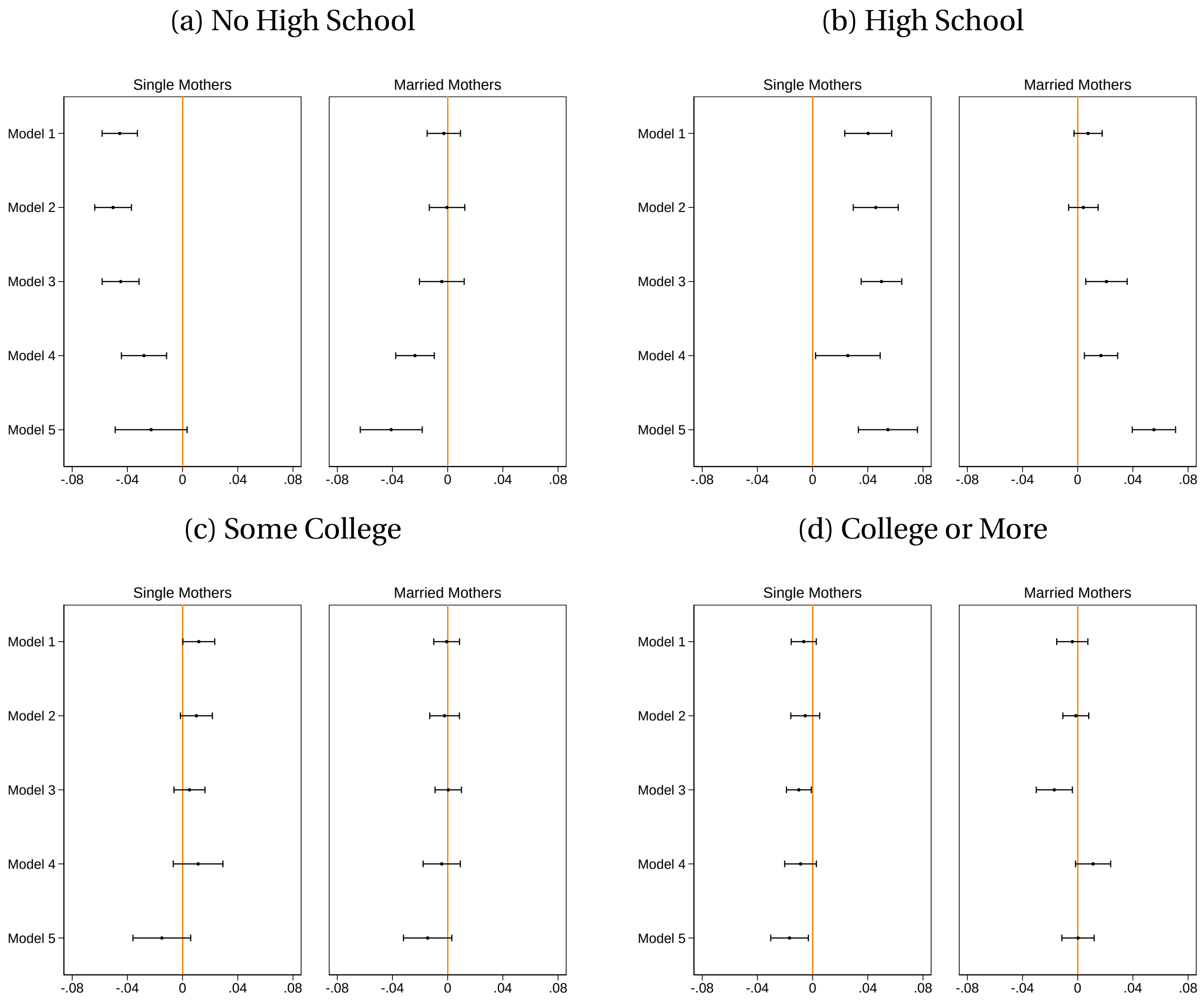

*Notes:* These figures show the coefficients and 95% confidence intervals from regressions estimating the effect of marital-status-specific total simulated eligibility on an indicator equals to one if the mother (a) has not graduated from high school, (b) finished high school, (c) attended some college, and (d) attained college or more. All models include maternal-level controls (indicators for maternal age, state of residence, calendar year, age of the youngest child, age of the oldest child, difference in age between oldest and youngest child, and number of children). All models except model 3 and 5 contain state-level controls (unemployment rate, minimum wage, inflation-adjusted maximum welfare benefit for a family of 4, state-level EITC, implementation of six types of welfare waivers, implementation of any waiver or TANF). Model 2 adds state-by-youngest child's age, state-by-oldest child's age, and state-by-difference in age between oldest and youngest child fixed effects to the baseline model. Model 3 adds state-by-year fixed effects to the baseline model. Model 4 adds year-by-youngest child's age, year-by-oldest child's age, and year-by-difference in age between oldest and youngest child fixed effects to the baseline model. Model 5 adds state-by-oldest child's age, state-by-youngest child's age, state-by-difference in age between oldest and youngest child, state-by-year, year-by-youngest child's age, year-by-oldest child's age, year-by-difference in age between oldest and youngest child fixed effects to the baseline model. The data is from CPS ASEC 1977-2018. The sample is restricted to mothers age 20-64 with children age 0-18.

Table A.1:
Effect of Total Simulated Eligibility on Maternal Occupational Choice

| | All | Single | Married |
|---|---|---|---|
| | | Managerial, Professional | |
| SIMT | -0.0071** | 0.0120*** | -0.0133*** |
| | (0.0034) | (0.0040) | (0.0042) |
| Observations | 785,386 | 188,438 | 596,948 |
| Adjusted $R^2$ | 0.0781 | 0.0614 | 0.0751 |
| Mean Y - Baseline | 0.1152 | 0.1028 | 0.1182 |
| Mean Y - Overall | 0.2187 | 0.1626 | 0.2377 |
| | | Technical, Sales, Administrative | |
| SIMT | 0.0205*** | 0.0291*** | 0.0178*** |
| | (0.0044) | (0.0058) | (0.0052) |
| Observations | 785,386 | 188,438 | 596,948 |
| Adjusted $R^2$ | 0.0250 | 0.0243 | 0.0248 |
| Mean Y - Baseline | 0.2559 | 0.2655 | 0.2536 |
| Mean Y - Overall | 0.2726 | 0.2868 | 0.2678 |
| | | Farming, Forestry, Fishing | |
| SIMT | -0.0007 | -0.0022* | -0.0002 |
| | (0.0011) | (0.0011) | (0.0012) |
| Observations | 785,386 | 188,438 | 596,948 |
| Adjusted $R^2$ | 0.0068 | 0.0036 | 0.0073 |
| Mean Y - Baseline | 0.0134 | 0.0068 | 0.0150 |
| Mean Y - Overall | 0.0067 | 0.0047 | 0.0074 |

continued on next page

Table A.2:
Effect of Total Simulated Eligibility on Maternal Occupational Choice

| | All | Single | Married |
|---|---|---|---|
| | | Service | |
| SIMT | -0.0101*** | -0.0116* | -0.0097*** |
| | (0.0032) | (0.0065) | (0.0031) |
| Observations | 785,386 | 188,438 | 596,948 |
| Adjusted $R^2$ | 0.0246 | 0.0130 | 0.0171 |
| Mean Y - Baseline | 0.1204 | 0.1591 | 0.1110 |
| Mean Y - Overall | 0.1439 | 0.2008 | 0.1246 |
| | | Precision Production, Craft, Repair | |
| SIMT | -0.0021** | -0.0021 | -0.0021** |
| | (0.0009) | (0.0017) | (0.0008) |
| Observations | 785,386 | 188,438 | 596,948 |
| Adjusted $R^2$ | 0.0019 | 0.0017 | 0.0017 |
| Mean Y - Baseline | 0.0120 | 0.0184 | 0.0104 |
| Mean Y - Overall | 0.0144 | 0.0166 | 0.0137 |
| | | Operators, Fabricators, Laborers | |
| SIMT | -0.0034 | -0.0005 | -0.0043* |
| | (0.0022) | (0.0048) | (0.0023) |
| Observations | 785,386 | 188,438 | 596,948 |
| Adjusted $R^2$ | 0.0270 | 0.0209 | 0.0271 |
| Mean Y - Baseline | 0.0836 | 0.1064 | 0.0781 |
| Mean Y - Overall | 0.0571 | 0.0747 | 0.0512 |

*Notes:* This table shows results from regressions estimating the effect of marital-status-specific total simulated eligibility on the probability of mothers reporting being in a one-digit occupation last year. All models include maternal-level controls (indicators for maternal age, state of residence, calendar year, age of the youngest child, age of the oldest child, difference in age between oldest and youngest child, and number of children) and state-level controls (unemployment rate, minimum wage, inflation-adjusted maximum welfare benefit for a family of 4, state-level EITC, implementation of six types of welfare waivers, implementation of any waiver or TANF). In models using the full sample a marital status indicator is included and all controls are interacted with a marital status indicator. Regressions are weighted with maternal survey weights. Standard errors in parentheses are clustered at the state level. The data is from CPS ASEC 1977-2018. The sample is restricted to mothers age 20-64 with children age 0-18. *** $p < 0.01$, ** $p < 0.05$, * $p < 0.10$.

Table A.3:
Effect of Total Simulated Eligibility on Maternal Occupational Choice

| | All | Single | Married |
|---|---|---|---|
| | Occupation with Wages above 25th Percentile | | |
| SIMT | 0.0111*** | 0.0494*** | -0.0013 |
| | (0.0038) | (0.0061) | (0.0044) |
| Observations | 785,386 | 188,438 | 596,948 |
| Adjusted $R^2$ | 0.0801 | 0.0895 | 0.0739 |
| Mean Y - Baseline | 0.3159 | 0.3325 | 0.3119 |
| Mean Y - Overall | 0.4440 | 0.3999 | 0.4589 |
| | Occupation with Wages above 50th Percentile | | |
| SIMT | -0.0032 | 0.0234*** | -0.0118** |
| | (0.0037) | (0.0042) | (0.0048) |
| Observations | 785,386 | 188,438 | 596,948 |
| Adjusted $R^2$ | 0.0799 | 0.0710 | 0.0770 |
| Mean Y - Baseline | 0.1374 | 0.1380 | 0.1373 |
| Mean Y - Overall | 0.2706 | 0.2185 | 0.2883 |
| | Occupation with Wages above 75th Percentile | | |
| SIMT | -0.0038 | 0.0107*** | -0.0085*** |
| | (0.0025) | (0.0028) | (0.0029) |
| Observations | 785,386 | 188,438 | 596,948 |
| Adjusted $R^2$ | 0.0506 | 0.0450 | 0.0479 |
| Mean Y - Baseline | 0.0742 | 0.0694 | 0.0754 |
| Mean Y - Overall | 0.1461 | 0.1105 | 0.1582 |

*Notes:* This table shows results from regressions estimating the effect of marital-status-specific total simulated eligibility on the probability of mothers reporting being last year in a three-digit occupation with average wage above 25, 50, and 75 percentile. All models include maternal-level controls (indicators for maternal age, state of residence, calendar year, age of the youngest child, age of the oldest child, difference in age between oldest and youngest child, and number of children) and state-level controls (unemployment rate, minimum wage, inflation-adjusted maximum welfare benefit for a family of 4, state-level EITC, implementation of six types of welfare waivers, implementation of any waiver or TANF). In models using the full sample a marital status indicator is included and all controls are interacted with a marital status indicator. Regressions are weighted with maternal survey weights. Standard errors in parentheses are clustered at the state level. The data is from CPS ASEC 1977-2018. The sample is restricted to mothers age 20-64 with children age 0-18. *** $p < 0.01$, ** $p < 0.05$, * $p < 0.10$.

Table A.4:
Effect of Total Simulated Eligibility on Maternal Occupational Mobility

| | Retrospective Occupational Mobility | | | Longitudinal Occupational Mobility | | |
|---|---|---|---|---|---|---|
| | All | Single | Married | All | Single | Married |
| | One-Digit Occupational Mobility | | | | | |
| SIMT | -0.0062* | -0.0102 | -0.0050 | 0.0097 | 0.0337** | 0.0020 |
| | ( 0.0034) | ( 0.0085) | ( 0.0030) | ( 0.0080) | ( 0.0151) | ( 0.0080) |
| Observations | 147,653 | 34,209 | 113,444 | 147,549 | 34,156 | 113,393 |
| Adjusted $R^2$ | 0.0104 | 0.0132 | 0.0051 | 0.0131 | 0.0152 | 0.0072 |
| Mean Y - Baseline | 0.0238 | 0.0350 | 0.0210 | 0.1506 | 0.1553 | 0.1494 |
| Mean Y - Overall | 0.0336 | 0.0465 | 0.0294 | 0.2023 | 0.2422 | 0.1896 |
| | Two-Digit Occupational Mobility | | | | | |
| SIMT | -0.0066** | -0.0150* | -0.0040 | 0.0111 | 0.0286* | 0.0055 |
| | ( 0.0032) | ( 0.0089) | ( 0.0031) | ( 0.0085) | ( 0.0150) | ( 0.0086) |
| Observations | 147,653 | 34,209 | 113,444 | 147,549 | 34,156 | 113,393 |
| Adjusted $R^2$ | 0.0120 | 0.0138 | 0.0065 | 0.0141 | 0.0145 | 0.0090 |
| Mean Y - Baseline | 0.0277 | 0.0423 | 0.0239 | 0.1853 | 0.1699 | 0.1893 |
| Mean Y - Overall | 0.0442 | 0.0608 | 0.0389 | 0.2735 | 0.3173 | 0.2596 |
| | Three-Digit Occupational Mobility | | | | | |
| SIMT | -0.0027 | 0.0026 | -0.0044 | 0.0054 | 0.0363** | -0.0045 |
| | ( 0.0049) | ( 0.0132) | ( 0.0043) | ( 0.0076) | ( 0.0172) | ( 0.0079) |
| Observations | 147,653 | 34,209 | 113,444 | 147,549 | 34,156 | 113,393 |
| Adjusted $R^2$ | 0.0142 | 0.0135 | 0.0074 | 0.0147 | 0.0142 | 0.0100 |
| Mean Y - Baseline | 0.0417 | 0.0735 | 0.0336 | 0.3622 | 0.3664 | 0.3611 |
| Mean Y - Overall | 0.0840 | 0.1146 | 0.0743 | 0.4399 | 0.4899 | 0.4240 |

*Notes:* This table shows results from regressions estimating the effect of marital-status-specific total simulated eligibility on maternal retrospective and longitudinal one-, two-, and three-digit occupational mobility. All models include maternal-level controls (indicators for maternal age, state of residence, calendar year, age of the youngest child, age of the oldest child, difference in age between oldest and youngest child, and number of children) and state-level controls (unemployment rate, minimum wage, inflation-adjusted maximum welfare benefit for a family of 4, state-level EITC, implementation of six types of welfare waivers, implementation of any waiver or TANF). In models using the full sample a marital status indicator is included and all controls are interacted with a marital status indicator. Regressions are weighted with maternal survey weights. Standard errors in parentheses are clustered at the state level. The data is from CPS ASEC 1977-2018. The sample is restricted to mothers age 20-64 with children age 0-18. *** $p < 0.01$, ** $p < 0.05$, * $p < 0.10$.

Table A.5:
Effect of Total Simulated Eligibility on Maternal Three-Digit Occupational Mobility

| | Retrospective Occupational Mobility | | | Longitudinal Occupational Mobility | | |
|---|---|---|---|---|---|---|
| | All | Single | Married | All | Single | Married |
| | Higher Paid Occupation | | | | | |
| SIMT | -0.0060* | 0.0033 | -0.0089*** | 0.0030 | 0.0285 | -0.0051 |
| | ( 0.0034) | ( 0.0098) | ( 0.0032) | ( 0.0052) | ( 0.0173) | ( 0.0057) |
| Observations | 147,653 | 34,209 | 113,444 | 147,534 | 34,153 | 113,381 |
| Adjusted $R^2$ | 0.0084 | 0.0086 | 0.0043 | 0.0062 | 0.0073 | 0.0038 |
| Mean Y - Baseline | 0.0229 | 0.0455 | 0.0172 | 0.1822 | 0.2021 | 0.1771 |
| Mean Y - Overall | 0.0435 | 0.0596 | 0.0383 | 0.2134 | 0.2386 | 0.2053 |
| | Occupation with Higher Variance of Wages | | | | | |
| SIMT | -0.0045 | -0.0035 | -0.0048 | 0.0020 | 0.0247** | -0.0052 |
| | ( 0.0035) | ( 0.0089) | ( 0.0030) | ( 0.0061) | ( 0.0117) | ( 0.0070) |
| Observations | 147,653 | 34,209 | 113,444 | 147,534 | 34,153 | 113,381 |
| Adjusted $R^2$ | 0.0078 | 0.0081 | 0.0040 | 0.0060 | 0.0076 | 0.0035 |
| Mean Y - Baseline | 0.0224 | 0.0398 | 0.0179 | 0.1831 | 0.1753 | 0.1851 |
| Mean Y - Overall | 0.0417 | 0.0566 | 0.0369 | 0.2130 | 0.2372 | 0.2053 |
| | Occupation with Higher Seperation Rates | | | | | |
| SIMT | 0.0019 | 0.0069 | 0.0004 | -0.0012 | 0.0082 | -0.0041 |
| | ( 0.0040) | ( 0.0104) | ( 0.0043) | ( 0.0078) | ( 0.0131) | ( 0.0083) |
| Observations | 147,653 | 34,209 | 113,444 | 147,534 | 34,153 | 113,381 |
| Adjusted $R^2$ | 0.0086 | 0.0088 | 0.0043 | 0.0055 | 0.0068 | 0.0034 |
| Mean Y - Baseline | 0.0203 | 0.0366 | 0.0161 | 0.1694 | 0.1981 | 0.1620 |
| Mean Y - Overall | 0.0410 | 0.0565 | 0.0361 | 0.2060 | 0.2265 | 0.1994 |
| | Occupation with Higher Educational Requirements | | | | | |
| SIMT | -0.0008 | -0.0025 | -0.0003 | 0.0062 | 0.0215 | 0.0012 |
| | ( 0.0036) | ( 0.0087) | ( 0.0032) | ( 0.0051) | ( 0.0140) | ( 0.0055) |
| Observations | 146,567 | 34,010 | 112,557 | 146,056 | 33,878 | 112,178 |
| Adjusted $R^2$ | 0.0085 | 0.0102 | 0.0041 | 0.0069 | 0.0079 | 0.0042 |
| Mean Y - Baseline | 0.0184 | 0.0145 | 0.0194 | 0.1692 | 0.1971 | 0.1621 |
| Mean Y - Overall | 0.0428 | 0.0568 | 0.0383 | 0.2083 | 0.2345 | 0.1999 |

*Notes:* This table shows results from regressions estimating the effect of marital-status-specific total simulated eligibility on different types of maternal retrospective and longitudinal three-digit occupational mobility. All models include maternal-level controls (indicators for maternal age, state of residence, calendar year, age of the youngest child, age of the oldest child, difference in age between oldest and youngest child, and number of children) and state-level controls (unemployment rate, minimum wage, inflation-adjusted maximum welfare benefit for a family of 4, state-level EITC, implementation of six types of welfare waivers, implementation of any waiver or TANF). In models using the full sample a marital status indicator is included and all controls are interacted with a marital status indicator. Regressions are weighted with maternal survey weights. Standard errors in parentheses are clustered at the state level. The data is from CPS ASEC 1977-2018. The sample is restricted to mothers age 20-64 with children age 0-18. *** $p < 0.01$, ** $p < 0.05$, * $p < 0.10$.

Table A.6:
Effect of Total Simulated Eligibility on Maternal Educational Attainment

| | All | Single | Married |
|---|---|---|---|
| | | **No High School** | |
| SIMT | -0.01*** | -0.05*** | -0.00 |
| | (0.00) | (0.01) | (0.01) |
| Observations | 789,581 | 187,976 | 601,605 |
| Adjusted $R^2$ | 0.12 | 0.11 | 0.12 |
| Mean Y - Baseline | 0.26 | 0.40 | 0.23 |
| Mean Y - Overall | 0.15 | 0.21 | 0.13 |
| | | **High School** | |
| SIMT | 0.02*** | 0.04*** | 0.01 |
| | (0.00) | (0.01) | (0.01) |
| Observations | 789,581 | 187,976 | 601,605 |
| Adjusted $R^2$ | 0.06 | 0.02 | 0.07 |
| Mean Y - Baseline | 0.45 | 0.38 | 0.47 |
| Mean Y - Overall | 0.35 | 0.37 | 0.34 |
| | | **Some College** | |
| SIMT | 0.00 | 0.01** | -0.00 |
| | (0.00) | (0.01) | (0.00) |
| Observations | 789,581 | 187,976 | 601,605 |
| Adjusted $R^2$ | 0.03 | 0.04 | 0.02 |
| Mean Y - Baseline | 0.17 | 0.16 | 0.17 |
| Mean Y - Overall | 0.27 | 0.29 | 0.26 |
| | | **College or More** | |
| SIMT | -0.00 | -0.01 | -0.00 |
| | (0.00) | (0.00) | (0.01) |
| Observations | 789,581 | 187,976 | 601,605 |
| Adjusted $R^2$ | 0.17 | 0.09 | 0.17 |
| Mean Y - Baseline | 0.11 | 0.06 | 0.13 |
| Mean Y - Overall | 0.23 | 0.13 | 0.27 |

*Notes:* This table shows results from regressions estimating the effect of marital-status-specific total simulated eligibility on maternal educational attainment (indicator for no high school, high school, some college, and college or more). All models include maternal-level controls (indicators for maternal age, state of residence, calendar year, age of the youngest child, age of the oldest child, difference in age between oldest and youngest child, and number of children) and state-level controls (unemployment rate, minimum wage, inflation-adjusted maximum welfare benefit for a family of 4, state-level EITC, implementation of six types of welfare waivers, implementation of any waiver or TANF). In models using the full sample a marital status indicator is included and all controls are interacted with a marital status indicator. Regressions are weighted with maternal survey weights. Standard errors in parentheses are clustered at the state level. The data is from CPS ASEC 1996-2018. The sample is restricted to mothers age 20-64 with children age 0-18. *** $p < 0.01$, ** $p < 0.05$, * $p < 0.10$.

Table A.7:
Effect of Total Simulated Eligibility on Medicaid Cost and Benefits & Taxes

| | All | Single | Married |
|---|---|---|---|
| | | Medicaid Cost | |
| SIMT | 993*** | 1,058*** | 971*** |
| | ( 122) | ( 77) | ( 142) |
| Observations | 597,688 | 145,437 | 452,251 |
| Adjusted $R^2$ | 0.80 | 0.80 | 0.79 |
| Mean Y - Baseline | 289 | 279 | 292 |
| Mean Y - Overall | 905 | 859 | 921 |
| | | Federal & State Net Tax | |
| SIMT | 378 | -574*** | 685 |
| | ( 569) | ( 132) | ( 721) |
| Observations | 785,386 | 188,438 | 596,948 |
| Adjusted $R^2$ | 0.13 | 0.06 | 0.09 |
| Mean Y - Baseline | 12,208 | 2,318 | 14,599 |
| Mean Y - Overall | 12,437 | 995 | 16,328 |
| | | Federal & State Net Tax (19% FICA) | |
| SIMT | 387 | -542*** | 686 |
| | ( 590) | ( 134) | ( 749) |
| Observations | 785,386 | 188,438 | 596,948 |
| Adjusted $R^2$ | 0.14 | 0.06 | 0.09 |
| Mean Y - Baseline | 13,310 | 2,676 | 15,881 |
| Mean Y - Overall | 14,303 | 1,634 | 18,611 |
| | | Federal & State Benefits | |
| SIMT | 871*** | 724*** | 918*** |
| | ( 141) | ( 58) | ( 188) |
| Observations | 785,386 | 188,438 | 596,948 |
| Adjusted $R^2$ | 0.23 | 0.23 | 0.19 |
| Mean Y - Baseline | 115 | 239 | 85 |
| Mean Y - Overall | 848 | 1,407 | 658 |
| | | Federal & State Tax | |
| SIMT | 1,249*** | 150 | 1,603*** |
| | ( 451) | ( 100) | ( 565) |
| Observations | 785,386 | 188,438 | 596,948 |
| Adjusted $R^2$ | 0.13 | 0.04 | 0.08 |
| Mean Y - Baseline | 12,322 | 2,557 | 14,684 |
| Mean Y - Overall | 13,285 | 2,402 | 16,986 |

*Notes:* This table shows results from regressions estimating the effect of marital-status-specific total simulated eligibility on child's total Medicaid cost ($2020) and maternal annual federal and state taxes ($2020) last year. All models include maternal-level controls (indicators for maternal age, state of residence, calendar year, age of the youngest child, age of the oldest child, difference in age between oldest and youngest child, and number of children) and state-level controls (unemployment rate, minimum wage, inflation-adjusted maximum welfare benefit for a family of 4, state-level EITC, implementation of six types of welfare waivers, implementation of any waiver or TANF). In models using the full sample a marital status indicator is included and all controls are interacted with a marital status indicator. Regressions are weighted with maternal survey weights. Standard errors in parentheses are clustered at the state level. The data is from CPS ASEC 1981-2012. The sample is restricted to mothers age 20-64 with children age 0-18. *** $p < 0.01$, ** $p < 0.05$, * $p < 0.10$.

Table A.8:
Effect of Total Simulated Eligibility on Family-Level Government Transfers

| | All | Single | Married |
|---|---|---|---|
| | | Public Assistance | |
| SIMT | -235.78*** | -683.37*** | -91.64*** |
| | (48.16) | (168.44) | (29.79) |
| Observations | 785,386 | 188,438 | 596,948 |
| Adjusted $R^2$ | 0.24 | 0.23 | 0.02 |
| Mean Y - Baseline | 876.56 | 3953.75 | 132.36 |
| Mean Y - Overall | 408.93 | 1314.74 | 100.92 |
| | | Disability Income | |
| SIMT | -4.09 | 14.33 | -10.05 |
| | (7.51) | (19.31) | (7.06) |
| Observations | 613,681 | 149,855 | 463,826 |
| Adjusted $R^2$ | 0.00 | 0.00 | 0.00 |
| Mean Y - Baseline | 34.07 | 37.67 | 32.84 |
| Mean Y - Overall | 60.82 | 80.53 | 53.93 |
| | | Supplemental Security Income | |
| SIMT | 25.84*** | 82.94*** | 7.45 |
| | (8.55) | (28.84) | (6.52) |
| Observations | 785,386 | 188,438 | 596,948 |
| Adjusted $R^2$ | 0.02 | 0.01 | 0.00 |
| Mean Y - Baseline | 39.34 | 155.75 | 11.19 |
| Mean Y - Overall | 126.71 | 329.66 | 57.70 |
| | | Unemployment Compensation | |
| SIMT | 16.57 | 9.76 | 18.76 |
| | (10.83) | (31.41) | (12.72) |
| Observations | 785,386 | 188,438 | 596,948 |
| Adjusted $R^2$ | 0.01 | 0.01 | 0.00 |
| Mean Y - Baseline | 216.98 | 437.77 | 163.58 |
| Mean Y - Overall | 187.53 | 283.55 | 154.88 |
| | | Education Assistance | |
| SIMT | -29.17*** | -26.12 | -30.16*** |
| | (10.48) | (26.45) | (9.10) |
| Observations | 613,681 | 149,855 | 463,826 |
| Adjusted $R^2$ | 0.01 | 0.01 | 0.01 |
| Mean Y - Baseline | 90.02 | 196.06 | 53.71 |
| Mean Y - Overall | 194.21 | 336.31 | 144.56 |

continued on next page

Table A.8:
Effect of Total Simulated Eligibility on Family-Level Government Transfers (Continued)

| | All | Single | Married |
|---|---|---|---|
| | Supplemental Nutrition Assistance Program | | |
| SIMT | 49.77 | -90.11* | 95.41* |
| | (40.33) | (48.37) | (47.55) |
| Observations | 747,470 | 181,122 | 566,348 |
| Adjusted $R^2$ | 0.22 | 0.22 | 0.08 |
| Mean Y - Baseline | 483.97 | 1468.35 | 200.62 |
| Mean Y - Overall | 559.37 | 1378.46 | 277.04 |
| | Monetary Value of School Lunch | | |
| SIMT | -77.98*** | -109.54*** | -67.76*** |
| | (15.55) | (20.00) | (18.56) |
| Observations | 556,717 | 135,157 | 421,560 |
| Adjusted $R^2$ | 0.36 | 0.43 | 0.29 |
| Mean Y - Baseline | 273.93 | 410.38 | 220.55 |
| Mean Y - Overall | 252.94 | 385.92 | 206.70 |
| | Housing Subsidy | | |
| SIMT | 12.09 | -7.65 | 18.64 |
| | (16.92) | (46.21) | (14.40) |
| Observations | 496,024 | 120,576 | 375,448 |
| Adjusted $R^2$ | 0.13 | 0.08 | 0.02 |
| Mean Y - Baseline | 243.63 | 690.63 | 68.75 |
| Mean Y - Overall | 187.06 | 566.40 | 54.85 |
| | Energy Subsidy | | |
| SIMT | -8.77** | -20.90** | -4.85 |
| | (4.10) | (9.56) | (3.50) |
| Observations | 711,868 | 173,267 | 538,601 |
| Adjusted $R^2$ | 0.05 | 0.05 | 0.02 |
| Mean Y - Baseline | 57.59 | 140.33 | 30.35 |
| Mean Y - Overall | 39.90 | 96.16 | 20.41 |

*Notes:* This table shows results from regressions estimating the effect of marital-status-specific total simulated eligibility on mother- and family-level annual income in 2020 dollars from public assistance, disability income, Supplemental Security Income, unemployment compensation, educational assistance, Supplemental Nutrition Assistance Program, monetary value of school lunch, housing subsidy, and energy subsidy. All models include maternal-level controls (indicators for maternal age, state of residence, calendar year, age of the youngest child, age of the oldest child, difference in age between oldest and youngest child, and number of children) and state-level controls (unemployment rate, minimum wage, inflation-adjusted maximum welfare benefit for a family of 4, state-level EITC, implementation of six types of welfare waivers, implementation of any waiver or TANF). In models using the full sample a marital status indicator is included and all controls are interacted with a marital status indicator. Regressions are weighted with maternal survey weights. Standard errors in parentheses are clustered at the state level. The data is from CPS ASEC 1996-2018. The sample is restricted to mothers age 20-64 with children age 0-18. *** $p < 0.01$, ** $p < 0.05$, * $p < 0.10$.

Table A.9:
Remaining Variation in Total Simulated Eligibility

| | All | Single | Married |
|---|---|---|---|
| Model 1 | 0.8856 | 0.9242 | 0.7832 |
| Model 2 | 0.8933 | 0.9289 | 0.7983 |
| Model 3 | 0.9141 | 0.9354 | 0.8442 |
| Model 4 | 0.9306 | 0.9605 | 0.8626 |
| Model 5 | 0.9675 | 0.9766 | 0.9401 |
| Observations | 785,390 | 188,439 | 596,951 |

*Notes:* This table shows adjusted $R^2$ from regressions estimating the effect of various fixed effects variations on marital-status-specific total simulated eligibility. Model 1 includes maternal-level controls (indicators for maternal age, state of residence, calendar year, number of children, youngest child's age, oldest child's age, and difference in age between oldest and youngest child) and state-level controls (unemployment rate, minimum wage, inflation-adjusted maximum welfare benefit for a family of 4, state-level EITC, implementation of six types of welfare waivers, implementation of any waiver or TANF). Model 2 adds interactions between state and youngest child's age, state and oldest child's age, state and difference in age between oldest and youngest child fixed effects to the baseline model. Model 3 adds interactions between state and year fixed effects to the baseline model. Model 4 adds interactions between year and youngest child's age, year and oldest child's age, year and difference in age between oldest and youngest child fixed effects to the baseline model. Model 5 adds state-by-oldest child's age, state-by-youngest child's age, state-by-difference in age between oldest and youngest child, state-by-year, year-by-youngest child's age, year-by-oldest child's age, and year-by-difference in age between oldest and youngest child fixed effects to the baseline model. In models using the full sample a marital status indicator is included and all controls are interacted with a marital status indicator. Regressions are weighted with maternal survey weights. Standard errors are clustered at the state level. The data is from CPS ASEC 1978-2018. The sample is restricted to mothers age 20-64 with children age 0-18.

Table A.10:
Effect of State-Level Characteristics on Medicaid Eligibility Limits

| | (1) | (2) | (3) |
|---|---|---|---|
| Labor Force Participation | -1.29 | -2.39 | -6.34 |
| | (40.31) | (41.02) | (41.94) |
| Hours Worked per Week | -1.86 | -1.78 | -1.14 |
| | (1.14) | (1.23) | (1.13) |
| State Earned Income Credit | -0.01 | -0.04 | -0.11 |
| | (0.26) | (0.26) | (0.27) |
| State Minimum Wage | -1.02 | 0.30 | 0.87 |
| | (3.04) | (3.01) | (2.98) |
| Welfare Benefit ($2020) | 0.03 | 0.02 | 0.01 |
| | (0.04) | (0.04) | (0.04) |
| Major Waiver or TANF | 7.36 | 9.24 | 10.05 |
| | (11.12) | (11.92) | (9.86) |
| Observations | 2,091 | 2,040 | 1,989 |
| Adjusted $R^2$ | 0.87 | 0.87 | 0.86 |
| Mean Y - Baseline | 67 | 67 | 67 |
| Mean Y - Overall | 165 | 165 | 165 |
| Demographic Controls | X | X | X |
| State Fixed Effects | X | X | X |
| Year Fixed Effects | X | X | X |

*Notes:* This table shows results from regressions estimating the effect of state-level characteristics on the maximum Medicaid eligibility limit for children age 0-18. Column 1, 2, and 3 show models using contemporaneous, first order lagged, and second order lagged state-level characteristics, respectively. Demographic controls include fraction of population white, black, married, without high school, with high school completion, with some college, age 0-18, age 20-64, with one child, and with multiple children, as well as income per capita. Standard errors in parentheses are clustered at the state level. The data is from 1977-2017. *** $p < 0.01$, ** $p < 0.05$, * $p < 0.10$.

Table A.11:
Effect of Total Simulated Eligibility on Maternal Labor Supply
Robustness to Marrital Status

| | Last Year Marrital Status | | | Current Year Marrital Status | | |
|---|---|---|---|---|---|---|
| | All | Single | Married | All | Single | Married |
| | Annual Total Earnings | | | | | |
| SIMT | -868 | 2,112** | -1,770** | -769 | 2,271** | -1,701** |
| | ( 612) | ( 964) | ( 776) | ( 610) | ( 945) | ( 745) |
| Observations | 232,770 | 50,740 | 182,030 | 232,770 | 51,217 | 181,553 |
| Adjusted $R^2$ | 0.07 | 0.09 | 0.07 | 0.07 | 0.09 | 0.07 |
| Mean Y - Baseline | 16,992 | 23,298 | 15,774 | 16,992 | 23,335 | 15,674 |
| Mean Y - Overall | 26,210 | 25,629 | 26,381 | 26,210 | 25,778 | 26,338 |
| | Usual Hours Worked per Week | | | | | |
| SIMT | -0.22 | 1.76*** | -0.81** | -0.20 | 1.76*** | -0.80** |
| | ( 0.32) | ( 0.61) | ( 0.36) | ( 0.33) | ( 0.56) | (0.35) |
| Observations | 232,344 | 50,497 | 181,847 | 232,344 | 51,104 | 181,240 |
| Adjusted $R^2$ | 0.09 | 0.11 | 0.07 | 0.09 | 0.11 | 0.07 |
| Mean Y - Baseline | 19.81 | 25.43 | 18.72 | 19.81 | 25.63 | 18.60 |
| Mean Y - Overall | 23.76 | 26.23 | 23.03 | 23.76 | 26.33 | 22.99 |
| | Weeks Worked per Year | | | | | |
| SIMT | -0.08 | 1.58* | -0.59 | -0.08 | 1.62** | -0.60 |
| | ( 0.39) | ( 0.83) | ( 0.40) | ( 0.39) | ( 0.73) | ( 0.40) |
| Observations | 232,770 | 50,740 | 182,030 | 232,770 | 51,217 | 181,553 |
| Adjusted $R^2$ | 0.08 | 0.11 | 0.07 | 0.08 | 0.11 | 0.07 |
| Mean Y - Baseline | 28.28 | 32.79 | 27.41 | 28.28 | 32.89 | 27.32 |
| Mean Y - Overall | 33.07 | 34.42 | 32.67 | 33.07 | 34.58 | 32.62 |
| | Labor Force Participation | | | | | |
| SIMT | -0.01 | 0.02 | -0.01 | -0.01 | 0.02 | -0.01 |
| | ( 0.01) | ( 0.01) | ( 0.01) | ( 0.01) | ( 0.01) | ( 0.01) |
| Observations | 228,229 | 49,612 | 178,617 | 228,229 | 50,102 | 178,127 |
| Adjusted $R^2$ | 0.06 | 0.07 | 0.06 | 0.06 | 0.07 | 0.06 |
| Mean Y - Baseline | 0.58 | 0.67 | 0.56 | 0.58 | 0.66 | 0.56 |
| Mean Y - Overall | 0.70 | 0.75 | 0.69 | 0.70 | 0.75 | 0.69 |

*Notes:* This table shows results from regressions estimating the effect of marital-status-specific total simulated eligibility on maternal usual hours worked per week last year, weeks worked last year, and labor force participation last week. Usual hours worked per week and weeks worked last year include zeros. All models include maternal-level controls (indicators for maternal age, state of residence, calendar year, age of the youngest child, age of the oldest child, difference in age between oldest and youngest child, and number of children) and state-level controls (unemployment rate, minimum wage, inflation-adjusted maximum welfare benefit for a family of 4, state-level EITC, implementation of six types of welfare waivers, implementation of any waiver or TANF). In models using the full sample a marital status indicator is included and all controls are interacted with a marital status indicator. Regressions are weighted with maternal survey weights. Standard errors in parentheses are clustered at the state level. The data is from CPS ASEC 1977-2018. The sample is restricted to mothers age 20-64 with children age 0-18. *** $p < 0.01$, ** $p < 0.05$, * $p < 0.10$.

Table A.12:
Effect of Total Simulated Eligibility on Maternal Labor Supply
Robustness to Imputation & Alternative Reference Period

| | All | Single | Married |
|---|---|---|---|
| | Non-Imputed Usual Hours Worked per Week | | |
| SIMT | 0.16 | 1.87*** | -0.39 |
| | (0.20) | (0.38) | (0.25) |
| Observations | 702,395 | 169,915 | 532,480 |
| Adjusted $R^2$ | 0.07 | 0.08 | 0.06 |
| Mean Y - Baseline | 22.29 | 25.08 | 21.37 |
| Mean Y - Overall | 25.84 | 28.09 | 25.06 |
| | Non-Imputed Weeks Worked per Year | | |
| SIMT | 0.12 | 1.45** | -0.31 |
| | (0.21) | (0.62) | (0.19) |
| Observations | 607,396 | 147,572 | 459,824 |
| Adjusted $R^2$ | 0.08 | 0.09 | 0.07 |
| Mean Y - Baseline | 29.49 | 29.04 | 29.64 |
| Mean Y - Overall | 32.79 | 34.03 | 32.35 |
| | Non-Imputed Labor Force Participation | | |
| SIMT | -0.00 | 0.03*** | -0.01*** |
| | (0.00) | (0.01) | (0.00) |
| Observations | 786,916 | 187,341 | 599,575 |
| Adjusted $R^2$ | 0.08 | 0.08 | 0.07 |
| Mean Y - Baseline | 0.51 | 0.63 | 0.49 |
| Mean Y - Overall | 0.68 | 0.74 | 0.66 |
| | Hours Worked Last Week | | |
| SIMT | -0.30** | 1.40*** | -0.86*** |
| | (0.15) | (0.28) | (0.23) |
| Observations | 788,824 | 187,830 | 600,994 |
| Adjusted $R^2$ | 0.09 | 0.09 | 0.08 |
| Mean Y - Baseline | 15.22 | 19.34 | 14.29 |
| Mean Y - Overall | 21.47 | 23.42 | 20.82 |

*Notes:* This table shows results from regressions estimating the effect of marital-status-specific total simulated eligibility on maternal usual hours worked per week last year, weeks worked last year, and labor force participation last week dropping observations with imputed values and reweighting the sample with inverse-probability weights as well as hours worked last week. Usual hours worked per week, hours worked last week, and weeks worked last year include zeros. All models include maternal-level controls (indicators for maternal age, state of residence, calendar year, age of the youngest child, age of the oldest child, difference in age between oldest and youngest child, and number of children) and state-level controls (unemployment rate, minimum wage, inflation-adjusted maximum welfare benefit for a family of 4, state-level EITC, implementation of six types of welfare waivers, implementation of any waiver or TANF). In models using the full sample a marital status indicator is included and all controls are interacted with a marital status indicator. Regressions are weighted with maternal survey weights. Standard errors in parentheses are clustered at the state level. The data is from CPS ASEC 1977-2018. The sample is restricted to mothers age 20-64 with children age 0-18. *** $p < 0.01$, ** $p < 0.05$, * $p < 0.10$.

## Table A.13:
## Effect of Total Simulated Eligibility on Maternal Annual Earnings
## Robustness to Imputation & Alternative Topcodes

| | All | Single | Married |
|---|---|---|---|
| | | Non-Imputed Observations | |
| SIMT | -515 | 1,900*** | -1,290** |
| | ( 386) | ( 360) | ( 567) |
| Observations | 703,388 | 167,105 | 536,283 |
| Adjusted $R^2$ | 0.07 | 0.08 | 0.07 |
| Mean Y - Baseline | 13,114 | 17,596 | 12,023 |
| Mean Y - Overall | 23,903 | 22,953 | 24,225 |
| | | Rank Proximity Swap Topcodes | |
| SIMT | -528 | 1,961*** | -1,331** |
| | ( 361) | ( 327) | ( 537) |
| Observations | 785,386 | 188,438 | 596,948 |
| Adjusted $R^2$ | 0.06 | 0.07 | 0.06 |
| Mean Y - Baseline | 13,531 | 18,097 | 12,426 |
| Mean Y - Overall | 24,912 | 23,712 | 25,320 |
| | | Cell Means Replacement Topcodes | |
| SIMT | -464 | 1,912*** | -1,229** |
| | ( 374) | ( 339) | ( 554) |
| Observations | 785,386 | 188,438 | 596,948 |
| Adjusted $R^2$ | 0.07 | 0.08 | 0.07 |
| Mean Y - Baseline | 13,514 | 18,011 | 12,427 |
| Mean Y - Overall | 24,918 | 23,690 | 25,336 |

*Notes:* This table shows results from regressions estimating the effect of marital-status-specific total simulated eligibility on maternal annual total earnings ($2020) last year dropping observations with imputed earnings and reweighting the sample with inverse-probability weights as well as using alternative topcodes. All models include maternal-level controls (indicators for maternal age, state of residence, calendar year, age of the youngest child, age of the oldest child, difference in age between oldest and youngest child, and number of children) and state-level controls (unemployment rate, minimum wage, inflation-adjusted maximum welfare benefit for a family of 4, state-level EITC, implementation of six types of welfare waivers, implementation of any waiver or TANF). In models using the full sample a marital status indicator is included and all controls are interacted with a marital status indicator. Regressions are weighted with maternal survey weights. Standard errors in parentheses are clustered at the state level. The data is from CPS ASEC 1977-2018. The sample is restricted to mothers age 20-64 with children age 0-18. *** $p < 0.01$, ** $p < 0.05$, * $p < 0.10$.

Table A.14:
Effect of Total Simulated Eligibility on Maternal Labor Supply
Robustness to Maternal Eligibility

| | Single Mothers | | | Married Mothers | | |
|---|---|---|---|---|---|---|
| | (1) | (2) | (3) | (1) | (2) | (3) |
| | Total Annual Earnings | | | | | |
| SIMT | 1,792*** | 1,792*** | 1,752*** | -1,120** | -1,120** | -1,226** |
| | ( 337) | ( 337) | ( 349) | ( 532) | ( 532) | ( 547) |
| Observations | 188,438 | 188,438 | 176,632 | 596,948 | 596,948 | 539,307 |
| Adjusted $R^2$ | 0.08 | 0.08 | 0.07 | 0.07 | 0.07 | 0.06 |
| Mean Y - Baseline | 17,985 | 17,985 | 18,536 | 12,417 | 12,417 | 12,607 |
| Mean Y - Overall | 23,658 | 23,658 | 24,534 | 25,291 | 25,291 | 25,744 |
| | Usual Hours Worked per Week | | | | | |
| SIMT | 1.63*** | 1.63*** | 1.59*** | -0.49** | -0.49** | -0.73*** |
| | (0.30) | (0.30) | (0.31) | (0.23) | (0.23) | (0.26) |
| Observations | 188,438 | 188,438 | 176,632 | 596,948 | 596,948 | 539,307 |
| Adjusted $R^2$ | 0.08 | 0.08 | 0.07 | 0.07 | 0.07 | 0.06 |
| Mean Y - Baseline | 24.74 | 24.74 | 25.15 | 19.11 | 19.11 | 19.14 |
| Mean Y - Overall | 28.05 | 28.05 | 28.54 | 24.74 | 24.74 | 24.94 |
| | Weeks Worked per Year | | | | | |
| SIMT | 1.52*** | 1.52*** | 1.48*** | -0.29 | -0.29 | -0.48** |
| | (0.44) | (0.44) | (0.44) | (0.18) | (0.18) | (0.21) |
| Observations | 188,438 | 188,438 | 176,632 | 596,948 | 596,948 | 539,307 |
| Adjusted $R^2$ | 0.11 | 0.11 | 0.09 | 0.09 | 0.09 | 0.08 |
| Mean Y - Baseline | 27.20 | 27.20 | 27.97 | 22.39 | 22.39 | 22.96 |
| Mean Y - Overall | 33.03 | 33.03 | 33.86 | 30.99 | 30.99 | 31.62 |
| | Labor Force Participation | | | | | |
| SIMT | 0.03*** | 0.03*** | 0.03*** | -0.01*** | -0.01*** | -0.02*** |
| | (0.01) | (0.01) | (0.01) | (0.00) | (0.00) | (0.00) |
| Observations | 187,830 | 187,830 | 175,992 | 600,994 | 600,994 | 542,868 |
| Adjusted $R^2$ | 0.08 | 0.08 | 0.06 | 0.07 | 0.07 | 0.06 |
| Mean Y - Baseline | 0.63 | 0.63 | 0.65 | 0.49 | 0.49 | 0.51 |
| Mean Y - Overall | 0.73 | 0.73 | 0.75 | 0.66 | 0.66 | 0.67 |

*Notes:* This table shows results from regressions estimating the effect of marital-status-specific total simulated eligibility on maternal annual total earnings ($2020) last year, usual hours worked per week last year, weeks worked last year, and labor force participation last week). Usual hours worked per week and weeks worked last year include zeros. Column 1 reports estimates using maternal eligibility (women age 15-44) for zero-year old children. Column 3 reports estimates using maternal eligibility (mothers age 15-44 with children of age zero) for zero-year old children. Column 4 reports estimates dropping children age zero. All models include maternal-level controls (indicators for maternal age, state of residence, calendar year, age of the youngest child, age of the oldest child, difference in age between oldest and youngest child, and number of children) and state-level controls (unemployment rate, minimum wage, inflation-adjusted maximum welfare benefit for a family of 4, state-level EITC, implementation of six types of welfare waivers, implementation of any waiver or TANF). In models using the full sample a marital status indicator is included and all controls are interacted with a marital status indicator. Regressions are weighted with maternal survey weights. Standard errors in parentheses are clustered at the state level. The data is from CPS ASEC 1977-2018. The sample is restricted to mothers age 20-64 with children age 0-18. *** $p < 0.01$, ** $p < 0.05$, * $p < 0.10$.

Table A.15:
Effect of Total Simulated Eligibility on Maternal Labor Supply
Robustness to Simulated Eligibility Type

| | Single Mothers | | | Married Mothers | | |
|---|---|---|---|---|---|---|
| | (1) | (2) | (3) | (1) | (2) | (3) |
| | | | Total Annual Earnings | | | |
| SIMT | 1,454*** | 1,437*** | 1,410*** | -1,048** | -1,039** | -1,037* |
| | ( 345) | ( 345) | ( 362) | ( 447) | ( 447) | ( 520) |
| Observations | 188,438 | 188,438 | 188,438 | 596,948 | 596,948 | 596,948 |
| Adjusted $R^2$ | 0.09 | 0.09 | 0.09 | 0.07 | 0.07 | 0.07 |
| Mean Y - Baseline | 17,985 | 17,985 | 17,985 | 12,417 | 12,417 | 12,417 |
| Mean Y - Overall | 23,658 | 23,658 | 23,658 | 25,291 | 25,291 | 25,291 |
| | | | Usual Hours Worked per Week | | | |
| SIMT | 1.90*** | 1.89*** | 1.75*** | -0.28 | -0.27 | -0.30 |
| | (0.29) | (0.29) | (0.31) | (0.19) | (0.19) | (0.23) |
| Observations | 188,438 | 188,438 | 188,438 | 596,948 | 596,948 | 596,948 |
| Adjusted $R^2$ | 0.09 | 0.09 | 0.09 | 0.07 | 0.07 | 0.07 |
| Mean Y - Baseline | 24.74 | 24.74 | 24.74 | 19.11 | 19.11 | 19.11 |
| Mean Y - Overall | 28.05 | 28.05 | 28.05 | 24.74 | 24.74 | 24.74 |
| | | | Weeks Worked per Year | | | |
| SIMT | 1.77*** | 1.77*** | 1.60*** | -0.22 | -0.21 | -0.28 |
| | (0.42) | (0.42) | (0.46) | (0.15) | (0.15) | (0.18) |
| Observations | 188,438 | 188,438 | 188,438 | 596,948 | 596,948 | 596,948 |
| Adjusted $R^2$ | 0.12 | 0.12 | 0.12 | 0.09 | 0.09 | 0.09 |
| Mean Y - Baseline | 27.20 | 27.20 | 27.20 | 22.39 | 22.39 | 22.39 |
| Mean Y - Overall | 33.03 | 33.03 | 33.03 | 30.99 | 30.99 | 30.99 |
| | | | Labor Force Participation | | | |
| SIMT | 0.04*** | 0.04*** | 0.03*** | -0.01** | -0.01** | -0.01** |
| | (0.01) | (0.01) | (0.01) | (0.00) | (0.00) | (0.00) |
| Observations | 187,830 | 187,830 | 187,830 | 600,994 | 600,994 | 600,994 |
| Adjusted $R^2$ | 0.08 | 0.08 | 0.08 | 0.07 | 0.07 | 0.07 |
| Mean Y - Baseline | 0.63 | 0.63 | 0.63 | 0.49 | 0.49 | 0.49 |
| Mean Y - Overall | 0.73 | 0.73 | 0.73 | 0.66 | 0.66 | 0.66 |

*Notes:* This table shows results from regressions estimating the effect of marital-status-specific total simulated eligibility on maternal annual total earnings ($2020) last year, usual hours worked per week last year, weeks worked last year, and labor force participation last week). Usual hours worked per week and weeks worked last year include zeros. Column 1,2, and 3 reports estimates using consumer price index, regional consumer price index, and wages respectively to construct the marital-status-specific total simulated fixed eligibility. All models include maternal-level controls (indicators for maternal age, state of residence, calendar year, age of the youngest child, age of the oldest child, difference in age between oldest and youngest child, and number of children) and state-level controls (unemployment rate, minimum wage, inflation-adjusted maximum welfare benefit for a family of 4, state-level EITC, implementation of six types of welfare waivers, implementation of any waiver or TANF). In models using the full sample a marital status indicator is included and all controls are interacted with a marital status indicator. Regressions are weighted with maternal survey weights. Standard errors in parentheses are clustered at the state level. The data is from CPS ASEC 1977-2018. The sample is restricted to mothers age 20-64 with children age 0-18. *** $p < 0.01$, ** $p < 0.05$, * $p < 0.10$.

# B Medicaid Eligibility

## B.1 Medicaid Legislation

This section describes the underlying legislative rules used to calculate Medicaid eligibility for children in the U.S. for the period 1977-2017. Medicaid eligibility is imputed using the calculator from East et al. (2023).[33] Eligibility calculations can be broadly categorized into two groups - before and after the Personal Responsibility and Work Opportunity Act (PRWORA). To determine Medicaid eligibility of children, rules under Aid to Families with Dependent Children (AFDC), state-optional programs (AFDC-Unemployed Parents (AFDC-UP), Ribicoff children, Medically Needy), Medicaid "Section 1931", State Children's Health Insurance Program (SCHIP), as well as federal and state Medicaid expansions are used.[34] In general eligibility is imputed using applicable rules based on date of eligibility determination, child's age, child's birthday, family structure, family income, and information on parental employment.

**Eligibility Calculations before PRWORA (1977-1996)**

***Eligibility under AFDC***

Historically, Medicaid eligibility was restricted to children in families receiving cash welfare benefits. To determine eligibility under AFDC, it is assumed that the child care dedication is not used by eligible families and that the parent has spent one month working. The first set of rules to assess whether the family is eligible for AFDC, are rules regarding family income and earned income disregards. Total family income is calculated by summing all sources of income except public assistance or welfare of each parent. To determine if any earned income disregards are applicable, months spent working are compared to number of months that disregards are allowed based on state rules. The applicable disregards are then calculated by using state rules. In order to be financially eligible for AFDC, the child's family has to satisfy three tests. First, the family must be eligible for a non-zero AFDC benefit amount based on state rules, monthly total family income, and family size. Second, total family income less applicable disregards must be below the state need's standard. Third, total family income must not exceed a given percentage of the state need's standard. Special rules are implemented for Connecticut and Minnesota. As a second set of rules, the family has to satisfy two family structure requirements. First, eligibility under AFDC requires the child to reside in a single-parent family. Second, the child has to be age 0-17 at date of eligibility determination and either a primary or subfamily member, but not a head or spouse of primary family or subfamily.

[33] This section summarizes the most important steps to determine Medicaid eligibility. See appendix and calculator documentation in Miller and Wherry (2019) and East et al. (2023) for more information about legislative information, data sources and methodology used to impute eligibility.

[34] Table B.1 shows the major mandatory and state optional legislations that affected Medicaid eligibility of children during the analysis period.

***Eligibility under AFDC-UP***

Prior to the federal mandate effective in October 1990, AFDC-UP - a state-optional program - extended eligibility to children in two-parent families where the primary earner was unemployed. To be classified as unemployed, the parent must work less than 100 hours per month. A child is assumed to be eligible under AFDC-UP if AFDC-UP program was effective in state and year, child's family is financially eligible for AFDC, maximum hours worked by any individual in the family do not exceed 1200 per year, and the child resides in a family with married parents.

***Eligibility under Medically Needy Program***

The medically needy program provides states the option to extend Medicaid eligibility to individuals with high medical expenses whose income exceeds the maximum income eligibility threshold, but who satisfy all other eligibility criteria for Medicaid. Income limits could be set no higher than 133% of the state's needs standard for AFDC. However families could use the medical expenditures to reduce the applicable income through spent-down provisions. Since there is no information about medical expenditures in the CPS, the eligibility limits are set to the Medically Needy levels in states with this program as an approximation. A child is eligible under the medically needy program if the family's income except public assistance or welfare is below the applicable eligibility thresholds.

### *Eligibility under the Ribicoff Children Program*

Under the Ribicoff Children Program, states are allowed to cover children who would qualify for cash welfare program given income criteria alone but who do not qualify based on family structure. Hence a child is eligible for Medicaid under the Ribicoff Children Program if the Ribicoff Children Program is present in state and year, child's family is income eligible for AFDC, and the child lives in a family with married parents. In addition, the child is eligible under the Ribicoff Children Program if federally-mandated expansions of Ribicoff Children Program are applicable (Deficit Reduction Act, 1984; Omnibus Budget Reconciliation Act, 1987).

### *Poverty-related Eligibility*

Beginning in 1984 states were required or given the option to expand Medicaid eligibility for children living in families with incomes below the eligibility limit. To impute eligibility under federal and state expansions child care deduction is assumed to be zero. All sources of income except public assistance or welfare of each parent are summed up to calculate the total family income. A child is eligible for Medicaid if the total family income minus the work expense deduction used in the net income calculations under AFDC is less than the applicable federal or state eligibility level in the given state, year, and age.

## Eligibility Calculations after PRWORA (1997-2017)

### *Eligibility under Medicaid "Section 1931"*

Medicaid "Section 1931" requires states to provide Medicaid coverage to children in families who meet eligibility requirements under AFDC and AFDC-UP effective on July 16, 1996 in the state of residence. To impute eligibility under "Section 1931", it is assumed that child care deduction is not used by eligible families and that the parent has spent one month working. Eligibility of a child is then determined using the eligibility rules for AFDC programs in effect for the state in July 1996. The procedure to calculate eligibility under AFDC and AFDC-UP is explained above. This calculator does not incorporate state optional "Section 1931" rules under which states have the option to set income and asset standards differently from those in effect under state AFDC program on July 16, 1996. Since eligibility requirements are presumably less restrictive under all other eligibility pathways after the welfare reform, omitting optional "Section 1931" eligibility will not bias the eligibility estimates.

***Eligibility under Separate State Programs under SCHIP***

Balanced Budget Act of 1997 allowed states to create separate state programs. Eligibility under separate state programs is imputed assuming that child care deductions and child support income are zero. To obtain the total family income all sources of income except public assistance or welfare of each parent are summed up. A child is eligible if the total family income minus the state- and SCHIP-specific work expense deduction per worker is less than SCHIP eligibility limit in the given state, year, and age.

***Poverty-related and Targeted Medicaid Eligibility***

After PRWORA two different pathways can determine eligibility under expansion-related rules. The poverty-related pathway is defined by a series of federal and state Medicaid expansions which extended eligibility for certain ages and income levels. The second path is given by targeted Medicaid expansions embedded in the Balanced Budget Act of 1997 which allow states to expand the state Medicaid programs. Technically, different income disregards are applied for the two pathways. However the calculator uses poverty-related disregards for both pathways. To impute eligibility, child care deductions and child support income are assumed to be zero. Total family income is calculated by summing all sources of income except public assistance or welfare of each parent. If the total family income minus the state- and Medicaid-specific work expense deduction per worker are less than the corresponding cutoffs in a given state, year and age, the child is assumed to be eligible.

## B.2 Eligibility Imputation

This section explains how Medicaid eligibility is imputed based on variables available in CPS ASEC. To impute Medicaid eligibility, I use a calculator that incorporates state and federal legislation based on rules for a given year, state of residence, age of the child, and family characteristics including family income and family structure.[35] A child is considered eligible for public health insurance if the child's family meets eligibility requirements for AFDC, one of the state-optional programs (AFDC-UP, Ribicoff children program, Medicaid's medically needy program), or federal and state-optional Medicaid expansions.

In survey year *t*, CPS ASEC provides information about income from calendar year *t-1*, family structure as of March of calendar year *t*, and age of the child as of March of calendar year *t*. Hence, to calculate eligibility in calendar year *t*, I use data on income from calendar year *t*, family structure from calendar year *t+1*, and adjust the age of the child accordingly. Depending on the birth month and calendar month of eligibility determination, some children are treated as if they were the same age and others are treated as if they were a year younger during the previous calendar year.[36] Child's age $a$ is therefore defined as age at calendar month of eligibility determination. I construct two Medicaid eligibility measures - contemporaneous to outcome variables measured as of previous calendar year ("last year eligibility") and contemporaneous to outcome variables measured as of interview month ("last week eligibility"). To obtain Medicaid eligibility contemporaneous to outcome variables measured as of previous calendar year (e.g., insurance coverage, usual hours worked, weeks worked), I calculate eligibility during each month of the given year and use the average eligibility across all months in that year. Medicaid eligibility contemporaneous to outcome variables measured as of survey date (e.g., marital status, educational attainment) or last week (e.g., labor force participation) is obtained by calculating eligibility during March of the given year.

[35] I use a calculator from East et al. (2023) which allows to impute eligibility for children age 0-18 for years 1977-2017.

[36] I randomly assign birth month because eligibility is imputed using the calendar month of eligibility determination and CPS does not provide the birth month of an individual.

To obtain the correct family structure and measure of total family income according to rules determining Medicaid eligibility, I construct nuclear families within a household. A family unit is defined as a parent, spouse if present, and children. First spouses within a household are linked and then parents are linked to their children within a household. I obtain family income by adding parental income within the nuclear family. To determine eligibility of a child in calendar year $t$, I follow the legislative rules to calculate Medicaid eligibility and first divide family income (except applicable disregards) by the corresponding poverty guideline for the family size of the nuclear family, in state $s$, and in calendar year $t$.[37] I then compare this ratio to the eligibility limit for a child of age $a$, in state $s$, and calendar year $t$. Since the eligibility cutoffs depend on the age of the child, the number of children in the family that are eligible for Medicaid may vary for families with the same income and number of children, but with children of different ages.

To check how well the calculator estimates eligibility of children, I examine the percentage of non-eligible children reporting Medicaid coverage or living in families receiving welfare payments since these children should be eligible for Medicaid. For the period 1977-2017, 3.78% of children imputed to be not eligible, report coverage by Medicaid and 0.56% of children live in families where a parent reports receiving cash benefits under the AFDC program, although these children are not eligible for Medicaid based on the calculator.

[37] Poverty guidelines depend on family size, year, and state of residence. All states except Alaska and Hawaii share the same poverty guidelines.

## B.3 Classic Simulated Eligibility Measure

The simulated eligibility is constructed by using all children of each age in each calendar year across the full sample period. Using this national data set, I construct simulated eligibility measures which vary at the state, calendar year, child's age, and maternal marital status. Following Ham and Shore-Sheppard (2005b), the simulated eligibility is obtained by using all children in the national data set except from the state for which the simulated eligibility is calculated. Each state-year-age-marital-status simulated eligibility measure is hence the fraction of children in the national data set in state *s* except children from state *s*, calendar year *t* when outcome of interest is measured, of age *a*, and maternal marital status *m* who would be eligible for Medicaid given the rules in each state *s*, calendar year *t*, and age *a*. Formally, the simulated eligibility for a given state, calendar year, child's age, and marital status of the mother is given by the following equation:

$$SIM_{stam} = \frac{\sum_{i=1}^{k_{\tilde{s}}} w_{i\tilde{s}tam} \cdot e_{i\tilde{s}tam}}{\sum_{i=1}^{k_{\tilde{s}}} w_{i\tilde{s}tam}} \tag{2}$$

where $k_{\tilde{s}}$ is the number of children in the national data set excluding children from state *s*, of age *a*, maternal marital status *m*, and in calendar year *t*. $e_{i\tilde{s}tar}$ and $w_{i\tilde{s}tar}$ are individual-level eligibility and CPS ASEC weight of child *i*, not residing in state *s*, in calendar year *t*, of age *a*, and maternal marital status *m*. Each child-specific simulated eligibility measure (last year and last week simulated eligibility) is then merged to each child based on child's state of residence, calendar year when outcome of interest is measured, age, and martial status of the mother.

## B.4 Family-level Simulated Eligibility Measure

To construct family's total simulated eligibility (last year and last week), I sum the simulated eligibility fractions $SIM_{stam}$ (last year and last week, respectively) across all chil-dren in a family.[38] To facilitate notation, I define a parent type by the number of children of any age between 0 and 18 of that parent - $j := (n_{j_0}, ..., n_{j_{18}})$ where $n_{j_a}$ is the number of children of age *a* of parent *j*.[39] Formally, the total simulated eligibility of a parent type *j*, state *s*, calendar year *t*, and martial status *m* is given by the following equation:

$$SIMT_{jstm} = SIMT_{(n_{j_0},...,n_{j_{18}})stm} = \sum_{a=0}^{18} SIM_{stam} * n_{j_a} \tag{3}$$

where $SIM_{stam}$ is the simulated eligibility measure defined in equation 2 and $n_{j_a}$ is the number of children of age *a* of parent *j*. Hence the level of variation of total simulated eligibility within a given state, year, marital status group, and number of children is the distribution of possible combinations of child's ages. In general, mothers in the same state, year, marital status group, and of the same type (same number of children and same age of each child) are characterized by the same total simulate eligibility measure.

[38]Consider for example a family with two children age 3 and 5. The child-specific simulated eligibility measure of the first and second child is 0.5 and 0.6 respectively. The family's total simulated eligibility is hence 1.1.

[39]Take for instance a parent with two one-year old, one three-year old, and two five-year old children. The parent type is hence given by the vector (0,2,0,1,0,2,0,...,0).

Table B.1:
Medicaid and CHIP Legislation Expanding Eligibility of Children 1977-2017

| Year | Legislation | Mandatory Expansion | State Option |
|---|---|---|---|
| 1984 | Deficit Reduction Act | Coverage of children under age 5 born after September 30, 1983 whose families are income and resource eligible for AFDC | |
| 1986 | Omnibus Budget Reconciliation Act | | Increase age level by 1 year each FY for all children under age 5 with incomes below 100% FPL. Infants in families with incomes below 100% FPL |
| 1987 | Omnibus Budget Reconciliation Act | Coverage of children under age 7 born after September 30, 1983 whose families are income and resource eligible for AFDC | Coverage of infants in families with incomes below 185% FPL and children under age 2, 3, 4, or 5 and born after September 30, 1983 in families with incomes below 100% FPL. Coverage of children under age 8 born after September 30, 1983 whose families are income and resource eligible for AFDC and children under age 8 born after September 30, 1983 with incomes below 100% FPL. |
| 1988 | Medicare Catastrophic Coverage Act | Coverage of infants in families with incomes below 75% FPL (1-Jul-89) and infants in families with incomes below 100% FPL (1-Jul-90) | Coverage of children up to eight years of age with family incomes below 75% FPL |
| 1988 | Family Support Act of 1988 | Extension to twelve months transitional Medicaid coverage to families leaving AFDC rolls due to earnings from work. Coverage of two-parent unemployed families meeting state AFDC income and resource standards. | |
| 1989 | Omnibus Budget Reconciliation Act | Coverage of children under age 6 with family incomes below 133% FPL | |
| 1990 | Omnibus Budget Reconciliation Act | Coverage of children under age 19 born after September 30, 1983 with incomes below 100% FPL. | |
| 1996 | Personal Responsibility and Work Opportunity Act | Coverage of families meeting AFDC eligibility standards as of July 16, 1996 ("Section 1931") | Coverage of higher-income families. |
| 1997 | Balanced Budget Act | | Coverage of children under age 19 in families with incomes below 200% FPL or higher |

*Notes:* Buchmueller et al. (2016) and Miller and Wherry (2019)

# C Occupation Classification

This section describes the occupational coding scheme. Following vom Lehn et al. (2022), I use a modified version of the 1990 Census Bureau occupational classification created by Autor and Dorn (2013). This coding scheme offers a consistent and balanced panel of occupations. The categories of one-, two- and three-digit occupations are shown in table C.1.

Table C.1:
Occupational Classification

| One-Digit Occupation | Two-Digit Occupation | Occupational Codes |
|---|---|---|
| Managerial and Professional Specialty Occupations | Executive, Administrative, and Managerial Occupations | 3-22 |
| | Management Related Occupations | 23-37 |
| | Professional Specialty Occupations | 43-199 |
| Technical, Sales and Administrative Support Occupations | Technicians and Related Support Occupations | 203-235 |
| | Sales Occupations | 243-283 |
| | Administrative Support Occupations | 303-389 |
| Service Occupations | Private Household Occupations | 405-408 |
| | Protective Service Occupations | 415-427 |
| | Other Service Occupations | 433-472 |
| Farming, Forestry and Fishing Occupations | Farm Operators and Managers | 473-475 |
| | Other Agricultural and Related Occupations | 479-498 |
| Precision Production, Craft and Repair Occupations | Mechanics and Repairers | 503-549 |
| | Construction Trades | 558-599 |
| | Extractive Occupations | 614-617 |
| | Precision Production Occupations | 628-699 |
| Operators, Fabricators and Laborers | Machine Operators, Assemblers, and Inspectors | 703-799 |
| | Transportation and Material Moving Occupations | 803-889 |